%&plain
%
\let\includefigures=\iftrue
%
% the following is to use blackboard bold fonts --
%\let\useblackboard=\iftrue
%
% activate this if you don't have them.
%\let\useblackboard=\iffalse
%
% You might also need to remove this line.
\newfam\black
\input harvmac.tex
\input rotate
\input epsf
\input xyv2
\noblackbox
\includefigures
\message{If you do not have epsf.tex (to include figures),}
\message{change the option at the top of the tex file.}
\def\figin{\epsfcheck\figin}\def\figins{\epsfcheck\figins}
\def\epsfcheck{\ifx\epsfbox\UnDeFiNeD
\message{(NO epsf.tex, FIGURES WILL BE IGNORED)}
\gdef\figin##1{\vskip2in}\gdef\figins##1{\hskip.5in}% blank space instead
\else\message{(FIGURES WILL BE INCLUDED)}%
\gdef\figin##1{##1}\gdef\figins##1{##1}\fi}
\def\DefWarn#1{}
\def\N{{\cal N}}
\def\figinsert{\goodbreak\midinsert}
\def\ifig#1#2#3{\DefWarn#1\xdef#1{fig.~\the\figno}
\writedef{#1\leftbracket fig.\noexpand~\the\figno}%
\figinsert\figin{\centerline{#3}}\medskip\centerline{\vbox{\baselineskip12pt
\advance\hsize by -1truein\noindent\footnotefont{\bf
Fig.~\the\figno:} #2}}
\bigskip\endinsert\global\advance\figno by1}
%%%
\else
\def\ifig#1#2#3{\xdef#1{fig.~\the\figno}
\writedef{#1\leftbracket fig.\noexpand~\the\figno}%
%\figinsert\figin{\centerline{#3}}\medskip\centerline{\vbox{\baselineskip12pt
%\advance\hsize by -1truein\noindent\footnotefont{\bf Fig.~\the\figno:} #2}}
%\bigskip\endinsert
\global\advance\figno by1} \fi
\def\yboxit#1#2{\vbox{\hrule height #1 \hbox{\vrule width #1
\vbox{#2}\vrule width #1 }\hrule height #1 }}
\def\fillbox#1{\hbox to #1{\vbox to #1{\vfil}\hfil}}
\def\ybox{{\lower 1.3pt \yboxit{0.4pt}{\fillbox{8pt}}\hskip-0.2pt}}

\def\rightarrowbox#1#2{
  \setbox1=\hbox{\kern#1{${ #2}$}\kern#1}
  \,\vbox{\offinterlineskip\hbox to\wd1{\hfil\copy1\hfil}
    \kern 3pt\hbox to\wd1{\rightarrowfill}}}

\def\half{{1\over 2}}
\def\Tr{{{\rm Tr~ }}}

\def\CF{{\cal F}}

\def\CM{{\cal M}}
\def\CN{{\cal N}}
\def\CO{{\cal O}}

\def\CS{{\cal S}}

\def\tilde{\widetilde}

\def\II{\relax{I\kern-.10em I}}

\def\bar{\overline}

\def\IZ{\relax\ifmmode\mathchoice
{\hbox{\cmss Z\kern-.4em Z}}{\hbox{\cmss Z\kern-.4em Z}}
{\lower.9pt\hbox{\cmsss Z\kern-.4em Z}} {\lower1.2pt\hbox{\cmsss
Z\kern-.4em Z}}\else{\cmss Z\kern-.4em Z}\fi}
\def\IB{\relax{\rm I\kern-.18em B}}
\def\IC{{\relax\hbox{$\inbar\kern-.3em{\rm C}$}}}
\def\ID{\relax{\rm I\kern-.18em D}}
\def\IE{\relax{\rm I\kern-.18em E}}
\def\IF{\relax{\rm I\kern-.18em F}}
\def\IG{\relax\hbox{$\inbar\kern-.3em{\rm G}$}}
\def\IGa{\relax\hbox{${\rm I}\kern-.18em\Gamma$}}
\def\IH{\relax{\rm I\kern-.18em H}}
\def\II{\relax{\rm I\kern-.18em I}}
\def\IK{\relax{\rm I\kern-.18em K}}
\def\IN{\relax{\rm I\kern-.18em N}}
\def\IP{\relax{\rm I\kern-.18em P}}
%\def\IX{\relax{\rm X\kern-.01em X}}
%this doesn't work

\def\hat{\widehat}

\def\inbar{\,\vrule height1.5ex width.4pt depth0pt}
\def\mod{{\rm\; mod\;}}

\font\cmss=cmss10 \font\cmsss=cmss10 at 7pt
\def\IR{\relax{\rm I\kern-.18em R}}

\def\tilde{\widetilde}

\def\lp10{l_P^{10}}
\def\lp11{l_P^{11}}
\def\R11{R_{11}}

\def\R{{\cal R}}
\def\Z{{\bf Z}}

%
% References
%

%\GopakumarKI
\lref\GopakumarKI{ R.~Gopakumar and C.~Vafa, ``On the gauge
theory/geometry correspondence,'' Adv.\ Theor.\ Math.\ Phys.\
{\bf 3}, 1415 (1999) [arXiv:hep-th/9811131].
%%CITATION = HEP-TH 9811131;%%
}

%\VafaWI
\lref\VafaWI{ C.~Vafa, ``Superstrings and topological strings at
large N,'' J.\ Math.\ Phys.\  {\bf 42}, 2798 (2001)
[arXiv:hep-th/0008142].
%%CITATION = HEP-TH 0008142;%%
}

%\FerrariKQ
\lref\FerrariKQ{ F.~Ferrari, ``Quantum parameter space and double
scaling limits in N = 1 super  Yang-Mills theory,''
arXiv:hep-th/0211069.
%%CITATION = HEP-TH 0211069;%%
}

%\VenezianoAH
\lref\VenezianoAH{ G.~Veneziano and S.~Yankielowicz, ``An
Effective Lagrangian For The Pure N=1 Supersymmetric Yang-Mills
Theory,'' Phys.\ Lett.\ B {\bf 113}, 231 (1982).
%%CITATION = PHLTA,B113,231;%%
}

%\cite{Aharony:1999ti}
\lref\agmoo{ O.~Aharony, S.~S.~Gubser, J.~M.~Maldacena, H.~Ooguri
and Y.~Oz, ``Large N field theories, string theory and gravity,''
Phys.\ Rept.\  {\bf 323}, 183 (2000) arXiv:hep-th/9905111.}
%%CITATION = HEP-TH 9905111;%%

%\NovikovEE
\lref\NovikovEE{V.~A.~Novikov, M.~A.~Shifman, A.~I.~Vainshtein and
V.~I.~Zakharov, ``Instanton Effects In Supersymmetric Theories,''
Nucl.\ Phys.\ B {\bf 229}, 407 (1983).
%%CITATION = NUPHA,B229,407;%%
}

%\ArgyresJJ
\lref\ArgyresJJ{ P.~C.~Argyres and M.~R.~Douglas, ``New phenomena
in SU(3) supersymmetric gauge theory,'' Nucl.\ Phys.\ B {\bf 448},
93 (1995) [arXiv:hep-th/9505062].
%%CITATION = HEP-TH 9505062;%%
}

%\ArgyresXN
\lref\ArgyresXN{ P.~C.~Argyres, M.~Ronen Plesser, N.~Seiberg and
E.~Witten, ``New N=2 Superconformal Field Theories in Four
Dimensions,'' Nucl.\ Phys.\ B {\bf 461}, 71 (1996)
[arXiv:hep-th/9511154].
%%CITATION = HEP-TH 9511154;%%
}

%\CachazoRY
\lref\CDSW{ F.~Cachazo, M.~R.~Douglas, N.~Seiberg and E.~Witten,
``Chiral rings and anomalies in supersymmetric gauge theory,''
arXiv:hep-th/0211170.
%%CITATION = HEP-TH 0211170;%%
}

%\ArgyresXH
\lref\ArgyresXH{ P.~C.~Argyres and A.~E.~Faraggi, ``The vacuum
structure and spectrum of N=2 supersymmetric SU(n) gauge theory,''
Phys.\ Rev.\ Lett.\  {\bf 74}, 3931 (1995) [arXiv:hep-th/9411057].
%%CITATION = HEP-TH 9411057;%%
}

%\KlemmQS
\lref\KlemmQS{ A.~Klemm, W.~Lerche, S.~Yankielowicz and
S.~Theisen, ``Simple singularities and N=2 supersymmetric
Yang-Mills theory,'' Phys.\ Lett.\ B {\bf 344}, 169 (1995)
[arXiv:hep-th/9411048].
%%CITATION = HEP-TH 9411048;%%
}

%\SeibergRS
\lref\SeibergRS{ N.~Seiberg and E.~Witten, ``Electric - magnetic
duality, monopole condensation, and confinement in N=2
supersymmetric Yang-Mills theory,'' Nucl.\ Phys.\ B {\bf 426}, 19
(1994) [Erratum-ibid.\ B {\bf 430}, 485 (1994)]
[arXiv:hep-th/9407087].
%%CITATION = HEP-TH 9407087;%%
}

%\deBoerAP
\lref\deBoerAP{ J.~de Boer and Y.~Oz, ``Monopole condensation and
confining phase of N = 1 gauge theories via  M-theory fivebrane,''
Nucl.\ Phys.\ B {\bf 511}, 155 (1998) [arXiv:hep-th/9708044].
%%CITATION = HEP-TH 9708044;%%
}

\lref\kovner{A. Kovner and M. Shifman, ``Chirally Symmetric Phase
Of Supersymmetric Gluodynamics,'' Phys. Rev. {\bf D56} (1997)
2396, hep-th/9702174.}

%\KonishiHF
\lref\konishione{ K.~Konishi, ``Anomalous Supersymmetry
Transformation Of Some Composite Operators In Sqcd,'' Phys.\
Lett.\ B {\bf 135}, 439 (1984).
%%CITATION = PHLTA,B135,439;%%
}

%\KonishiTU
\lref\konishitwo{ K.~i.~Konishi and K.~i.~Shizuya, ``Functional
Integral Approach To Chiral Anomalies In Supersymmetric Gauge
Theories,'' Nuovo Cim.\ A {\bf 90}, 111 (1985).
%%CITATION = NUCIA,A90,111;%%
}

\lref\arkmur{N. Arkani-Hamed and H. Murayama, hep-th/9707133.}

%\BrezinSV
\lref\bipz{ E.~Brezin, C.~Itzykson, G.~Parisi and J.~B.~Zuber,
``Planar Diagrams,'' Commun.\ Math.\ Phys.\  {\bf 59}, 35 (1978).
%%CITATION = CMPHA,59,35;%%
}

%\BershadskyCX
\lref\BershadskyCX{ M.~Bershadsky, S.~Cecotti, H.~Ooguri and
C.~Vafa, ``Kodaira-Spencer theory of gravity and exact results for
quantum string amplitudes,'' Commun.\ Math.\ Phys.\  {\bf 165},
311 (1994) [arXiv:hep-th/9309140].
%%CITATION = HEP-TH 9309140;%%
}

%\GinspargIS
\lref\mmreview{ P.~Ginsparg and G.~W.~Moore, ``Lectures On 2-D
Gravity And 2-D String Theory,'' arXiv:hep-th/9304011.
%%CITATION = HEP-TH 9304011;%%
}

%\GorskyUK
\lref\gorsky{ A.~Gorsky, ``Konishi anomaly and N = 1 effective
superpotentials from matrix models,'' arXiv:hep-th/0210281.
%%CITATION = HEP-TH 0210281;%%
}

%\CachazoJY
\lref\CIV{ F.~Cachazo, K.~A.~Intriligator and C.~Vafa, ``A large N
duality via a geometric transition,'' Nucl.\ Phys.\ B {\bf 603},
3 (2001) [arXiv:hep-th/0103067].
%%CITATION = HEP-TH 0103067;%%
}

%\KutasovVE
\lref\KutasovVE{ D.~Kutasov, ``A Comment on duality in N=1
supersymmetric nonAbelian gauge theories,'' Phys.\ Lett.\ B {\bf
351}, 230 (1995) [arXiv:hep-th/9503086].
%%CITATION = HEP-TH 9503086;%%
}

%\FerrariJP
\lref\FerrariJP{ F.~Ferrari, ``On exact superpotentials in
confining vacua,'' arXiv:hep-th/0210135.
%%CITATION = HEP-TH 0210135;%%
}

%\WittenXI
\lref\WittenXI{ E.~Witten, ``The Verlinde Algebra And The
Cohomology Of The Grassmannian,'' arXiv:hep-th/9312104, and in
{\it Quantum Fields And Strings: A Course For Mathematicians}, ed.
P. Deligne et. al. (American Mathematical Society, 1999), vol. 2,
pp. 1338-9.
%%CITATION = HEP-TH 9312104;%%
}

%\SeibergPQ
\lref\SeibergPQ{ N.~Seiberg, ``Electric - magnetic duality in
supersymmetric nonAbelian gauge theories,'' Nucl.\ Phys.\ B {\bf
435}, 129 (1995) [arXiv:hep-th/9411149].
%%CITATION = HEP-TH 9411149;%%
}

%\IntriligatorID
\lref\IntriligatorID{ K.~A.~Intriligator and N.~Seiberg,
``Duality, monopoles, dyons, confinement and oblique confinement
in supersymmetric SO(N(c)) gauge theories,'' Nucl.\ Phys.\ B {\bf
444}, 125 (1995) [arXiv:hep-th/9503179].
%%CITATION = HEP-TH 9503179;%%
}

%\FujiWD
\lref\FujiWD{ H.~Fuji and Y.~Ookouchi, ``Comments on effective
superpotentials via matrix models,'' arXiv:hep-th/0210148.
%%CITATION = HEP-TH 0210148;%%
}

%\AtiyahQF
\lref\AtiyahQF{ M.~Atiyah and E.~Witten, ``M-theory dynamics on a
manifold of G(2) holonomy,'' arXiv:hep-th/0107177.
%%CITATION = HEP-TH 0107177;%%
}

%\FriedmannCT
\lref\FriedmannCT{ T.~Friedmann, ``On the quantum moduli space of
M theory compactifications,'' Nucl.\ Phys.\ B {\bf 635}, 384
(2002) [arXiv:hep-th/0203256].
%%CITATION = HEP-TH 0203256;%%
}

%\KutasovNP
\lref\KutasovNP{ D.~Kutasov and A.~Schwimmer, ``On duality in
supersymmetric Yang-Mills theory,'' Phys.\ Lett.\ B {\bf 354}, 315
(1995) [arXiv:hep-th/9505004].
%%CITATION = HEP-TH 9505004;%%
}

%\KutasovSS
\lref\KutasovSS{ D.~Kutasov, A.~Schwimmer and N.~Seiberg, ``Chiral
Rings, Singularity Theory and Electric-Magnetic Duality,'' Nucl.\
Phys.\ B {\bf 459}, 455 (1996) [arXiv:hep-th/9510222].
%%CITATION = HEP-TH 9510222;%%
}

%\IntriligatorAU
\lref\IntriligatorAU{ K.~A.~Intriligator and N.~Seiberg,
``Lectures on supersymmetric gauge theories and electric-magnetic
duality,'' Nucl.\ Phys.\ Proc.\ Suppl.\  {\bf 45BC}, 1 (1996)
[arXiv:hep-th/9509066].
%%CITATION = HEP-TH 9509066;%%
}

%\DijkgraafFC
\lref\DijkgraafFC{ R.~Dijkgraaf and C.~Vafa, ``Matrix models,
topological strings, and supersymmetric gauge theories,''
arXiv:hep-th/0206255.
%%CITATION = HEP-TH 0206255;%%
}

%\DijkgraafVW
\lref\DijkgraafVW{ R.~Dijkgraaf and C.~Vafa, ``On geometry and
matrix models,'' arXiv:hep-th/0207106.
%%CITATION = HEP-TH 0207106;%%
}

%\DijkgraafDH
\lref\DijkgraafDH{ R.~Dijkgraaf and C.~Vafa, ``A perturbative
window into non-perturbative physics,'' arXiv:hep-th/0208048.
%%CITATION = HEP-TH 0208048;%%
}

%\IntriligatorJR
\lref\IntriligatorJR{ K.~A.~Intriligator, R.~G.~Leigh and
N.~Seiberg, ``Exact superpotentials in four-dimensions,'' Phys.\
Rev.\ D {\bf 50}, 1092 (1994) [arXiv:hep-th/9403198].
%%CITATION = HEP-TH 9403198;%%
}

%\SeibergBZ
\lref\SeibergBZ{ N.~Seiberg, ``Exact results on the space of vacua
of four-dimensional SUSY gauge theories,'' Phys.\ Rev.\ D {\bf
49}, 6857 (1994) [arXiv:hep-th/9402044].
%%CITATION = HEP-TH 9402044;%%
}

\lref\vari{A. Abouelsaood, ``Are There Chromodyons?'' Nuc. Phys.
{\bf B226} (1983) 309; P. Nelson, ``Excitations Of $SU(5)$
Monopoles,'' Phys. Rev. Lett. {\bf 50} (1983) 939; P. Nelson and
A. Manohar, ``Global Color Is Not Always Defined,'' Phys. Rev.
Lett. {\bf 50} (1983) 943; A. P. Balachandran, G. Marmo, N.
Mukunda, J. S. Nilsson, E. C. G. Sudarshan, and F. Zaccaria,
``Nonabelian Monopoles Break Color: 1. Classical Mechanics, 2.
Field Theory'' Phys. Rev. {\bf D29} (1984) 2919, 2936; S. Coleman
and P. Nelson, ``What Becomes Of Global Color?'' Nucl. Phys. {\bf
B237} (1984) 1. }

%\DijkgraafPP
\lref\DijkgraafPP{ R.~Dijkgraaf, S.~Gukov, V.~A.~Kazakov and
C.~Vafa, ``Perturbative analysis of gauged matrix models,''
arXiv:hep-th/0210238.
%%CITATION = HEP-TH 0210238;%%
}

%\GopakumarWX
\lref\GopakumarWX{ R.~Gopakumar, ``${\cal N}=1$ Theories and a
Geometric Master Field,'' arXiv:hep-th/0211100.
%%CITATION = HEP-TH 0211100;%%
}

\lref\Schnitzer{S.G.~ Naculich, H.J.~ Schnitzer and N.~Wyllard,
``The $\CN=2$ $U(N)$ gauge theory prepotential and periods from a
perturbative matrix model calculation,'' arXiv:hep-th/0211123.}

%\SeibergVC
\lref\SeibergVC{ N.~Seiberg, ``Naturalness versus supersymmetric
nonrenormalization theorems,'' Phys.\ Lett.\ B {\bf 318}, 469
(1993) [arXiv:hep-ph/9309335].
%%CITATION = HEP-PH 9309335;%%
}

%\DouglasNW
\lref\DouglasNW{ M.~R.~Douglas and S.~H.~Shenker, ``Dynamics of
SU(N) supersymmetric gauge theory,'' Nucl.\ Phys.\ B {\bf 447},
271 (1995) [arXiv:hep-th/9503163].
%%CITATION = HEP-TH 9503163;%%
}

%\CachazoPR
\lref\CachazoPR{ F.~Cachazo and C.~Vafa, ``N = 1 and N = 2
geometry from fluxes,'' arXiv:hep-th/0206017.
%%CITATION = HEP-TH 0206017;%%
}

%\IntriligatorJR
\lref\IntriligatorJR{ K.~A.~Intriligator, R.~G.~Leigh and
N.~Seiberg, ``Exact superpotentials in four-dimensions,'' Phys.\
Rev.\ D {\bf 50}, 1092 (1994) [arXiv:hep-th/9403198].
%%CITATION = HEP-TH 9403198;%%
}

%\DijkgraafXD
\lref\DijkgraafXD{ R.~Dijkgraaf, M.~T.~Grisaru, C.~S.~Lam, C.~Vafa
and D.~Zanon, ``Perturbative Computation of Glueball
Superpotentials,'' arXiv:hep-th/0211017.
%%CITATION = HEP-TH 0211017;%%
}
%\GatesNR
\lref\superspace{ S.~J.~Gates, M.~T.~Grisaru, M.~Rocek and
W.~Siegel, ``Superspace, Or One Thousand And One Lessons In
Supersymmetry,'' Front.\ Phys.\  {\bf 58}, 1 (1983)
[arXiv:hep-th/0108200].
%%CITATION = HEP-TH 0108200;%%
}

%\NicolaiNR
\lref\nicolai{ H.~Nicolai, ``On A New Characterization Of Scalar
Supersymmetric Theories,'' Phys.\ Lett.\ B {\bf 89}, 341 (1980).
%%CITATION = PHLTA,B89,341;%%
}

%\MigdalGJ
\lref\migdal{ A.~A.~Migdal, ``Loop Equations And 1/N Expansion,''
Phys.\ Rept.\  {\bf 102}, 199 (1983).
%%CITATION = PRPLC,102,199;%%
}

%\StaudacherXY
\lref\staudacher{ M.~Staudacher, ``Combinatorial solution of the
two matrix model,'' Phys.\ Lett.\ B {\bf 305}, 332 (1993)
[arXiv:hep-th/9301038].
%%CITATION = HEP-TH 9301038;%%
}

\lref\voiculescu{{\it Free Probability Theory}, ed. D. Voiculescu,
pp. 21--40, AMS, 1997.}

%\CeresoleZS
\lref\CeresoleZS{ A.~Ceresole, G.~Dall'Agata, R.~D'Auria and
S.~Ferrara, ``Spectrum of type IIB supergravity on AdS(5) x T(11):
Predictions on N  = 1 SCFT's,'' Phys.\ Rev.\ D {\bf 61}, 066001
(2000) [arXiv:hep-th/9905226].
%%CITATION = HEP-TH 9905226;%%
}

\lref\ofer{O.~Aharony, unpublished.}

\newbox\tmpbox\setbox\tmpbox\hbox{\abstractfont }
\Title{\vbox{\baselineskip12pt\hbox to\wd\tmpbox{\hss
hep-th/0301006}}}
{\vbox{\centerline{Phases of $\CN=1$ Supersymmetric Gauge Theories}
\smallskip
\centerline{and Matrices}}}
\smallskip
\centerline{ Freddy Cachazo, Nathan Seiberg and Edward Witten}
\smallskip
\bigskip
\centerline{School of Natural Sciences, Institute for Advanced
Study, Princeton NJ 08540 USA}
\medskip
\bigskip
\vskip 1cm
 \noindent
$\CN=1$ supersymmetric $U(N)$ gauge theory with adjoint matter
$\Phi$ and a polynomial superpotential $\Tr W(\Phi)$ has been much
studied recently. The classical theory has several vacua labeled
by integers $(N_1,N_2,...,N_k)$, with the classical unbroken gauge
group $\prod_i U(N_i)$. Quantum mechanically, each classical
vacuum leads to $\prod_i N_i$ different vacua.  As the parameters
of $W(\Phi)$ are varied, these vacua change in a continuous (and
holomorphic) fashion. We find that vacua associated with
$(N_1,N_2,...,N_k)$ can be continuously transformed to vacua with
$(\tilde N_1,\tilde N_2,...,\tilde N_k)$, thus leading to a new kind of duality.
Traditional order parameters, like the Wilson loop and 't Hooft
loop, sometimes distinguish different phases.  We also find phases
that are not distinguished by conventional order parameters. The
whole picture of the phase diagram is reminiscent of the phase
diagram of $M$-theory.

\Date{December, 2002}

\newsec{Introduction}

Dynamics of four-dimensional  supersymmetric gauge theories has
proved to be a remarkably rich subject --  one that hopefully will
ultimately be important for the understanding of nature! Among
other attractions of these theories, models with ${\cal N}=1$
supersymmetry exhibit many subtle properties, such as dynamical
generation of mass, confinement of charge, and dynamical symmetry
breaking, that are seen in the world of strong interactions and
are still not fully understood (for a review, see e.g.\
\IntriligatorAU).  One example that has been much-studied in
recent years is a $U(N)$ gauge theory with a chiral superfield
$\Phi$ in the adjoint representation and a general single-trace
superpotential $\Tr\,W(\Phi)$, for some polynomial function $W$
that we will take to be of degree $k+1$ and suitably generic. This
model has been studied by its relation to geometric transitions
and mirror symmetry in string theory
\refs{\GopakumarKI\VafaWI-\CIV} and more recently by its
surprising relation \refs{ \DijkgraafFC\DijkgraafVW-\DijkgraafDH}
to a bosonic matrix model -- a zero-dimensional model with an
$\hat N\times \hat N$ matrix $M$ and potential $\Tr W(M)$, where
$\hat N\to\infty$.  In this paper, we will investigate some
additional fascinating properties of this example.

Let us first explain the original question that motivated our
investigation. In this model, if the function $W'(x)$ has critical
points $a_i$, $i=1,\dots, k$, then a classical vacuum is chosen by
taking $\Phi$ to be a diagonal matrix whose eigenvalues are the
$a_i$, taken with multiplicity $N_i$; the $N_i$ are any
non-negative integers such that $\sum_iN_i=N$. We let $n$ be the
number of choices of $i$ with $N_i>0$. Clearly, $n\le k$. At the
classical level, the choice of $\Phi$ spontaneously breaks the
gauge group from $U(N)$ to $\prod_iU(N_i)$ (where we include in
the product only the positive $N_i$).  Locally, this group is
isomorphic to $U(1)^n\times\prod_iSU(N_i)$.  At low energies, the
$SU(N_i)$ gauge theories become confining and massive, so quantum
mechanically, the gauge group that is actually observed at low
energies is $U(1)^n$.

Since this is independent of $N_i$, the question now arises of
whether it is possible quantum mechanically to distinguish in a
precise way between vacua that have the same value of $n$ but
different $N_i$.  To make this question clear, start with a choice
of $W(\Phi)$ such that the $a_i$ are far apart and the classical
description of the $\Phi$ field is a good approximation, and put
the theory in a vacuum characterized by a given choice of
integers $N_i$. Now vary the parameters in $W(\Phi)$, possibly
passing through a region in which $a_i$ are not far apart, but
then return to a semiclassical region in which the $a_i$ are
widely separated.  In such a process, can one interpolate from a
vacuum with one set of integers $N_i$ to a vacuum with another set
of integers $\tilde N_i$?

Let us provide a more systematic framework for this question. For
each choice of $W(\Phi)$, the theory under investigation has only
a finite set of vacua.  We will allow the parameters in $W$ to
vary, keeping only the leading coefficient (and the microscopic
gauge coupling parameter of the underlying $U(N)$ theory) fixed.
The varying parameters describe a complex manifold ${\cal T}\cong
{\bf C}^k$.\foot{In this counting, we discard an irrelevant
additive constant in $W(\Phi)$ and set the highest power in $W(x)$
to ${g_k \over k+1 } x^{k+1}= {1 \over k+1 } x^{k+1}$.} As we vary
the parameters in ${\cal T}$, the vacua fit into a moduli space
${\cal M}$ which because of supersymmetry is a complex manifold.
${\cal M}$ is a union of components (irreducible complex
submanifolds) ${\cal M}_\alpha$ which may possibly intersect at
singularities. Each component ${\cal M}_\alpha$ is a finite cover
of ${\cal T}$, and hence by going to infinity in ${\cal T}$, we
reach in each ${\cal M}_\alpha$ at least one semiclassical ``end''
with a definite set of $N_i$.  Our question is now whether the
same ${\cal M}_\alpha$ can have different ends with different sets
of $N_i$.

To avoid undue suspense, let us assert that the answer to this
question has turned out to be ``yes.''  In considering the
question, however, it soon becomes clear that  one really should
ask all the questions that are usually asked in string theory and
$M$-theory. What are the components of ${\cal M}$ and what
distinguishes them? What semi-classical ends do they have and to
what extent do they intersect at singularities? What is the
physics at these singularities?  (Some examples of singularities
have been studied in \FerrariKQ.)  When ${\cal M}$ does have
distinct components ${\cal M}_\alpha$, what kind of natural
``off-shell'' interpolations can one make between them? Without
going back to the underlying $U(N)$ gauge theory (which really
gives the most incisive definition of the whole problem), what
kind of a low energy phenomenological Lagrangian or a ``string
field theory'' can one find that describes all of the ${\cal
M}_\alpha$ at once?

Smooth interpolations between classical limits with different
gauge groups have been found in some other four-dimensional
models.  For example, in \AtiyahQF, in the context of $M$-theory
compactification on a $G_2$ manifold, an interpolation from
$SO(8+2n)$ to $Sp(n)$ was argued, along with several other
examples. This was generalized by Freedman \FriedmannCT, who
considered some additional examples, including models with $SU(N)$
gauge symmetry broken by Wilson lines to $\prod_{i=1}^nSU(N_i)\times U(1)^{n-1}$, and proposed interpolation between different
sets of $N_i$ in that case. Because the symmetry breaking pattern is the
same, one may wonder if the model is actually equivalent by some
string duality to the one considered in the present paper. One obstacle
to finding such a duality is that it is not
clear what the superpotential $W(\Phi)$ would correspond to in the
model considered in \FriedmannCT. At any rate, the present paper
is devoted to studying a field theory problem by field theory
methods; our interpolations and dualities are purely field
theoretic and do not depend on stringy phenomena.

It will become clear that the subject we are exploring is very
rich, and in the
present paper we will really only describe some general properties
and explore some examples.  Most likely, many interesting
phenomena remain to be uncovered.

Since it turns out that ${\cal M}$ does have many components, one
basic question is how to distinguish them. One fundamental
invariant that we have already identified is the rank $n$ of the
true low energy gauge group. In many physical theories, another
important invariant comes from the realization of symmetries:
which subgroup of the global symmetry group is unbroken in a given
component of the moduli space of vacua?  In the present paper,
this criterion will be somewhat less useful, because for generic
$W$ there generally are no global symmetries; still, the
realization of global symmetries, and more generally,
transformations of $\Phi$ that must also be taken to act on the
parameters in ${\cal T}$, will give some important information.
Additional important information will come from a more subtle but
still standard type of order parameter that is provided by
confinement: for an external Wilson loop in a given representation
of $U(N)$, does one observe a Coulomb law or an area law?

All these standard order parameters will play a major role in the
present paper, but we will see that they do not suffice.  We also
will encounter a more subtle sort of order parameter that only can
be defined using supersymmetry.  It may be characterized as
follows.  Suppose that a theory with ${\cal N}=1$ supersymmetry
has $r$ independent chiral superfields ${\cal O}_s$,
$s=1,\dots,r$.\foot{When we say that they are independent, we mean
that they obey no relations in the chiral ring of the theory that are
independent of the parameters $g_k$ in the superpotential.  Since we
are working over the parameter space ${\cal T}$, we are treating the
$g_k$ as variables rather than as fixed complex numbers.}  By
supersymmetry, the ${\cal O}_s$ (or more precisely their
expectation values)  are holomorphic functions on ${\cal M}$.  Now
suppose that $r$ exceeds the complex dimension of the moduli space
${\cal M}$.  Then as functions on any given component ${\cal
M}_\alpha$, the ${\cal O}_s$ inevitably obey some algebraic
equations; if these equations are different for different ${\cal
M}_\alpha$, this gives an order parameter of sorts by which the
${\cal M}_\alpha$ can be distinguished.  We will find many
examples of this situation, where the ${\cal M}_\alpha$ cannot be
distinguished by any conventional order parameter that we can see
but can be distinguished by the chiral equations that are obeyed
on them. When this occurs, we do not know if the
$\alpha$-dependence of these chiral  equations should be regarded
as an explanation for the existence of different branches, or
merely as a description of the phenomenon.

\bigskip\noindent{\it Organization Of The Paper}

This paper is organized as follows.   Section 2 is devoted to
general considerations. In section 2.1, we describe what can be
predicted about the moduli space of vacua from a ``weak coupling''
(actually weak gauge coupling) point of view in which the effects
of the superpotential $W(\Phi)$ are assumed to set in at much
higher energies than the effects of the gauge coupling.  We focus
on understanding what can be said about the moduli space of vacua
using the expected confinement of the low energy gauge theory.

In section 2.2, following previous literature
\refs{\CIV,\CachazoPR}, we consider a strong (gauge) coupling
point of view, in which the superpotential is considered as a
small perturbation of a strongly coupled gauge theory with $W=0$;
this theory has ${\cal N}=2$ supersymmetry and a vacuum structure
determined in \refs{\SeibergRS\ArgyresXH-\KlemmQS}. In previous
work, the strong coupling point of view was used to analyze the
vacua with a given rank $n$ of the unbroken symmetry group. That
analysis assumed that the $N_i$ defined above are all nonzero and
that the degree of $W$ is not too large.  We generalize the
analysis to remove these restrictions.  (This analysis is
completed in Appendix A.)

In section 2.3, we review a construction originally described in
\CIV\ whereby, given vacua for the $U(N)$ gauge theory with any
given set of $N_i$, and any integer $t>1$, one can construct vacua
for the $U(tN)$ gauge theory with $N_i'=tN_i$.  Comparing to our
analysis of confinement in section 2.1, we will see that the vacua
arising this way are precisely the confining vacua in which the
confinement index (defined in section 2.1) is $t$.  Conversely,
all vacua with confinement index $t$ arise by this construction.
Since the theory with $N_i'=tN_i$ also has vacua without
confinement or with lower confinement index (as will be clear in
section 2.1), there must be additional vacua with $N_i'=tN_i$;
they will be studied (in examples) in section 3.

In section 2.4, we present some ways in which different branches
$\CM_\alpha$ can arise and some constraints on the possible
classical limits of a given branch.   We also describe the
simplest mechanism by which different branches may intersect.

In section 3, we will explore the question of interpolating from
one set of $N_i$ to another.  For this, we want to consider the
simplest case in which $n$  is as small as possible. For $n=1$,
the gauge group is completely unbroken at the classical level, so
the question of interpolating between classical limits with
different unbroken gauge groups does not arise.  So the first case
is really that the rank of the low energy gauge group is $n=2$
which (in the minimal case with all $N_i$ non-vanishing) arises
for a cubic superpotential, $k=2$. Since we do not know how to
make the analysis for general $N$, we will explore in detail the
cases of $N=2,3,4,5$, and 6. For these values of $N$ and $k=n=2$,
we determine in detail all branches of the moduli space of vacua,
recovering the confining vacua that were found in section 2.3 and
describing the vacua without confinement.  For the vacua without
confinement, we do get the promised smooth continuations between
different values of $N_i$, including interpolation from
$(N_1,N_2)=(1,3)$ to $(2,2)$ for $N=4$,  from $(1,4) $ to $(2,3)$
for $N=5$, and from $(1,5)$ to $(2,4)$ to $(3,3)$ for $N=6$. We
conjecture that for any $N$, smooth interpolations are possible on
the Coulomb branch from any $(N_1,N_2)$ to any other
$(\tilde N_1,\tilde N_2)$.

In this detailed description for $2\leq N\leq 6$, we also see how
branches with $n=2$ meet branches with $n=1$ at singularities. In
addition to these intersections at finite points on the moduli
space, we find for $2\leq N\leq 6$ that all branches meet the
Coulomb branch at infinity, where the semiclassical description is
valid.

In section 4, we consider the opposite limit of large $n$.  For
$n=N$, that is all $N_i=1$, there is a unique vacuum for each
function $W(\Phi)$ and so our question about the different
branches does not arise.  So we consider the case that $N-n$ is
small but positive.  Here, the simple strong coupling analysis of
section 2.2 predicts that there must be at least $N-n+1$ branches
of vacua.  As far as we know, they are not distinguished by
conventional order parameters.  We show, as suggested above, that
they can be distinguished by determining which holomorphic
functions of the chiral superfields vanish on each branch.  We do
not know how best to interpret this result.

In two appendices, we give more details of the strong gauge
coupling analysis.  In appendix A, we extend the proof that
relates the matrix model curve to the $\CN=2$ curve to $W$ of
high degree. In appendix B, we show how the generalized Konishi
anomaly \refs{\konishione,\konishitwo} which was derived and used
in \CDSW\ arises in the strong gauge coupling regime where the
elementary gauge fields are not visible.  In appendix C we list
some useful properties of Chebyshev polynomials.  Appendix D
defines a magnetic index $\nu$ which characterizes the confinement
in a theory with massless photons.  It is related to the
confinement index $t$ of section 2.3 through $N=t\nu$.

\newsec{General Considerations}

\subsec{Confinement}

Here, we will see what we can say about the $U(N)$ gauge theory
with adjoint superfield $\Phi$ by using confinement as an order
parameter.  First we consider the matter directly in field
theory, and then we re-examine the issues using the matrix model.

We consider a classical vacuum in which $N_i$ eigenvalues of
$\Phi$ are placed at the $i^{th}$ critical point of the
superpotential $W$.  To keep the exposition simple, we will
assume that the $N_i$ are all positive. We also assume that the
critical points of $W$ are distinct, so that all components of
$\Phi$ are massive classically.
 Supposing that the underlying $U(N)$ gauge theory is weakly
coupled at the scale set by those masses, the low energy physics
is simply that of the pure supersymmetric gauge theory with
classical gauge group $G^{cl}_{low}=U(1)^n\times \prod_{i=1}^n
SU(N_i)$.

If $N_i>1$, each of the $SU(N_i)$ theories becomes confining at
exponentially smaller energies. For any $N_i\geq 1$, the $SU(N_i)$
theory has $N_i$ vacua, each with a mass gap. The total number of
vacua for fixed $W$ and fixed $N_i$ is hence $\prod_{i=1}^n N_i$.
If only $W$ and $G^{cl}_{low}$ are specified, and not the
individual number $N_i$ of eigenvalues at the $i^{th}$ critical
point, then the total number of vacua is larger, as one must allow
for permutations of eigenvalues among the critical points.

In the limit of small gauge coupling, the different $SU(N_i)$
factors decouple and each of them confines.  This confinement can
be diagnosed by the possible area law of the Wilson loops $W_i$ of
the various $SU(N_i)$.
 In the full theory, these $W_i$ are not well-defined.
Correspondingly, the full $SU(N)$ theory is not necessarily
confining even if the sub-theories are.
 (In this discussion, the
$U(1)$ factor in the underlying gauge group $U(N)=U(1)\times
SU(N)$ is unimportant, as it decouples and does not contribute to
the dynamics.)  For a precise criterion for confinement, we have
to use Wilson lines of the full $SU(N)$ theory.

To probe confinement, we consider a large Wilson loop in some
representation ${\cal R}$ of $SU(N)$, and ask if it exhibits an
area law or a Coulomb law. All that matters about ${\cal R}$ is
how it transforms under the center of $SU(N)$, which is $\Z_N$,
generated by $\omega={\rm diag}(e^{2\pi i/N},e^{2\pi i/N},
\dots,e^{2\pi i/N})$.  The reason that only the center matters is
that if ${\cal R}$ and ${\cal R}'$ are two irreducible
representations that transform in the same way under the center,
then an external charge in the representation ${\cal R}$ can
combine with gluons to make an external charge in the
representation $\R'$, so one of these external charges is confined
(has infinite energy) if and only if the other is. Turning one
representation into another by combining with gluons is a process
that we may call electric screening.

To give an example with any desired action of the center, we can
simply take ${\cal R}$ to be the tensor product of $r$ copies of
the fundamental representation, for some $r\geq 0$. If $W$ is a
Wilson loop in the fundamental representation, the Wilson loop
for the representation $\R$ is just $W^r$.  Only the value of $r$
modulo $N$ matters.

If two representations ${\cal R}$ and ${\cal R}'$ are unconfined,
then so is the tensor product $\R\otimes \R'$.  So the set of
values of $r$ for which there is no confinement is closed under
addition. There is no area law for $r=N$ (since $r=N$ is
equivalent to $r=0$), so the smallest positive value of $r$ for
which there is no area law is always a divisor of $N$.  We denote
this number as $t$ and call it the confinement index. If the
theory is completely confining (all Wilson loops with non-trivial
action of the center have an area law), then $t=N$, and if the
theory is completely unconfining (no Wilson loop shows an area
law), then $t=1$.

Let us now give a few examples of the behavior of the $SU(N)$
theory.  Suppose that $SU(N)$ is broken classically to
$SU(N-1)\times U(1)$.  All vacua will be unconfining simply
because there is no confinement in $U(1)$.  The fundamental
representation of $SU(N)$ has a component which is an $SU(N-1)$
singlet, and this component feels no confining gauge forces.  When
we evaluate the expectation value of a Wilson line in the
fundamental representation ${\cal R}$,
  \eqn\evrep{\left\langle\Tr_{\cal R}P\exp\oint_CA\right\rangle,}
the gauge field is effectively an $(N-1)\times (N-1)$ matrix and a
commuting $1\times 1$ component associated with the $U(1)$ factor
(plus massive fields that will not influence the question of
confinement). In computing the expectation value in \evrep,
assuming that $C$ is very large compared to the scale at which
$SU(N-1)$ becomes strongly coupled, the dominant contribution will
come from the ``bottom'' component of $\R$. This component does
not interact with the $SU(N-1)$ gauge fields, so there is no area
law.  Since this component interacts with the massless $U(1)$
gauge field, the Wilson loop \evrep\ exhibits the behavior of a
Coulomb phase rather than a Higgs phase.

\bigskip\noindent{\it Computation Of Confinement Index In An
Example}

For a simple example with a non-trivial confinement index, we will
explore the case
 that $SU(N)$ is broken to $SU(N_1)\times
SU(N_2)\times U(1)$. The fundamental representation of $SU(N)$
decomposes as $({\bf N_1},{\bf 1})\oplus ({\bf 1},{\bf N_2})$
under $SU(N_1)\times SU(N_2)$.  (We ignore the $U(1)$ charges,
which will not affect confinement.)  The tensor product of $N_1$
copies of $({\bf N_1},{\bf 1})$ contains a singlet. So, as in the
above example, the Wilson line $W^{N_1}$ will show no area law.
Likewise, $W^{N_2}$ will show no area law.

Let $t'$ be the greatest common divisor of $N_1$ and $N_2$.
Absence of confinement for $r=N_1$ and for $r=N_2$ implies that
there is no confinement for $r=l_1N_1+l_2N_2$ for any positive
integers $l_1,l_2$.  We can pick $l_1$ and $l_2$ so that
$l_1N_1+l_2N_2$ is congruent mod $N$ to $t'$. So the confinement
index $t$ can be no bigger than $t'$, and if electric screening,
which we have considered so far, were the only mechanism, it would
equal $t'$ precisely.

However, there is another mechanism, magnetic screening. Let us
first recall 't Hooft's description of confined phases of $SU(N)$
gauge theory.   One introduces an 't Hooft loop, which we will
call $H$; $H$ is constructed using a twist by an element of the
center of $SU(N)$.  The general loop order parameter is $W^rH^s$,
where $r,s$ both take values from $0$ to $N-1$ (alternatively,
their values only matter modulo $N$). Massive phases are described
by saying for which values of $r,s$ there is no area law; one says
that charges with this value of $r$ and $s$ are ``condensed.'' (In
some situations, for example a confining theory obtained by a
small perturbation of an ${\cal N}=2$ theory, one can make the
intuition behind this language precise \SeibergRS.)  In
particular, there are $N$ possible confining phases; in the
$r^{th}$ such phase, $W^rH$ has no area law, or equivalently a
charge with quantum numbers $W^rH$ is ``condensed.''

${\cal N}=1$ super Yang-Mills theory with gauge group $SU(N)$ has
$N$ confining vacua, with each type appearing precisely once. In
fact, an adiabatic increase of the theta angle by $2\pi$ has the
effect of increasing $r$ by 1.

So when we break the underlying $SU(N)$ theory to $SU(N_1)\times
SU(N_2)\times U(1)$, we get a low energy theory that indeed has
$N_1N_2$ vacua, but they are not equivalent.  They can be
distinguished by the type of confinement.  For each pair $r_1,r_2$
(with $0\leq r_i\leq N_i-1$), there is one vacuum in which
$W_1^{r_1}H_1$ and $W_2^{r_2}H_2$ are unconfined. (Here we write
$W_1$ and $W_2$ for Wilson lines in the fundamental
representations of $SU(N_1)$ and $SU(N_2)$, respectively, and
similarly $H_1$ and $H_2$ are the 't Hooft loops of the two
factors of the low energy gauge group.  As we remarked above,
this makes sense only near the weak gauge coupling limit.)

This theory also has 't Hooft-Polyakov magnetic monopoles.  If
$N=2$, $N_1=N_2=1$, we take the basic $SU(2)$ monopole whose
magnetic field at infinity is (in a unitary gauge) a multiple of
\eqn\harmo{\left(\matrix{1 & 0 \cr 0 & -1 \cr}\right).} For any
$N_1,$ $N_2$, and $N$, we just embed the $SU(2)$ monopole in
$SU(N)$ in such a way that it does not fit entirely in either
$SU(N_1)$ or $SU(N_2)$ -- we take the ``1'' in the upper left to
act in $SU(N_1)$ and the ``$-1$'' in the lower right to act in
$SU(N_2)$.  (We could also embed the monopole entirely in
$SU(N_1)$ or entirely in $SU(N_2)$, but this would give nothing
interesting for our present purposes.)

Now suppose we probe the $(r_1,r_2)$ vacuum of the full $SU(N)$
theory with  the external Wilson loop $W^{r_1-r_2}$. Although
$r_1-r_2$ is not necessarily a multiple of $t'$, we claim that
this Wilson loop will not show an area law. The external $SU(N)$
charge $r_1-r_2$ can, after symmetry breaking to $SU(N_1)\times
SU(N_2)$, be divided between the two factors in various ways, and
confinement will be avoided if there is any way to divide the
charge to make the energy finite.  We simply put $r_1$ units of
charge in $SU(N_1)$ and $-r_2$ units in $SU(N_2)$. (The charges
are only well-defined modulo $N_1$ and $N_2$.)  So in other words,
$W^{r_1-r_2}$ can behave in the low energy theory as $\tilde W=
W_1^{r_1} W_2^{-r_2}$. But the external Wilson line $\tilde W$ can
be screened by the spontaneous nucleation from the vacuum of the
magnetic monopole described in the last paragraph. Since the
monopole has magnetic charge $1$ in $SU(N_1)$ and $-1$ in
$SU(N_2)$, it can be represented as $H_1H_2^{-1}$.  So in
conjunction with a monopole, $\tilde W$ can be represented in the
low energy theory as $W_1^{r_1}H_1 W_2^{-r_2}H_2^{-1}$, and has no
area law since (by the definition of $r_1$ and $r_2$),
$W_1^{r_1}H_1$ has no area law in $SU(N_1)$, and
$W_2^{-r_2}H_2^{-1}$ (which is obtained from $W_2^{r_2}H_2$ by
reversing the orientation of the loop) has no area law in
$SU(N_2)$.

The net effect is that there is no area law for $W^{N_1}$, for
$W^{N_2}$, or for $W^{r_1-r_2}$.  Hence the confinement index $t$
is at most the greatest common divisor of $N_1$, $N_2$, and $b =
r_1-r_2$.  We claim that this is the correct value, since by now
we have considered screening by all of the electric and magnetic
objects that exist in the theory.

It is easy to see why the confinement index depends not on $r_1$
and $r_2$ separately but only on their difference $b$.  In fact,
under $\theta\to\theta+2\pi$, one has $r_i\to r_i+1$, for $i=1,2$,
while external Wilson loops are unaffected.  The confinement index
must be invariant under this operation, so can only depend on the
difference of the $r_i$.

\bigskip\noindent{\it General Case}

We can easily extend this analysis to the general case that $U(N)$
is broken down to $\prod_{i=1}^nU(N_i)$.  Electric screening
ensures that there is no area law for $W^{N_i}$, $i=1,\dots,n$. In
addition, we must consider magnetic screening. The low energy
theory  has $\prod_{i=1}^nN_i$ vacua, which are characterized by
giving integers $r_i$, $i=1,\dots ,n$ (with $0\leq r_i\leq N_i-1$)
such that in the low energy $SU(N_i)$ theory, $W_i^{r_i}H_i$ has
no area law. A magnetic monopole can be shared between $SU(N_i)$
and $SU(N_{i+1})$ in such a way that it has charges
$H_iH_{i+1}^{-1}$.\foot{One can more generally share a monopole
between $SU(N_i)$ and $SU(N_j)$ for any $i,j$, but this monopole
leads to nothing new because its charges are a linear combination
of the charges of monopoles considered in the text.} It follows,
just as in the case with $n=2$ that was considered above, that the
external Wilson line $W^{r_i-r_{i+1}}$ has no area law. (To make
the argument, we simply place $r_i$ units of charge in $SU(N_i)$
and $-r_{i+1}$ units in $SU(N_{i+1})$ and then screen this
configuration using the monopole.)

So if we let $b_i=r_i-r_{i+1}$, the confinement index $t$ is now
the greatest common divisor of the $N_i$ and the $b_i$.

There are $t$ different types of branch with confinement index
$t$. For a branch of the $u^{th}$ type, the operators that do not
have an area law are $W^t$ and $W^uH$.

\bigskip\noindent{\it Interpretation In The Matrix Model}

The matrix model is described \CIV\ by a complex curve $\Sigma$:
 \eqn\guffy{y_m^2= W'(x)^2+f_{n-1}(x),}
where $f$ depends on the gauge coupling and the $N_i$. We assume
for simplicity that $k=n$; i.e.\ no $N_i$ vanishes. $\Sigma$ is a
double cover of the $x$-plane. With $W'$ being of degree $n$, the
right hand side of \guffy\ has $2n$ zeroes; the projection of
$\Sigma$ to the complex $x$-plane is branched at these $2n$
points.

\bigskip
\centerline{\epsfxsize=0.75\hsize\epsfbox{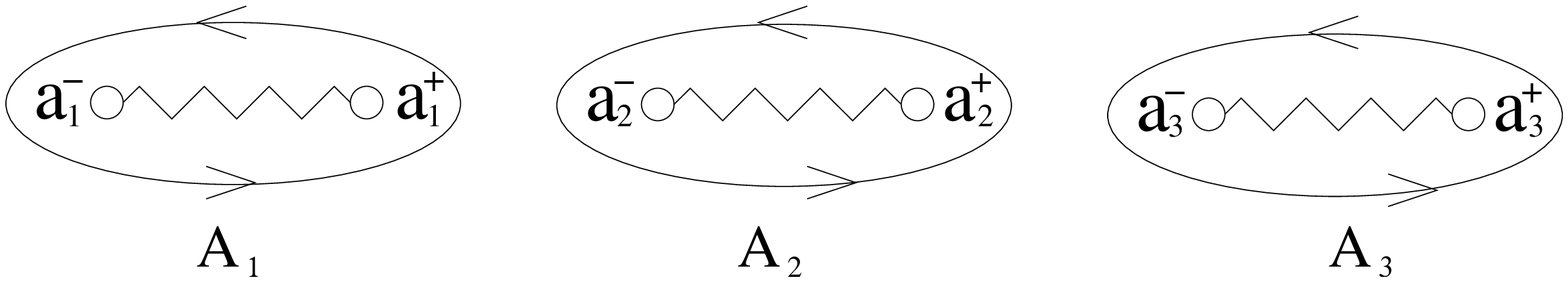}}
\noindent{\ninepoint\sl \baselineskip=8pt {\bf Figure 1}:{\sl $\;$
Branch cuts, depicted by zigzags connecting the $a_i^{\pm}$, and
 cycles $A_i$ surrounding the cuts, are depicted here
for $n=3$.}}
\bigskip

The situation can be described particularly simply for weak
coupling, where $f$ is small. The polynomial $W'$ has $n$ roots
$a_i$, and for small $f$, the polynomial $(W')^2+f$ has, for each
$i$, a pair of roots $a_i^{\pm}$ that are near $a_i$.\foot{When
the parameters are such that the roots $a_i^\pm$ are real, we ask
that $a_i^+>a_i^-$.   More generally, the symbols $\pm$ are just
labels.} We connect the $a_i^\pm$ with branch cuts, as in figure
1, so that $y_m$ is a single-valued function away from these cuts.

In \CDSW, the operator-valued differential
\eqn\tox{T(x)=\Tr{dx\over x-\Phi}} was introduced.  It was shown
that over a cycle $A_i$ surrounding the $i^{th}$ cut (also shown
in figure 1), the period of $T(x)$ was
 \eqn\uvu{N_i=\oint_{A_i} T.}
(To keep the formulas simple, in this paper we define the contour
integral $\oint $ to include a factor of $1/2\pi i$.) In the
present paper, we want to go away from the weak coupling limit, so
we cannot assume that the $2n$ zeroes of $(W')^2+f$ are neatly
paired up.  Likewise, we will not have a distinguished set of
$A_i$ cycles.  So in developing the theory, we will have to
include, along with the $A_i$, the other compact cycles with which
the $A_i$ will mix. There are $n-1$ of these, sketched in figure
2.  The new cycles, which we call $C_i$, roughly connect $a_i^+$
to $a_{i+1}^-$, for $i=1,\dots,n-1$.  We define
 \eqn\nuvu{c_i=\oint_{C_i}T.}
One might guess that since the $N_i$ are integers, and will mix
with the $c_i$ under strong coupling monodromies, the $c_i$ are
also integers.  We will prove this at the end of this section,
using results from \CDSW.  It turns out that the periods of $T$
\nuvu\ are the integers $c_i$ only on shell; i.e.\ only after all
the equations of motion have been solved. Off shell, before
solving the equations of motion of $S_i$, the periods around the
$A_i$ cycles \uvu\ are given by the integers $N_i$ but the periods
around the $C_i$ cycles are not constrained. A more detailed
discussion of this point will appear below and in a separate
publication.

\bigskip
\centerline{\epsfxsize=0.75\hsize\epsfbox{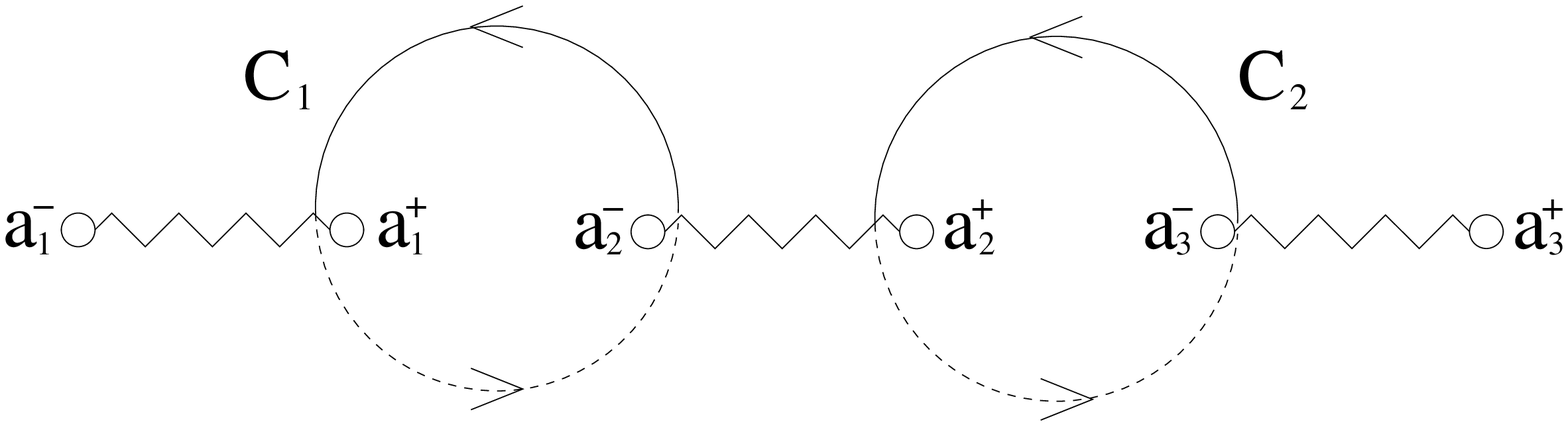}}
\noindent{\ninepoint\sl \baselineskip=8pt {\bf Figure 2}:{\sl $\;$
Choice of compact  cycles $C_i$  for $n=3$. The part of the contour
depicted by a dashed line is on the second sheet below the cut.
}}
\bigskip

Now let us recall the matrix model construction of the gauge
theory  effective superpotential. Here one uses another
differential, which in the matrix model is
 \eqn\boxo{R(x)={g_m\over \widehat N}\Tr {dx\over x-M}}
where $g_m$ is the finite 't Hooft coupling of the matrix model.
\boxo\ has been interpreted in gauge theory in \CDSW. Its periods
are \eqn\noggo{\eqalign{\oint_{A_i}R & = S_i \cr
                    \oint_{B_i}R & = \Pi_i={1\over 2\pi i}
                    {\partial \CF\over \partial S_i}.}}
Here $B_i$ are the noncompact cycles sketched in figure 3.  The
$A_i$ and $B_i$ are a canonical basis of the homology of the
compact surface made by adding points at infinity to $\Sigma$;
their intersection pairings are $A_i\cap A_j=B_i\cap B_j=0$,
$A_i\cap B_j=\delta_{ij}$.  The $B_i$ are related to the compact
cycles $C_i$ introduced in the last paragraph by
 \eqn\toggo{C_i=B_{i+1}-B_{i}.}
Furthermore, $S_i=-{1\over 32\pi^2}\Tr W_{\alpha i}W^{\alpha i}$
is the gluino bilinear of $U(N_i)$  (a more rigorous definition of
$S_i$ is given in \CDSW), and $\CF$ is called the
``prepotential.''

\bigskip
\centerline{\epsfxsize=0.75\hsize\epsfbox{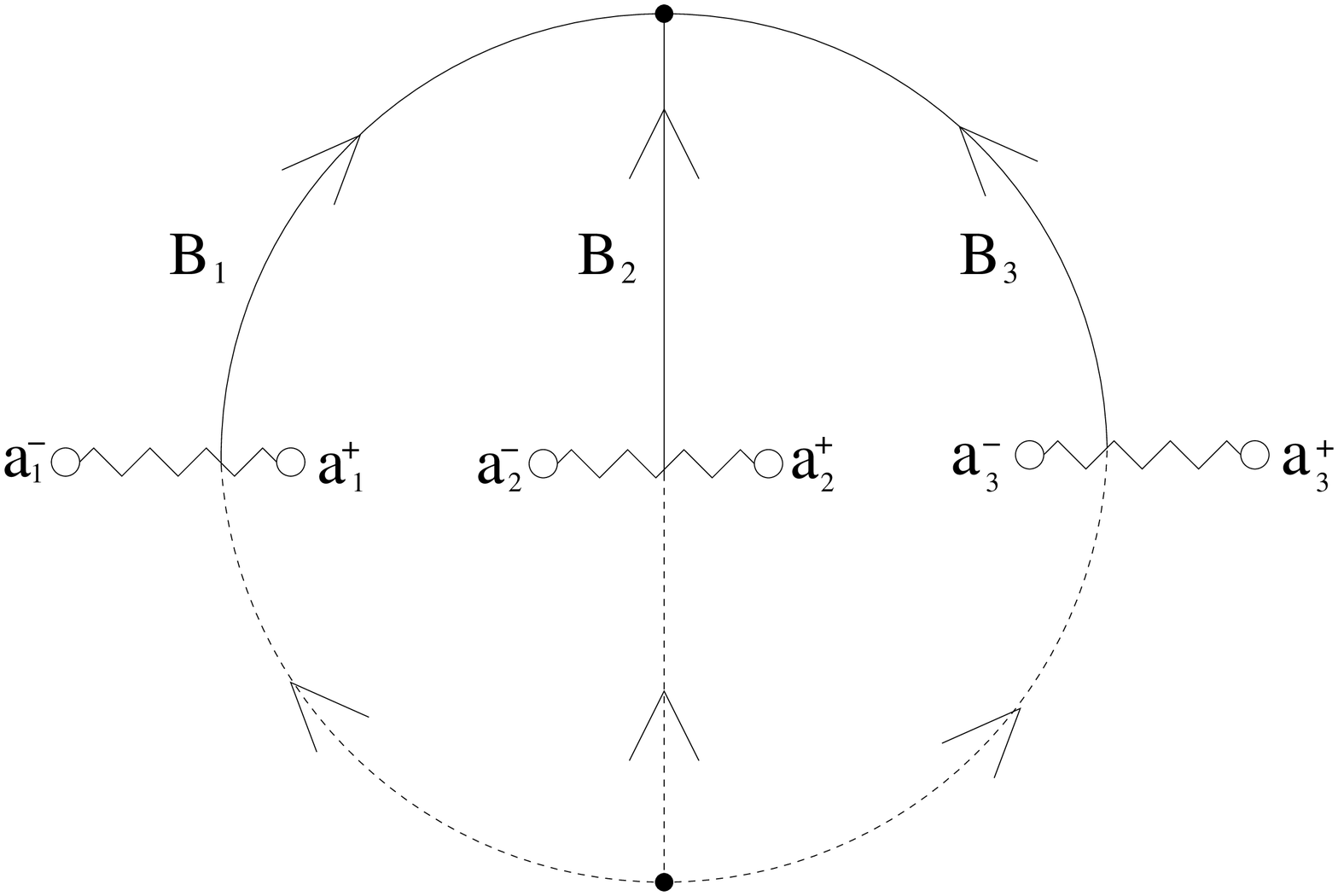}}
\noindent{\ninepoint\sl \baselineskip=8pt {\bf Figure 3}:{\sl $\;$
Non-compact cycles $B_i$ for $n=3$. The bullets represent the two
punctures of the Riemann surface (that is, the two points lying
above $x=\infty$). The  cycle $B_i$ passes through the $i^{th}$ cut
and connects the two
punctures, which are on different sheets.}}
\bigskip

To obtain the matrix model expression for the superpotential, one
starts
\CIV\ with the fact that the model is dual to string theory on
a certain non-compact Calabi-Yau threefold $X$.
On the threefold, the superpotential is
${1\over 2\pi i}\int_X H\wedge \Omega$, where  $\Omega$ is the   covariantly constant
 holomorphic
three-form of $X$,
and the flux $H$ has integral periods on compact cycles.
By integrating out the extra dimensions, one reduces
this expression to an integral over a Riemann surface (subsequently
understood as the
matrix model curve).  In this reduction, $H$ descends to a differential
form $\bar T_0$ whose periods on compact cycles are integers, and
$\Omega$ descends to the differential $R$.  The superpotential becomes
$W={1\over 2\pi i}
\int_\Sigma \bar T_0\wedge R.$\foot{$\bar T_0$, like $H$, is real;
we have denoted it $\bar T_0$ because $R$ is a differential of
type $(1,0)$, so only the $(0,1)$ part of $\bar T_0$ actually
contributes to the integral.}
Evaluating this in the usual way in terms of contour integrals
over one-cycles $A_i$ and $B_i$ (and remembering our factor of $1/2\pi i$ in the
definition of a contour integral), we get the  usual matrix model formula
for the effective superpotential for the $S_i$:
 \eqn\luggu{W_{eff}(S_i)=2\pi i \sum_i\left(\oint_{A_i}\bar T_0
 \oint_{B_i} R- \oint_{B_i}\bar T_0\oint_{A_i}R\right)}
We will modify previous treatments only in not assuming that
$\oint_{B_i}\bar T_0$ is independent of $i$.

According to the familiar analyses \CIV\ of the superpotential,
$\oint_{A_i}\bar T_0=N_i$.  In addition, we set $\oint_{C_i}\bar
T_0=c_i'$. Let us compare these to the periods of $T$ as given in
equations \uvu\ and \nuvu. Since $\bar T_0$ and $T_0$ have the
same periods on the $A$-cycles, and the $A$-cycles will mix with
the $C$-cycles under strong coupling monodromies, it is natural to
conjecture that also their periods on the $C$-cycles are equal,
$c_i=c_i'$. As will be explained in detail in a separate
publication, this holds on-shell, that is after imposing the
equations of motion of the $S_i$. (Off-shell, the $C_i$ periods of
$T$ depend on the $S_i$, while $\bar T_0$ has integral periods
even before imposing the $S_i$ equations of motion.)

The integrals of $\bar T_0$ over the noncompact cycles $B_i$ must
then be as follows.  $\oint_{B_1}\bar T_0=-\tau_0$ for some
complex number $\tau_0$, and more generally
 \eqn\mugugu{\oint_{B_i}\bar T_0 = -\tau_0 -b_i,}
with
 \eqn\muggu{b_i=-\sum_{j=1}^{i-1}c_j.}
So we get for the effective superpotential
 \eqn\tuggu{W_{eff}(S_i)=\sum_{i=1}^n N_i{\partial \CF\over
 \partial S_i} +2\pi i \tau_0 \sum_{i=1}^n S_i +2\pi i
 \sum_{i=2}^nb_iS_i.} The derivation of  \tuggu\ does not require
 the knowledge that $\bar T_0=T$ on shell.

The only addition that we are making to previous studies of this
problem is to include the last term, proportional to the $b_i$.
Let us explain what this term does.  The Veneziano-Yankielowicz
superpotential for $SU(M)$ supersymmetric gluodynamics is
\VenezianoAH
 \eqn\fory{W(S)=S\left[\log(\Lambda^{3M}/S^M) +M\right].}
The logarithm is only defined modulo $2\pi i$, so $W$ is not
defined on the $S$-plane but on an infinite cover thereof where
for the time being we allow all possible branches of the
logarithm. On this infinite cover, the number of critical points
of $W$ is $M$; the equation for a critical point is indeed
 \eqn\rito{\log(\Lambda^{3M}/S^M)=0,}
and this equation implies that $S^M=\Lambda^{3M}$, an equation
that has the familiar $M$ roots. (Moreover, after picking $S$ so
$S^M=\Lambda^{3M}$, we must, to obey \rito, pick the correct
branch of the logarithm, so each solution of $S^M=\Lambda^{3M}$
leads to only one vacuum.) Now consider, instead of \fory, the
superpotential
 \eqn\mory{\tilde W(S)=M\left[ S\log(\Lambda^3/S)+S\right].}
Again, this superpotential is defined on an infinite cover of the
$S$ plane, but now on this infinite cover, the equation for a
critical point is $M\log(\Lambda^3/S)=0$, which implies $\log
(\Lambda^3/S)=0$. There is only one root, namely $S=\Lambda^3$.

What has happened?  Mathematically, there are really $M$ different
ways to define an infinite cover of the $S$-plane on which the
Veneziano-Yankielowicz superpotential \fory\ is defined. The
reason is that when the argument of $S$ shifts by $2\pi$,
$\log(\Lambda^{3M}/S^M)$ shifts by $-2\pi iM$.  We do not need to
allow all branches of the logarithm to get a space on which $W$ is
defined; it suffices to consider $1/M$ of all of the branches.  We
could pick an arbitrary integer $b$ and say that when
$\Lambda^3/S$ is positive, we want $\log(\Lambda^{3M}/S^M)$ to
have an imaginary part congruent to $2\pi ib$ $(\mod \,2\pi iM)$.
So the Veneziano-Yankielowicz superpotential can be defined on
each of $M$ different spaces ${\cal S}_b$, $b=0,\dots, M-1$, each
of which is an infinite cover of the $S$-plane.  Chiral symmetry
permutes these, so physically we should include all of them, which
is equivalent to including all branches of
$\log(\Lambda^{3M}/S^M).$

The superpotential $\tilde W(S)$ is equivalent to $W(S)$ defined
only on ${\cal S}_0$.  It therefore only describes one of the
vacua.  To describe the others, in this language, we should
explicitly include additional branches on which the
superpotential is
 \eqn\nory{\tilde W_b=M\left[ S\log(\Lambda^3/S)+S\right]
 +2\pi i b S.}
The physical meaning of $b$ is very simple.  A superpotential
term which is an imaginary multiple of $S$ is just a shift in the
theta angle.  So including $b$ just shifts the theta angle by
$2\pi b$, which will rotate confinement (condensation of an 't
Hooft loop) to oblique confinement (condensation of a mixture of
't Hooft and Wilson loops). Thus, for given $b$, the vacuum that
is realized has condensation of  a $b$-dependent combination of
't Hooft and Wilson loops $W^bH$.

Since in this paper we will be interested in studying $U(M)$ gauge
theories rather than $SU(M)$, we would like to extend \nory\ to
this case.  As in \CDSW, we define $S=\hat S - {1 \over
2M}(w_\alpha)^2$.  Here $\hat S$ is the trace over the $SU(M)$
fields and it is taken to be an independent chiral superfield in
the effective theory.  $w_\alpha$ is the field strength superfield
of the $U(1) \subset U(M)$.  A fact which was useful in \CDSW\ is
the decoupling of this $U(1)$.  This decoupling is implemented by
considering the ``superfield'' $\CS= S + \psi^\alpha w_\alpha
-\psi^1\psi^2 M$, where $\psi^\alpha$ is an auxiliary
anticommuting Lorentz spinor.  The first term in \nory\ is then
written as $\int d^2 \psi \left[\half
\CS^2\log(\Lambda^3/\CS)+{3\over 4} \CS^2\right]=M\left[ \hat
S\log(\Lambda^3/\hat S)+\hat S\right]$, which is independent of
$w_\alpha$. (In fact, any expression of the form $\int d^2\psi
H(\CS)$ is independent of $w_\alpha$.)   The $2\pi i bS$ term in
\nory\ can also be written equally well in terms of $S$ in (in
$U(M)$) or $\hat S$ (in $SU(M)$), since $b$ is an integer and
abelian gauge theory is invariant under a $2\pi$ shift in the
theta angle.

Now let us go back to the matrix model superpotential \tuggu,
replacing $M$ in the above discussion with any of the $N_i$.  For
small $S_i$, $\partial \CF/\partial S_i\approx S_i\left[
\log(\Lambda_i^3/S_i)+1\right]$.  So our effective superpotential
for small $S_i$ is (we drop the terms which involve two
derivatives of the prepotential $\CF$)
 \eqn\tycon{W_{eff}(S_i)=\sum_i \left( 2\pi i \tau_0S_i + N_iS_i
 \left[ \log(\Lambda_i^3/ S_i) +1\right] + 2\pi ib_iS_i\right)
 + \CO(S_i S_j).}
The first term is the bare coupling.  The second term includes the
one loop renormalization of the coupling constant and the strong
$SU(N_i)$ IR dynamics.  The last term $\CO(S_i S_j)$ represents
the perturbative contribution of the high energy theory.  Our new
term $2\pi ib_iS_i$ arises also from the strong IR dynamics of
$SU(N_i)$.  It should be independent of the photon of $U(1)
\subset U(N_i)$ and therefore the constants $b_i$ are quantized.

{}From \tycon\ it is clear that the integers $b_i$ represent
relative shifts by $2\pi b_i$ of the theta angles of the various
$SU(N_i)$. More explicitly, relative to the theta angle in
$SU(N_1)$, the theta angle in $SU(N_k)$ is greater by $2\pi b_k$.
So if $SU(N_1)$ has condensation of $W^{r_1}H$,\foot{The value of
$r_1$ depends on the bare coupling $\tau_0$ and on how $H$ is
defined.} then $SU(N_k)$ has condensation  of $W^{r_k}H$, where
$r_k-r_1=b_k$. (Our definitions are such that $b_1=0$.)

Thus, the $b_k$ that we have defined in \muggu\ are the same as
the ones we introduced in our general discussion of confinement.
So the confinement index $t$ is the greatest common divisor of
the $N_i$ and the $b_k$, or equivalently, the greatest common
divisor of the $N_i$ and $c_i$.  Since $N_i$ and $c_i$ are a
complete set of periods of $T$ (integrated over  compact cycles),
we can describe this by saying that the confinement index is the
greatest integer $t$ such that all compact periods of $T/t$  are
integral.  This gives, in the context of the matrix model, a
manifestly ``modular invariant'' definition of the confinement
index.

In section 2.3, we will (following \CIV) describe an operation
that multiplies $N$ by an arbitrary positive integer $t$, also
multiplying the individual $N_i$ by $t$ and multiplying $T$ by
$t$. This operation only generates confining vacua.  In fact, it
multiplies the confinement index by $t$.  We will see that all
confining vacua arise by applying this operation, starting with a
non-confining vacuum with a smaller value of $N$.  So for any
given $N$, the ``new'' vacua that cannot be predicted based on
what happened for smaller $N$ are the ``Coulomb'' vacua, the ones
without confinement.  For every $N$, and every set of $N_i$,
there are Coulomb vacua -- for example, those vacua in which any
of the $b_j$ is 1.

Finally, we explain how to deduce from the results of \CDSW\ that
the  $c_i$ are integers.  Instead of studying the present problem
using the matrix model curve, we can use the ${\cal N}=2 $ curve
 \eqn\gryf{y^2=P_N(x)^2-4\Lambda^{2N}.}
(According to \CIV, these curves are closely related, with
$P_N^2-4\Lambda^{2N}=H_{N-n}^2((W')^2+f)$; we will not need here
to explicitly use this relation.) In appendix A of \CDSW, it was
argued that $T=(P_N'/y)dx$.  This can alternatively be written
 \eqn\nuryf{T={1\over P_N+y}d(P_N+y)=d\log(P_N+y).}
Like any logarithmic derivative, $T$ has integer periods (recall
that we include a factor of $1\over 2\pi i$ in the definition of
$\oint$). In other words, $\oint T$ over a compact cycle is the
change in ${1\over 2\pi i} \log(P_N+y)$ around that cycle, and
this is an integer.

\subsec{Considerations based on strong gauge coupling}

The opposite of the weak gauge coupling approach of section 2.1
is a strong gauge coupling approach in which the superpotential
is regarded as a small perturbation of the ${\cal N}=2$ gauge
dynamics that one would have if $W=0$. This approach was
developed in \CIV\ by using methods of \deBoerAP.  Here we will
review this analysis and extend it to allow some of the $N_i$ to
vanish and to allow superpotentials of any degree.

We write the superpotential as
\eqn\supp{W = \sum_{r=0}^k {g_r\over r+1} \Tr\, \Phi^{r+1}  .}
In our present discussion, the superpotential is considered a
small perturbation of an $\N=2$ $U(N)$ theory. We interpret $W$
as an effective superpotential on the $\N=2$ Coulomb moduli space
by replacing  $\Tr\, \Phi^r$ by its vev $\langle \Tr\,
\Phi^{r}\rangle$, regarded as a function on the  moduli space.  In
order to look for vacua in which the low energy gauge group is
$U(1)^n$, we must extremize the superpotential \supp\ constrained
to submanifolds of the Coulomb branch where $N-n$ monopoles of
the ${\cal N}=2$ theory are massless (physically, the constraint
is imposed because the massless monopoles get masses and vevs
when $W$ is turned on).  The perturbation by $W$ lifts all  vacua
of the $\N=2$ Coulomb moduli space, except for a finite number
which survive.

It is convenient to introduce, for each ${\cal N}=2$ vacuum, a
classical $N\times N$ matrix $\Phi_{cl}$ such that $\langle \Tr\,
\Phi^r\rangle = \Tr\, \Phi_{cl}^r$ for $r=1,\dots,N$.   Moreover,
we set  $u_r={1\over r}\Tr\,\Phi_{cl}^r$ for all positive
integers $r$. For $r=1,\dots,N$, the $u_r$ are independent and
are the usual order parameters of the $\CN=2$ theory.  For $r>N$,
both $\langle\Tr\,\Phi^r\rangle$ and $\Tr\, \Phi_{cl}^r$ can be
expressed in terms of the $u_r $ of $r\leq N$, but for $r\geq
2N$, as shown in Appendix A of \CDSW, they are unequal.

 The classical
vacua of this theory are obtained by setting all eigenvalues of
$\Phi$ and $\Phi_{cl}$ equal to
 roots of $W'(z) = \sum_{r=0}^k g_r z^r$.  In the present section,
we will  take the degree of the superpotential to be $k+1 \le N$,
so that the $u_r$ that appear in the superpotential are
independent and $\langle\Tr\,\Phi^r\rangle =\Tr\,\Phi_{cl}^r$.  In
Appendix A, we generalize to arbitrary $k$.

At a generic point on the Coulomb branch of the ${\cal N}=2$
theory, the low energy gauge group is $U(1)^N$.   We want to
study vacua in which the perturbation by $W$ leaves only $U(1)^n$
gauge group  at low energies, with $n\leq k$. This occurs if the
remaining degrees of freedom become massive when $W\not= 0$
because of the condensation of $N-n$ mutually local magnetic
monopoles. This can happen only at points where (at $W'=0$) the
monopoles are massless. Monopole condensation for $W\not= 0$  can
be seen by including the $N-n$ monopole hypermultiplets in the
superpotential,
\eqn\hyp{ W_{eff} = \sum_{l=1}^{N-n} M_l(u_k) q_l \tilde q_l +
\sum_{r=0}^k g_r u_{r+1}. }
Here $q_l$, $\tilde q_l$ are the monopole fields and $M_l(u_k)$
is the mass of $l^{th}$ monopole as a function of the $u_k$.  In a
supersymmetric vacuum,  the variation of $W_{eff}$ with respect to
$q_l$ and $\tilde q_l$ is zero. However, for suitable $g_r$, the
variation with respect to the $u_k$'s does not allow $q_l \tilde
q_l$ to be zero. Therefore the mass of the monopoles has to
vanish, i.e. $M_l(\langle u_k\rangle ) = 0 $ for $l=1,\ldots
,N-n$.

The masses $M_l$ are known to be equal to periods of a certain
meromorphic one-form over some cycles of the $\N=2$ hyperelliptic
curve,
\eqn\ntwo{ y^2 = P_N^2(x) - 4\Lambda^{2N} }
where $P_N(x) = \det (x-\Phi_{cl} )$ is a polynomial of degree
$N$.

In \CIV, it was shown that it is more convenient to use the fact
that at points with $N-n$ mutually local massless monopoles, the
${\cal N} =2$ curve degenerates as follows:
\eqn\fact{  y^2 = P_N^2(x) - 4\Lambda^{2N} = F_{2n}(x)
H_{N-n}^2(x). }
This factorization is satisfied on an $n$-dimensional subspace of
the Coulomb branch on which the superpotential should be
extremized to find the points that preserve ${\cal N}=1$
supersymmetry.

The condition \fact\ can easily be incorporated by means of
Lagrange multipliers \CIV. Using such a superpotential, it was
shown in \CIV\ that on shell and when the degree $k$ of $W'(x)$ is
equal to $n$, the highest $n+1$ coefficients of $F_{2n}(x)$ are
given  in terms of $W'(x)$ as follows,
\eqn\simp{F_{2n}(x) ={1\over g^2_n} W'(x)^2 + {\cal O}(x^{n-1}).}
Assuming that $W$ is given and the problem is to determine the
${\cal N}=2$ vacuum or equivalently $P_N(x)$, \simp\ is used as
follows. \simp\ determines $F_{2n}(x)$ in terms of $n$ unknown
coefficients. These are then determined, together with the
desired $P_N$, by asking for existence of a polynomial $H_{N-n}$
such that $P_N^2-4\Lambda^{2N}=F_{2n}H_{N-n}^2$, as in \fact.  It
is also often convenient to study the inverse problem: starting
with an ${\cal N}=2$ vacuum with $N-n$ massless monopoles -- in
other words, a $P_N$ that enables the factorization in \fact\ for
suitable $H_{N-n}$ and $F_{2n}$ -- one asks what superpotential
$W$ would lead to this vacuum.  For $k=n$, this problem has a
unique solution (modulo an irrelevant additive constant) since if
$F_{2n}(x)$ is known, \simp\ determines $W'(x)$.

The main result of this subsection is to give the generalization
of \simp\ to $k>n$, or in other words to the case in which some
of the $N_i$ are zero. We  repeat the proof of \simp\ given in
\CIV, using a slightly different way to introduce the
constraints. In Appendix A, we make a
further generalization to arbitrary $k$, dropping the restriction
$k\leq N-1$.

Let us add to the superpotential constraints imposing the
factorization \fact. It is useful to write $H_{N-n}(x) =
\prod_{i=1}^{N-n}(x-p_i)$; i.e.\ $p_i$ are the locations of the
double roots of $P_N^2-4\Lambda^{2N}$. Then we take
 \eqn\const{ \eqalign{
 W_{eff} =&  \sum_{r=0}^k g_r u_{r+1} +
 \sum_{i=1}^{N-n} \left( L_i \oint {P_N(x)
 -2\epsilon_i \Lambda^{N}\over x-p_i} dx
 + B_i\oint {P_N(x)-2\epsilon_i \Lambda^{N}
 \over (x-p_i)^2}dx \right)\cr =&
 \sum_{r=0}^k g_r u_{r+1} +
 \sum_{i=1}^{N-n} \left( L_i(\oint {P_N(x)\over x-p_i}
 dx -2\epsilon_i \Lambda^{N})
 + B_i\oint {P_N(x)\over (x-p_i)^2}dx \right) }}
Here $P_N(x)$ is a function of $u_r$. $L_i$ and $B_i$ are Lagrange
multipliers imposing the constraints, $\epsilon_i = \pm 1$ and
the contour of integration is a large curve around $x=\infty$ that
encloses all $p_i$'s. We also have included a factor of ${1\over
2\pi i}$ in the definition of the symbol $\oint$ in order to avoid
cluttering of equations.

The variation with respect to $B_i$ leads to
 \eqn\biva{ 0=\oint{P_N(x) \over (x-p_i)^2} dx= P'_N(p_i) =
 P_N(p_i)\Tr {1\over p_i-\Phi_{cl}}
 }
(The last step in \biva\ follows from the fact that
$P_N(x)=\det(x-\Phi_{cl})=\prod_i(x-\phi^i_{cl})$, where
$\phi^i_{cl}$ are the eigenvalues of $\Phi_{cl}$.  Since
$P_N(p_i)\not=0$ (given \fact\ and the fact that
$H_{N-n}(p_i)=0$), we learn that
 \eqn\pecid{\Tr {1\over p_i-\Phi_{cl}} =0}
We can use either \biva\ or \pecid\ to solve for $p_i$ in terms of
the $u_r$, but let us not do it now.  Equation \biva\ or \pecid\
will be used in Appendix B.

Minimization of $W_{eff}=W_{eff}(u_r,p_i,L_i,B_i;g_r,\Lambda)$
leads to $W_{low}=W_{low}(g_r,\Lambda)$. However, we will not
attempt to carry this out. Instead, we will try to get information
about $F_{2n}$ at the minimum by continuing to study the field
equations.

Let us consider variations of \const\ with respect to $p_i$,
\eqn\Qcons{ \oint P_N(x) \left( \sum_{i=1}^{N-n}
{L_i\over (x-p_i)^2} +{2 B_i\over (x-p_i)^3}\right) dx=2 B_i
 \oint { P_N(x) \over (x-p_i)^3} dx =0}
where we used \biva.  Assuming that \fact\ does not have any
triple or higher roots, \Qcons\ implies that $B_i=0$.

Let us now consider variations with respect to $u_r$.  We use
 \eqn\pniden{P_N(x) = \exp\left(\Tr\, \ln (x-\Phi_{cl} ) \right)
 = x^N \exp \left(-\sum_{r=0}^\infty {u_r \over x^r}\right) .}
Since $P_N(x)$ is a polynomial in $x$, \pniden\ can be used to
express $u_r$ with $r>N$ in terms of $u_r$ with $r\le N$ by
imposing the vanishing of negative powers of $x$. (See Appendix
A.) We can also write \pniden\ as
 \eqn\pnideni{P_N(x) = \left[ x^N
 \exp(-\sum_{r=0}^\infty {u_r \over x^r})\right]_+}
only in terms of the independent $u_r$, those with $r\leq
N$.\foot{The symbol $[A]_+$ denotes the polynomial part of a
Laurent series $A$.} The derivative of \pnideni\ with respect to
$u_r$ is
 \eqn\pderu{{\partial P_N(x) \over \partial u_r} = -
 \left[ {P_N(x) \over x^r }\right]_+}
We are now ready to differentiate $W_{eff}$ with respect to $u_r$:
 \eqn\Ucons{ g_{r-1} = \oint  \left[ {P_N(x) \over x^r }\right]_+
 \sum_{i=1}^{N-n} {L_i\over x-p_i} dx = \oint  {P_N(x) \over x^r}
 \sum_{i=1}^{N-n} {L_i\over x-p_i} dx}
It is convenient to multiply \Ucons\ with $z^{r-1}$ and to sum
over $r$ ($z$ is inside the contour of integration)
 \eqn\Wcons{ W'(z) =  \oint  {P_N(x) \over x-z}
 \sum_{i=1}^{N-n} {L_i\over x-p_i} dx.}
We define the polynomial $Q(x)$ in  terms of
 \eqn\Qdefi{\sum_{i=1}^{N-n} {L_i\over x-p_i} = {Q(x) \over
 H_{N-n}(x)}.}
and write \Wcons\ as
\eqn\grs{  W'(z) =  \oint {Q(x)\over H_{N-n}(x)} {P_N(x) \over
x-z}\ dx.}
Since $W'(z)$ is a polynomial of degree $k$, the polynomial
$Q(x)$ is of degree $k-n$, and we will denote it as $Q_{k-n}$
(its definition \Qdefi\ determined it as a polynomial of degree $N
- n-1\ge k-n$).

Finally, notice from \fact\ that,
$$ P_N(x) = \sqrt{F_{2n}(x)} H_{N-n}(x) +{\cal O}(x^{-N}) .$$
So we can freely replace $P_N(x)$ by $ \sqrt{F_{2n}(x)}
H_{N-n}(x)$ in \grs\ to get,
\eqn\rec{ W'(z) =  \oint {Q_{k-n}(x) \sqrt{F_{2n}(x)} \over x-z
}dx .}
We can easily recognize \rec\ if it is written
\eqn\mat{ W'(z) = \oint { y_m(x) \over x-z} dx}
with
 \eqn\ymmq{ y_m^2(x) = F_{2n}(x) Q^2_{k-n}(x). }
\ymmq\ is the  matrix model curve (for a case in which only $n$
of a possible $k$ cuts have ``opened up''), and \mat\ is the
corresponding ``equation of motion'' for the matrix model
eigenvalues.

{}From \mat\ we conclude that \simp\ has to be generalized to
\eqn\geon{ F_{2n}(x) Q^2_{k-n}(x) = W'_k(x)^2 + {\cal O}(x^{k-1}).
}
In the case when $k=n$, we recover \simp\ with $Q_0 = g_n$.
\geon\ can be used just like \simp\ to determine the ${\cal N}=2$
vacuum if $W$ is given.  With $W$ known, \geon\ determines
$F_{2n}$, assuming its leading coefficient is normalized to 1, in
terms of $n$ parameters, which are then determined, along with
$P_N$ and $H_{N-n}$, by requiring the factorization \fact.  For
the inverse problem, things are a little different.  If we start
with an ${\cal N}=2$ vacuum, so that $P_N$, $F_{2n}$, and
$H_{N-n}$ are known, and we want to constrain $W$, then \geon\
actually leaves $k-n$ free parameters in the determination of $W$.
In the classical limit, the meaning of this is that if all
eigenvalues of $\Phi$ are placed at $n$ zeroes of $W'$, then the
expectation values $\langle\Tr\,\Phi^r\rangle$ are entirely
independent of the other $n-k$ zeroes.

In appendix A, we prove that the generalization of \geon\ for
arbitrarily large $k$ is
\eqn\appfin{ F_{2n}(x){\tilde Q}_{k-n}^2(x) = W'_k(x)^2 + {\cal
O}(x^{k-1})  }
where
$${\tilde Q}_{k-n}(x) = V_{k-N}(x) H_{N-n}(x) + Q_{N-n-1}(x) $$
and $V_{k-N}(x)$ is a polynomial of degree $k-N$ whose
coefficients are new Lagrange multipliers.

Note that \geon\ and \appfin\ have the same structure. Moreover,
they provide enough equations, as in the case of \simp, to
determine the vacuum if $W$ is given, or to determine $W$ modulo
a physically sensible redundancy in the case of the inverse
problem.

It is interesting to note that this result is consistent with the
fact that the matrix model does not have information about the
rank of the field theory gauge group $U(N)$. It only knows about
the number of  cuts, i.e.\ the number of non-zero $N_i$.

See Appendix A for the generalization to superpotentials of
arbitrary degree.

\subsec{The Multiplication Map And The Confinement Index }

In \CIV, a construction was described that, for any positive
integer $t$, maps vacua of the $U(N)$ theory with a given
superpotential $W$ to vacua of the $U(tN)$ theory with the same
superpotential.
 This operation maps classical limits with unbroken
$\prod_{i=1}^nU(N_i)$ to classical limits with unbroken
$\prod_{i=1}^nU(tN_i)$ and so in particular leaves fixed the rank
$n$ of the low energy gauge group. We now review this construction
and show that it multiplies the confinement index by $t$ and that
all vacua of $U(tN)$ with confinement index $t$ arise from the
Coulomb vacua of $U(N)$ under this multiplication map. This
implies in particular that, for any $N$, the only ``new'' vacua,
which cannot be deduced from a $U(N/t)$ vacuum (for some $t$)  via
the multiplication map, are the Coulomb vacua.

A key role is played in the analysis by the polynomials
\DouglasNW\ that describe those ${\cal N}=2$ vacua of the $U(t)$
theory that have the maximum number of massless monopoles, namely
$t-1$. These are the vacua that survive when the ${\cal N}=2$
theory is perturbed by a quadratic superpotential
$W=\half\Tr\,\Phi^2$; this perturbation removes $\Phi$ from the
low energy theory, whereupon the $SU(t)$ gauge theory becomes
massive at low energies, with $t$ vacua.  From the ${\cal N}=2$
point of view, this mass generation in the $SU(t)$ sector arises
from  $t-1$ condensed monopoles. To have $t-1$ massless monopoles
before the superpotential is turned on, the right hand side of
the equation of the ${\cal N}=2$ curve
$y^2=P_t(x)^2-4\Lambda^{2t}$ must admit a special factorization,
\eqn\unbr{ P^2_t(x) - 4\Lambda^{2t} = F_2(x)H_{t-1}^2(x), }
where we can assume that $P_t$, $F_2$, and $H_{t-1}$ all have
leading coefficient 1.  Moreover, setting $F_2=(W')^2+f_0$, we see
that if $W=x^2/2$, we want $F_2=x^2-a\Lambda^2$ for some constant
$a$. This problem leads to $t$ possibilities for $P_t$ and
$H_{t-1}$, found by using Chebyshev polynomials.

Chebyshev polynomials of the first and second kind of degree $t$
and $t-1$ respectively are defined as follows.  (See Appendix C
for some explicit formulas.) Set $x=\cos\theta$ and let
\eqn\chb{ {\cal T}_t(x) =\cos (t\theta )  \qquad  {\cal
U}_{t-1}(x) ={1\over t} {d{\cal T}_t\over dx}(x) = {\sin
(t\theta)\over \sin \theta}, }
the point being that $\cos(t\theta)$ and
$\sin(t\theta)/\sin\theta$ are both polynomials in $x$. Some
identities that can easily be checked are
\eqn\prop{\eqalign{ & {\cal T}_t(x)^2 - 1  = (x^2 - 1){\cal
U}_{t-1}(x)^2 \cr
  & {\cal T}_t(x)   ={1\over 2}\left( (x+\sqrt{x^2-1})^t +
 (x-\sqrt{x^2-1})^t \right).
} }
{}From the second identity\foot{To prove this, write $\cos
(t\theta) = {1\over 2}(e^{t\theta i}+e^{-t\theta i})$} it is clear
that the coefficient of $x^t$ in $T_t(x)$ is $2^{t-1}$.

The solutions to \unbr\ are\foot{We are making a slight refinement
of the formulas as they have been presented previously.}
\eqn\ntop{ P_t(x) = 2\Lambda^t\eta^t {\cal T}_t\left( {x\over
2\eta\Lambda}\right),  ~~ F_2(x)=x^2-4\eta^2\Lambda^2,~~
H_{t-1}(x) = \eta^{t-1}\Lambda^{t-1} {\cal U}_{t-1}\left( {x\over
2\eta\Lambda}\right) ,}
with $\eta$ a $2t$-th root of unity, i.e.\ $\eta^{2t}=1$. Using
the fact that $P_t(-x)=(-1)^tP_t(x)$, one can see that \ntop\ is a
function only of $\eta^2$ and hence  gives us precisely the
expected $t$ solutions of the factorization problem.

Using this, we can go back to the original problem of mapping
$U(N)$ solutions with a given superpotential, and some
renormalization scale $\Lambda_0$, to $U(tN)$ solutions with the
same superpotential and renormalization scale $\Lambda$.

 By assumption,
we have polynomials $P_N(x)$, $F_{2n}(x)$ and $H_{N-n}(x)$, all
with leading coefficient 1, obeying the following relations:
\eqn\peq{\eqalign{P_N^2(x) - 4\Lambda_0^{2N}&
=F_{2n}(x)H^2_{N-n}(x)\cr F_{2n}(x)&=(W')^2+f_{n-1}(x).\cr}}
{}From \prop, one can show that $2\Lambda^{tN}\eta^t{\cal
T}_t(\tilde x)$ with $\tilde x={1\over 2\eta\Lambda^N} P_N(x)$
satisfies \eqn\timest{ \eqalign{ 2\Lambda^{tN}\eta^t{\cal
T}_t(\tilde x) & = x^{tN} + {\cal O}(x^{tN-1}) \cr \left(
2\Lambda^{tN}\eta^t {\cal T}_t(\tilde x)\right)^2 - 4\Lambda^{2tN}
& = \left(P^2_N(x) -4\eta^2\Lambda^{2N}\right) \left(
\eta^{t-1}\Lambda^{N(t-1)} {\cal U}_{t-1}(\tilde x)\right)^2  .} }
Using \peq\ in the last equation with the identification
$\Lambda_0^{2N}=\eta^2\Lambda^{2N}$, one finds
\eqn\ffchev{\left( 2\Lambda^{tN}\eta^t {\cal T}_t(\tilde
x)\right)^2 - 4\Lambda^{2tN} = F_{2n}(x)
H^2_{N-n}(x)\left(\eta^{t-1}\Lambda^{N(t-1)} {\cal U}_{t-1}(\tilde
x)\right)^2. }

The factorization problem for vacua of the $U(tN)$ problem with
low energy gauge group $U(1)^n$ is
\eqn\cond{P_{tN}^2(x) - 4\Lambda^{2tN} = \tilde
F_{2n}(x)H_{tN-n}^2(x). }
Comparison of \ffchev\ and \cond\ leads to the identification of
\eqn\sol{\eqalign{P_{tN}(x) & =  2\Lambda^{tN}\eta^t
 {\cal T}_t
 \left( {P_N(x)\over 2\eta\Lambda^{N}}\right) \cr \tilde
F_{2n}(x)& = F_{2n}(x)\cr H_{tN-n}(x) & =
\eta^{t-1}\Lambda^{N(t-1)}H_{N-n}(x) {\cal U}_{t-1} \left(
{P_N(x)\over 2\eta\Lambda^{N}}\right)\cr }}
as solutions of \cond.  Since $\tilde F_{2n}=F_{2n}$, the vacua
constructed this way for the $U(tN)$ theory have the same
superpotential as the vacua of the $U(N)$ theory.

For a given superpotential $W$, the $U(N)$ theory has a finite
number of vacua with given $n$.  This number is independent of
$\Lambda_0$. For $t$ different values of the $U(N)$ parameter
$\Lambda_0^{2N}$, namely
\eqn\pulpo{\Lambda_0^{2N}=\eta^2\Lambda^{2N},} we have been able
to construct  $U(tN)$ vacua.  Thus, this construction gives $t$
times as many $U(tN)$ vacua as there are $U(N)$ vacua. The
relation \pulpo\ shows that, to obtain $U(tN)$ vacua in which the
$\theta$-angle equals $\theta_{tN}$, the analogous angle
$\theta_N$ of the $U(N)$ theory must be $\theta_N=\theta_{tN}+2\pi
l/t$ for some integer $l$.

This construction gives $t$ branches in $U(tN)$ for every branch
in $U(N)$, since the construction depends on the choice of
$\eta^2$. Since we will show that the construction multiplies the
confinement index by $t$, the counting of branches is in accord
with the observation in section 2.1 that Wilson and 't Hooft
loops can distinguish $t$ types of branch with confinement index
$t$.

\bigskip
\noindent {\it Unbroken Gauge Group}

As a check of our formalism, let us see how the $tN$ vacua with
unbroken $U(tN)$ arise by applying the multiplication by $t$ map
to the unbroken vacua of $U(N)$.

In this case, $P_N(x)$ satisfies \peq\ with $n=1$, i.e.\ it is
given by,
$$ P_N(x) =  2\Lambda_0^N \epsilon^N {\cal T}_N \left( x\over
2\epsilon \Lambda_0 \right) $$
with $\epsilon^{2N}=1$.

Using this in \sol, we get
\eqn\mult{ P_{tN}(x) = 2\Lambda^{tN}\eta^t {\cal T}_t\left(
 {\Lambda_0^N \epsilon^N \over \eta\Lambda^N} {\cal T}_N
\left( x\over 2\epsilon \Lambda_0 \right)  \right)  }
Using $\Lambda_0^{2N} = \eta^2 \Lambda^{2N}$, one can easily check
that the combination ${\Lambda_0^N \epsilon^N \over
\eta\Lambda^N}$ squares to one and can be pulled out of ${\cal
T}_t$ giving a factor of $\left( {\Lambda_0^N \epsilon^N \over
\eta\Lambda^N} \right)^t$. This implies that \mult\ can be written
as
\eqn\multS{ P_{tN}(x) = 2(\epsilon\Lambda_0)^{tN} {\cal T}_t\left(
 {\cal T}_N
\left( x\over 2\epsilon \Lambda_0 \right)  \right)  }
Let us introduce $\tilde \sigma$ obeying
$\tilde\sigma^{2N}=\eta^2$ as well as $\Lambda_0 =\tilde\sigma
\Lambda$, and define $\sigma =\epsilon\tilde\sigma$. Finally,
using the identity ${\cal T}_{tN}(x)  ={\cal T}_t({\cal T}_N(x))$
in \multS, one gets
$$ P_{tN}(x) = 2\Lambda^{tN}\sigma^{tN} {\cal T}_{tN} \left(
{x\over 2\sigma \Lambda} \right). $$ Recalling that $\eta$ is a
$2t$-root of unity we get that $\tilde\sigma$ is a $2tN$-root of
unity, and so is $\sigma$ since $\epsilon$ is a $2N$-root of
unity. This, as we wished to show, is the expected solution of the
$U(tN)$ problem.

\bigskip
\noindent {\it Generating Function of Observables $\Tr\,\Phi^k$}

In section 2.1, a criterion for confinement was given in terms of
the operator-valued one-form constructed from the generating
function of $\Tr\, \Phi^k$,
\eqn\ghh{T(x) =  \left\langle \Tr {dx \over x-\Phi }
\right\rangle .}
Let us denote by $T_{tN}(x)$ and $T_N(x)$ the one forms of $U(tN)$
and $U(N)$ respectively.

Recall that (according to Appendix A of \CDSW)
\eqn\obsgf{ \left\langle \Tr {1\over x-\Phi } \right\rangle =
{P'_{tN}(x) \over \sqrt{P^2_{tN}(x) - 4\Lambda^{2tN}}}.}
{}From \chb\ and \sol,  we get
\eqn\der{ P'_{tN}(x) = 2\eta^t\Lambda^{tN} {d\tilde x\over dx}
{\cal T}'_t(\tilde x) = t \eta^{t-1} P'_N(x) \Lambda^{N(t-1)}{\cal
U}_{t-1}(\tilde x).}
Using this, \timest, and \sol, we get \eqn\vevgf{\left\langle \Tr
{1\over x-\Phi } \right\rangle = t {P'_N(x)\over
\sqrt{P_N^2(x)-4\Lambda_0^{2N}}}. } If we denote by $ \Phi_0$ the
adjoint field in the $U(N)$ theory, then \vevgf\ can be written as
\eqn\vevrel{\left\langle \Tr {1\over x-\Phi } \right\rangle =
t\left\langle \Tr {1\over x-{ \Phi_0} } \right\rangle }
or equivalently $T_{tN}(x) = t T_N(x)$.

So the multiplication by $t$ map multiplies all periods of $T$ by
$t$.  Since the confinement index is the greatest common divisor
of the periods, it follows that this map multiplies the
confinement index by $t$.  Since the $N_i$ are certain periods of
$T$, it follows also that the map multiplies the $N_i$ by $t$, as
originally shown in \CIV.

\bigskip
\noindent {\it All The Confining Vacua}

As a special case, if we start with a Coulomb vacuum in $U(N)$,
the multiplication by $t$ map produces vacua with confinement
index $t$ in $U(tN)$.

The converse is also true: all vacua in  $U(tN)$ with confinement
index $t$ arise in this way from Coulomb vacua in $U(N)$. This can
be shown by counting vacua.  Because of the freedom to choose
$\eta$, the multiplication by $t$ map produces $t$ times as many
vacua  with confinement index $t$ in $U(tN)$ as there are Coulomb
vacua in $U(N)$.  Let us show that this is the correct number.

As explained in section 2.1, a Coulomb vacuum in $U(N)$ with a
classical limit based on $\prod_iU(N_i)$ is specified by picking
some integers $r_i $ with $0\leq r_i\leq N_i-1$ and such that the
collection of the  $N_i$ and $b_i=r_i-r_{i+1}$ are relatively
prime. Instead, to give a vacuum in $U(tN)$ with classical limit
based on $\prod_iU(tN_i)$ and with confinement index $t$, one must
pick some integers $\tilde r_i$, $0\leq \tilde r_i\leq tN_i-1$,
such that the collection $tN_i$ and $\tilde b_i =\tilde r_i-\tilde
r_{i+1}$ have greatest common divisor $t$.  Any such $\tilde r_i$
are uniquely of the form $\tilde r_i =u+tr_i$ with $0\leq u\leq
t-1$ (and that formula works for any $u$ and for any $r_i$ that
solve the problem in $U(N)$).  As there are $t$ choices of $u$,
the number of $U(tN)$ vacua with confinement index $t$ should
indeed be $t$ times the corresponding number of $U(N)$ Coulomb
vacua.

\subsec{Further Strong Coupling Generalities}

Here we will consider some additional issues from the strong
coupling point of view, in which one starts with the solution of
the ${\cal N}=2$ theory and treats the superpotential as a small
perturbation.  The $SU(N)$ dynamics is described by the curve
$\Sigma$: \eqn\ntwocurve{y^2=P_N(x)^2-4\Lambda^{2N}.} Here
$P_N(x)=\det(x-\Phi_{cl})$.  Generically, the expectation value of
$\Phi$ breaks $SU(N)$ to $U(1)^{N-1}$, and $\Sigma$ has genus
$N-1$. Its Jacobian determines the effective couplings of the
$N-1$ low energy abelian vector multiplets.  Including the $U(1)$
center of $U(N)$, which is free and not described by $\Sigma$, the
low energy unbroken gauge group is $U(1)^N$.\foot{In this paper,
we will not require $\Phi$ to be traceless. So $\Phi$ describes
all chiral superfields, including the ${\cal N}=2$ partner of the
center of $U(N)$.  But the Jacobian of $\Sigma$ describes the
gauge couplings only for the $SU(N)$ gauge fields. When a
superpotential is turned on, the trace part of $\Phi$ interacts
with the other fields, but the $U(1)$ gauge field remains free.}

If the right hand side of $\Sigma$ has $w$ double roots, this
corresponds to existence of $w$ independent massless monopoles.
After small perturbation by a generic superpotential, reducing
the supersymmetry to ${\cal N}=1$, the monopoles will condense,
reducing the low energy gauge group from $U(1)^N$ to $U(1)^{N-w}$.
So to describe a vacuum of the ${\cal N}=1$ theory with low
energy gauge group $U(1)^n$, we need $N-n$ double zeroes of the
polynomial $P_N(x)^2-4\Lambda^{2N}$.

This polynomial factorizes
 \eqn\factpo{P_N^2-4\Lambda^{2N}=
 (P_N+2\Lambda^{N})(P_N-2\Lambda^{N}).}
The two factors $P_N+2\Lambda^N$ and $P_N-2\Lambda^N$ never vanish
simultaneously, so a double root is a double root of one factor or
the other.  To get $w=N-n$ double roots, we therefore place $s_+$
double roots in one factor and $s_-$ in the other factor, with
$w=s_++s_-$.  If $w$ is given, there are $w+1$ ways to pick $s_+$
and $s_-$ with $w=s_++s_-$, so if all possibilities actually
occur, there are (at least) $w+1$ branches of the moduli space of
vacua on which the rank of the unbroken group is $n=N-w$.  In the
examples that we have looked at, all values of $s_+$ and $s_-$
with $2s_+,2s_-\leq N$ are possible -- they all are realized for
some $N^{th}$ order polynomial $P_N$.  (Note that every $P_N(x)$
is of the form $P_N(x)=\det(x-\Phi_{cl})$ for some $\Phi_{cl}$.)
We actually will meet cases with several branches for the same
$s_+$ and $s_-$.

This multiplicity of branches cannot, as far as we know, be
explained using conventional order parameters.  The exchange
$s_+\leftrightarrow s_-$ can be understood as a discrete chiral
symmetry of the $\CN=2$ theory under which $\Lambda^N$ changes
sign. In some cases (with $w$ close to $N$) that we will study in
section 3, confinement goes a long way in explaining why there are
several branches. However, in general, chiral symmetries and
confinement (and any conventional order parameters we know) do not
suffice to explain why there are so many branches.  In section 4,
we study examples in which $w$ is small compared to $N$ and show
that the branches with different $s_+$ and $s_-$ can be
characterized according to which holomorphic functions of the
chiral order parameters vanish on a given branch.

Now let us explain something that will be useful background for
our computations in section 3.  Consider a family of ${\cal N}=2$
vacua with $N-n$ massless monopoles and given $s_+,s_-$ (and
$s_++s_-=N-n$). As we take $\Lambda\to 0$, can the theory converge
to a classical limit in which the unbroken gauge group is
$\prod_{i=1}^nU(N_i)?$

If such a limit does exist, then for very small $\Lambda$, the
$U(N_i)$ theories approximately decouple from each other.  Each
has its own polynomial $P_{N_i}$ and its own mass scale
$\Lambda_i$, and its own curve
$y_i^2=P_{N_i}(x)^2-4\Lambda_i^{2N_i}$. If each $U(N_i)$ theory
has $w_i$ massless monopoles, we need $\sum_iw_i=N-n$ so that the
full theory has $N-n$ massless monopoles.  But in $U(N_i)$, the
maximum possible number of massless monopoles is $N_i-1$, and
since $\sum_{i=1}^n(N_i-1)=N-n$, we must in fact have $w_i=N_i-1$
for all $i$.

The particular polynomials $P_{N_i}(x)$ such that
$P_{N_i}^2-4\Lambda_i^{2N_i}$ has $N_i-1$ double roots are known
explicitly in terms of Chebyshev polynomials \DouglasNW\ and were
used in section 2.3. Even without recalling the details, we can
determine what numbers $s_{+i}$ and $s_{-i}$ of double roots of
$P_{N_i}\pm 2\Lambda_i^{2N_i}$ are possible if
$s_{+i}+s_{-i}=N_i-1$. A polynomial of degree $N_i$ has at most
$N_i/2$ double roots, so $s_{+i},s_{-i}\leq N_i/2$. If $N_i$ is
odd, the only possibility is $s_{+i}=s_{-i}=(N_i-1)/2$, and if
$N_i$ is even, there are two cases
$(s_{+i},s_{-i})=(N_i/2,N_i/2-1)$ or $(N_i/2-1,N_i/2)$.

Finally we can answer our question of which classical limits can
exist for given $s_+,s_-$.  A branch of given $s_+$ and $s_-$ can
have a classical limit with unbroken $\prod_{i=1}^nU(N_i)$ only if
one can write $s_+=\sum_is_{+i},$ $s_-=\sum_is_{-i}$, with $0\leq
s_{+i},s_{-i}\leq N_i/2$ and $s_{+i}+s_{-i}=N_i-1$. The above
argument shows that this condition is necessary; in examples that
we have studied, it is also sufficient.  For example, in section
3, we will examine in detail the case of $N=6$ with $n=2$, so
$s_++s_-=4$. For $(s_+,s_-)=(3,1)$ or $(1,3)$, our conditions only
allow the classical limit $U(4)\times U(2)$, in agreement with
what we will find. But for $(s_+,s_-)=(2,2)$, they allow
$U(5)\times U(1)$, $U(4)\times U(2)$, and $U(3)\times U(3)$, all
of which will duly appear.

\bigskip\noindent{\it Intersections of Branches}

Now let us consider whether different branches might meet at
singularities. For example, can a branch $\CM_1$ with
$(s_+,s_-)=(a,b)$ meet a branch $\CM_2$ with
$(s_+,s_-)=(a+1,b-1)$? This will naturally occur at a point with
$w+1$ massless monopoles and $(s_+,s_-)=(a+1,b)$. Such a point is
on both $\CM_1$ and $\CM_2$, as well as being on a branch $\CM$
with $w+1$ massless monopoles. So in general, distinct branches
with ``adjacent'' values of $s_+,s_-$ will meet at singular
points with an extra massless monopole.

Such a singularity corresponds to shrinking a cycle in the
Riemann surface of the $\CN=2$ theory.  From \nuryf\ it is
clear that the differential $T$ cannot have a pole.
Therefore, a cycle can shrink only if  the period of $T$
around it vanishes.  This leads to restrictions on the
possible singularities in a branch.  Similar restrictions
have been pointed out in \FerrariKQ.   They do not affect
the examples we study below because even if $\oint T$ around
a certain cycle is not zero, we can find another cycle with
vanishing period which can be shrunk.

Here we have described the situation in an ${\cal N}=2$
language.  From an ${\cal N}=1$ point of view, the situation is
different.  For a given branch with given $w$, the superpotential
that leads to a given ${\cal N}=2$ curve on the branch can be
found by the recipe of \CIV, reviewed in section 2.2.  (It
generically triggers condensation of all the massless monopoles.)
For simplicity, we assume that on the branches $\CM_1$ and
$\CM_2$, $k=n$ so the superpotential is uniquely determined. The
superpotential needed to get a given ${\cal N}=2$ curve on a
branch with $(s_+,s_-)=(a,b)$ is generically different from the
superpotential needed to get the same curve on a branch with
$(s_+,s_-)=(a+1,b-1)$.  So this type of intersection of branches
is not really relevant for ${\cal N}=1$, which is our main
interest in the present paper.

What is more relevant for ${\cal N}=1$ is that a branch $\CM_1$
with $(s_+,s_-)=(a,b)$ meets a branch $\tilde {\cal M}$ with
$(s_+,s_-)=(a+1,b)$ at a point with an extra double root.  (An
example was discussed in \FerrariKQ.)\foot{The branch $\tilde
{\cal M}$ likewise meets the branch $\CM_2$ with
$(s_+,s_-)=(a,b-1)$; as we have just explained, this occurs with
the same ${\cal N}=2$ curve but a different superpotential from
the one that leads to $\CM_1$ intersecting $\tilde {\cal M}$.}

The notion of $\tilde \CM$ and $\CM_1$ ``meeting'' may seem
confusing since from an ${\cal N}=2$ point of view, $\tilde\CM$
is just a subspace of $\CM_1$ singled out by existence of an
extra double root.  So let us elaborate.
 $\tilde\CM$ actually has a reduced rank of the low energy gauge
group $n=k-1$, so on this branch the superpotential is not
uniquely determined by the ${\cal N}=2$ curve, but depends on one
extra complex parameter.  This parameter controls the
condensation of the massless monopole that appeared when
$P_N+2\Lambda^{2N}$ developed an extra double root.  This
parameter should be included in describing the $\tilde \CM$
branch from an $\N=1$ point of view.  Thus, from an $\N=1$ point
of view, $\tilde\CM$ is not a subspace of $\CM_1$, and in fact
they have the same dimension.

Starting on $\CM_1$, the intersection with $\tilde\CM$ is
obtained by adjusting the $\N=2$ curve to get an extra double
root.  Starting from $\tilde\CM$ (at a generic point where  the
extra monopole is condensed), the intersection is reached by
varying the superpotential until it becomes the $\CM_1$
superpotential for the same ${\cal N}=2$ curve.

For this to make sense, we need to know that the $\CM_1$
superpotential of any ${\cal N}=2$ curve with $(s_+,s_-)=(a+1,b)$
is always one of the allowed superpotentials that can lead to
that curve on $\tilde\CM$.  This can readily be shown from the
generalized \CIV\ recipe for the superpotential that was
presented in section 2.2 (see \geon\ and the discussion following
it).

Sometimes it will happen that the branch $\tilde\CM$ is confining
while $\CM_1$ is not confining (or has a smaller confinement
index).  If so, then as one approaches the intersection of the
two branches starting on $\tilde\CM$, the coefficient of the area
law must vanish (as noted in an example in \FerrariKQ).

Actually, a point with $(s_+,s_-)=(a+1,b)$ is on an $(a+1)$-fold
self-intersection of $\CM_1$, since there are $a+1$ different ways
to forget a double root of $P_N+2\Lambda^N$, leaving $a$ of them.
Such a point is likewise on a $b$-fold self-intersection of
$\CM_2$.   However, the different branches do not really meet as
${\cal N}=1$ theories, since they would require different
superpotentials.

Though in the present paper we focus on double roots of
$P_N^2-4\Lambda^{2N}$, it is also possible to consider the case of
triple or higher order roots.  The first example was studied by
Argyres and Douglas \ArgyresJJ.  At the ${\cal N}=2$ level,
before turning on a superpotential, these configurations are
believed to correspond to non-trivial critical points.  What
happens if a superpotential is turned on, breaking to ${\cal
N}=1$?  A generic superpotential will not be extremal at a point
with a triple root, but by tuning parameters to get a
superpotential that is extremal at such a point, one can
presumably get a nontrivial critical point with ${\cal N}=1$
supersymmetry.

\bigskip
\noindent{\it Off-shell Interpolation}

Most of this paper is devoted to the question of branches of
physical vacua.  As the parameters in $W(\Phi)$ are varied, the
vacua change and we explore the various branches and the smooth
interpolation from one vacuum to another within a branch.  These
are on-shell interpolations.  It is interesting to ask whether
there exists a Lagrangian which allows us to extrapolate between
the different branches.

This question can be answered at various levels. The most
complete description of the theory is in terms of the microscopic
gauge theory.  It describes all the vacua of the theory.

A more macroscopic description is the effective Lagrangian of
\CIV\ for the fields $S_i$.  The vacua are found by solving the
equations of motion of $S_i$.  As we vary the parameters in
$W(\Phi)$, these vacua change and we find the on-shell
interpolation.  One of our results is that we can continuously
interpolate between vacua with one value of $N_i$ to vacua with
other values of $N_i$.  Therefore, the superpotential of \CIV\
has many stationary points corresponding to the different vacua
with different $N_i$ on the same branch.  We will see examples of
this in the next section.

Another type of an effective Lagrangian we can consider is the
following.  The effective Lagrangian for a given branch has the
form \tuggu\
 \eqn\tugguu{\sum_i N_i S_i \left(\log\left({\Lambda^3 \over
 S_i}\right)+ 1\right ) + 2\pi i b_i S_i + \sum_iN_i\CO(S_jS_k)}
(we absorbed $\tau_0$ into $\Lambda$ and $b_1=0$). We can replace
it with
 \eqn\effwall{\sum_i S_i \left(\log\left({\Lambda^{3N_i } \over
 S_i^{N_i }}\right)+ N_i \right) +\sum_iN_i\CO(S_jS_k).}
As we explained in section 2.1, this superpotential has more
stationary points than \tugguu.  For small $S_i$ it has $\prod_i
N_i$ vacua, rather than a single vacuum.  Therefore, it describes
different branches, some of them can be confining and others can
be in a Coulomb phase.  A superpotential like \effwall\ might be
a good effective Lagrangian for all vacua for a given $N_i$ for
small $S_i$, i.e.\ at weak coupling.  However, it is unlikely to
describe the proper interpolation to large $S_i$ where vacua with
different values of $N_i$ are present.

\newsec{Examples}

In this section, we will analyze several explicit examples
demonstrating the phenomena we have discussed.  In the first five
subsections, we will study examples with gauge group $U(N)$ for
$N=2,\dots,6$ and with low rank of the low energy gauge group. The
case that the rank is $1$ is too trivial for our purposes; this
means that $U(N)$ is unbroken classically.  The corresponding
${\cal N}=2$ curves are described by Chebyshev polynomials and
were reviewed in section 2.3.  We focus therefore on the next case
that the low energy group has rank $n=2$; thus, classically $U(N)$
is broken to $U(N_1)\times U(N_2)$, and the question arises of
whether smooth interpolation can occur between different pairs
$(N_1,N_2)$.  In section 3.6, we present some special examples
with $n>2$.

For each $N$, the ``new'' vacua are the Coulomb vacua, since
according to the discussion in section 2.3, the confining vacua
with confinement index $t$ are determined by what happened for
$U(N/t)$.  In our systematic construction of all vacua with $n=2$
for given $N$, we will come across all vacua, both Coulomb and
confining.  For the range of $N$ we consider, interpolation
between different pairs $(N_1,N_2)$ occurs only for Coulomb vacua,
simply because (if $t>1$) $N/t$ is always too small for such
interpolation to be possible in $U(N/t)$.

We will analyze the theory along the following steps.  We start
by considering the $\CN=2$ $U(N)$ gauge theory whose Coulomb
branch is described by the hyper-elliptic curve
 \eqn\utg{y^2=P_N(x)^2 -4 \Lambda^{2N}=(P_N(x)+ 2\Lambda^N)
 (P_N(x)- 2 \Lambda^N )}
where the $N^{th}$ order polynomial $P_N(x)=\det(x-\Phi_{cl})$
parameterizes the point in the moduli space.  This point is
labeled by the $N$ eigenvalues of the matrix $\Phi_{cl}$ modulo
permutations.

To get $n=2$, the $\CN=2$ theory must have  $N-2$ massless
magnetic monopoles, so the polynomial \utg\ must have $N-2$ double
roots.  Since $P_N$ depends on $N$ complex parameters, the
subspace on which $P_N^2-4\Lambda^{2N}$ has $N-2$ double roots is
two-dimensional. As discussed in section 2.4, the distribution of
double roots between the two factors in \utg\ is labeled by two
integers. There are $s_+$ double roots in the first factor and
$s_-$ double roots in the second factor. The two integers $s_+$
and $s_-$ must satisfy $s_++s_-=N-2$ and $s_+,s_-\le N/2$.
Different values of $s_+$ and $s_-$ correspond to different
branches.

After picking values of $s_+$ and $s_-$ we solve the factorization
problem
 \eqn\facg{\eqalign{
 &P_N(x)+ 2\Lambda^N=H_{s_+}^2(x) R_{N-2s_+}(x) \cr
 &P_N(x) - 2 \Lambda^N=\tilde H_{s_-}^2(x) \tilde R_{N-2s_-}(x)}}
with polynomials $H_{s_+}$, $\tilde H_{s_-}$, $R_{N-2s_+}$ and
$\tilde R_{N-2s_-}$, where the subscript of the polynomial denotes
its degree.  We normalize the coefficient of the highest power of
$x$ in each polynomial to be one, and solve for the other
coefficients such that
 \eqn\facgp{H_{s_+}^2(x)R_{N-2s_+}(x)-4\Lambda^N=\tilde
 H_{s_-}^2(x) \tilde R_{N-2s_-}(x).}
We will see in the examples below that often there are several
distinct branches for the same $(s_+, s_-)$.

The moduli space $\CM$ is parametrized by the coefficients in the
polynomials $H_{s_+}$, $\tilde H_{s_-}$, $R_{N-2s_+}$ and $\tilde
R_{N-2s_-}$. As we mentioned above, $\CM$ is two dimensional. In
solving the factorization problem \facg, we will sometimes shift
$x$ by a constant in order to simplify the equations. This
constant can trivially be reinstated at the end of the
calculation.  Ignoring this constant, our moduli space is one
dimensional.

We then examine the semiclassical limit $\Lambda \to 0$. Since we
limit ourselves to the case $n=2$, in the classical limit the
microscopic $U(N)$ gauge group is broken to $U(N_1) \times
U(N_2)$.  We want to determine for each branch $\CM_\alpha$ of the
moduli space what values of $N_1$ and $N_2$ are possible.

We will also study singular points on $\CM_\alpha$. Triple roots
of $y^2$ lead to Argyres-Douglas points \ArgyresJJ. We will not
determine them here.  We will focus on singularities in which
$y^2$ has a new double root. These occur when either
$R_{N-2s}(x)$ or $\tilde R_{N-2t}(x)$ acquire a double root. This
means that these points have $(s_++1,s_-)$ or $(s_+,s_-+1)$
double roots in the two branches. At these subspaces, there are
$N-1$ massless monopoles. As explained in section 2.4, at these
points $\CM_\alpha$ meets a branch with $n=1$, i.e., the fully
confined branch that classically has unbroken $U(N)$.

We then break $\CN=2$ to $\CN=1$ by turning on a tree level
superpotential which leads to a vacuum at a point on $\CM_\alpha$;
this lifts the degeneracy of $\CM_\alpha$, and only a finite set
of vacua survive for given superpotential. We want to determine
what superpotential can lead to a given point on the moduli space.
Since $n=2$, the superpotential must have at least two critical
points, and to keep things simple we will consider the case of a
cubic superpotential, which has exactly two critical points.  In
section 2.2, we reviewed the recipe of \CIV\ for determining what
cubic superpotential, normalized to have leading term $x^3/3$,
leads to a given point on the $\CN=2$ moduli space. We take the
matrix model curve, which in this case is
 \eqn\mmcgo{y_m^2=R_{N-2s_+}(x) \tilde R_{N-2s_-}(x),}
and we write \eqn\mmcg{R_{N-2s_+}(x) \tilde R_{N-2s_-}(x) =W'(x)^2
+ f_1(x).} Recall that $s_++s_-=N-2$, and therefore the left hand
side of \mmcg\  is a quartic polynomial. \mmcg\ determines $W'$,
and hence determines $W$ up to an irrelevant additive constant.
Moreover, the value of the gluino bilinear $S$ can be read off
from $f_1(x)$; if $f_1(x)=Ax+B$, then $S=S_1+S_2=-A/4$.

The intersection  points of different branches are points where
$y_m^2$ has a double root and therefore equation \mmcgo\
describes a genus zero curve. This is consistent with the fact
that such points are also on the $n=1$ branch because, as
explained in section 2.2, when $n<k$ (here $n=1<k=2$), the matrix
model curve has double zeros. The physics of these points in the
$\CN=1$ theory is interesting. The low energy spectrum at these
points consists of the two $U(1)$ multiplets which exist at
generic points in $\CM_\alpha$ as well as another massless
monopole. Moving away from this point along $\CM_\alpha$, this
monopole acquires a mass.  Moving away from this point through
the condensation of this monopole takes us to the fully confining
vacua with unbroken $U(N)$ and only a single massless $U(1)$
multiplet.

If we denote the critical points of $W$ as $x_1$ and $x_2$, we
have
 \eqn\inco{W'=(x-x_1)(x-x_2).}
The physics is essentially unchanged by adding a constant to $x$,
which shifts $x_1$ and $x_2$ by a constant. The invariant
information contained in the choice of $W'$, modulo shifting $x$
and exchanging the two critical points, is contained in the
parameter
 \eqn\deltadefg{\Delta= (x_1-x_2)^2.}
It is possible and sometimes convenient to shift $x$ so that a
superpotential of given $\Delta$ takes the form
 \eqn\superex{W={x^3\over 3}-{\Delta\over 4} x.}
$W$ is odd under the transformation $x\to -x$, which thus
corresponds to an $R$-symmetry.  (Of course, the symmetry is
present whether or not we shift $x$ to put $W$ in the form
\superex.)  The critical points $x_1,x_2$ are exchanged by  the
symmetry, so a choice of classical vacuum with $N_1\not= N_2$
spontaneously breaks the symmetry.  In some examples, the
symmetry exchanges different branches $\CM_\alpha$, and in other
cases, it acts within a fixed branch.

The matrix model formulas can be used to compute $T(x)$.  Though
we will not do this in detail, it could be done as follows.
Following \CDSW,
 \eqn\tcom{T(x)=\left\langle \Tr {dx \over x-\Phi}\right\rangle
 ={P_N'(x)\over \sqrt{P_N^2(x) -4\Lambda^{2N}}}dx.}
{}From \facg\
 \eqn\Pprime{\eqalign{
 P_N'(x)\ &=H_{s_+}(x)\left(2 H_{s_+}'(x)R_{N-2s_+}(x)+H_{s_+}
 (x)R_{N-2s_+}'(x)\right) \cr
 &\ =\tilde H_{s_-}(x)\left(2 \tilde H_{s_-}'(x)\tilde
 R_{N-2s_-}(x)+ \tilde H_{s_-} (x)\tilde R_{N-2s_-}'(x)\right)\cr
 &\ = -{1\over 4} c(x) H_{s_+}(x)\tilde H_{s_-}(x).}}
Since we learn from the first two expressions that $P_N'(x)$ is
divisible by $H_{s_+}(x)$ and $\tilde H_{s_-}(x)$, $c(x)$ must be
a polynomial (the factor of $-1/4$ is in order to agree with the
conventions of \CDSW).  Therefore \tcom\ can be written as
 \eqn\tcomn{T(x)=-{c(x) \over 4y_m}dx}
which leads to the identification $c(x) = -4\left[W'(x)
T(x)\right]_+ = -4\Tr {W'(x) - W'(\Phi)\over x-\Phi}$ in \CDSW.

In some of our examples, we will find $S=0$, and find that this
can be interpreted as a consequence of the symmetry. However, the
symmetry is special to a cubic superpotential. A generic
superpotential of higher degree does not have such a symmetry and
it can still lead on certain branches to $S=0$, as we will
demonstrate in section 3.6. Lacking such a symmetry, vanishing of
$S$ on a branch of vacua would not be natural in a
nonsupersymmetric  field theory. With supersymmetry, such
``unnatural'' phenomena can be dictated by holomorphy. In section
4, we discuss more elaborate examples of such phenomena.

Let us consider the examples.

\subsec{$U(2)$}

Our first example is a $U(2)$ gauge theory.  The hyper-elliptic
curve \utg\ is
 \eqn\utoct{y^2=(P_2(x) + 2\Lambda^2)(P_2(x) - 2 \Lambda^2 ).}
The factorization problem \facg\ in this case is quite trivial,
since to get $n=2$ we do not need double roots at all -- just the
classical symmetry breaking $U(2)\to U(1)\times U(1)$. We write
 \eqn\utod{P_2(x)  = (x-a) (x-b)  }
Our moduli space $\CM$ with $n=2$ is parametrized by $a$ and $b$
modulo the exchange $a\leftrightarrow b$. In other words, it is
the whole $\CN=2$ Coulomb branch.

The semiclassical limit $\Lambda \to 0$ is also trivial since
$P_2(x)= (x-a) (x-b) $ is independent of $\Lambda$. Setting
$P_2(x)=\det(x- \Phi_{cl})$, we find $\langle \Phi_{cl} \rangle =
\pmatrix{a&0\cr 0&b}$, breaking $U(2)$ to $U(1)\times U(1)$.

The subspace with a massless monopole is determined by looking for
points where $y^2$ has a double roots.  This happens when
 \eqn\utopt{(a-b)^2 = \pm 8 \Lambda^2.}
These subspaces are also on the unbroken $U(2)$ branch
corresponding to $n=1$. If we shift $x$ such that $a=-b$ (as in
$SU(2)$), at these points
 \eqn\topto{P_2(x)=x^2\mp 2\Lambda^2,~~y^2=x^2(x^2\mp
 4\Lambda^4).}
Note that  $P_2(x)=x^2-2\rho^2 \Lambda^2$ with $\rho^4=1$; this is
equal to $2\Lambda^2\rho^2{\cal T}_2({x\over 2\rho \Lambda})$
from section 2.3.  After a small perturbation causing the
monopoles to condense, one of these points has confinement and
one has oblique confinement.

We now break $\CN=2$ to $\CN=1$ by turning on a cubic
superpotential chosen to put the system at the point $(a,b)$. We
find it through the matrix model curve
 \eqn\wpfto{y_m^2 = P_2^2-4\Lambda^4=\left((x-a) (x-b)\right)^2
   -4 \Lambda^4}
from which we derive
 \eqn\suft{\eqalign{
 W'(x) &=(x-a)(x-b) \cr
 f(x)&=-4\Lambda^4\cr
  S &=0 . }}
In this case, vanishing of $S$ can be attributed to the discrete
chiral symmetry.

If $W'$ is given, \suft\ determines $a$ and $b$ modulo the
exchange $a\leftrightarrow b$.  Since the moduli space is obtained
by dividing out by this exchange, there is in this example only
one vacuum for given $W$.  This occurs whenever all $N_i$ are 1,
as then there is no strong dynamics at low energies.

\subsec{$U(3)$}

Our next example is somewhat more interesting.  It is based on
the gauge group $U(3)$.

We start by finding the subspace of the $\CN=2$ moduli space with
one monopole.  The $U(3)$ curve
 \eqn\utct{y^2=(P_3(x) + 2\Lambda^3)(P_3(x) - 2 \Lambda^3 )}
should have a single double root.  The solution of this problem
has two branches labeled by $\eta=\pm 1$ depending on which factor
of \utct\ has the double root; i.e.\ $(s_+, s_-)=(0,1)$ or
$(1,0)$.  In either case, the factorization we need is
 \eqn\utd{\eqalign{
 &P_3(x) + 2 \eta \Lambda^3 = (x-a)^2 (x-b), \cr}}
 implying also
\eqn\nutd{ P_3(x) - 2 \eta \Lambda^3 = (x-a)^2 (x-b)
-4\eta\Lambda^3.} As in the previous example, $a$ and $b$ are the
parameters on the moduli space; now there is no symmetry
exchanging them.

The semiclassical limit $\Lambda \to 0$ leads to $P_3(x) \to
(x-a)^2 (x-b) $ and therefore it corresponds to $U(3) $ broken to
$U(2) \times U(1)$.  The existence of two branches -- via the
choice of $\eta$ -- has a simple explanation: the $SU(2)$ that is
unbroken classically is strongly coupled at low energies and has
two supersymmetric vacua, corresponding in our problem to the two
branches.

Each branch has three one-dimensional subspaces with one more
massless monopole.  They are determined by looking for points
where $y^2$ has two double roots.  This happens when
 \eqn\utpt{(a-b)^3 = 27 \eta \Lambda^3.}
Since these points have $(s_+,s_-)=(1,1)$, at these points the two
branches labeled by $\eta$ meet each other, and they also meet
the $n=1$ subspace of the $\CN=2$ moduli space.  Therefore, they
can be interpreted as unbroken $U(3)$ vacua, in keeping with the
discussion in section 2.4.  In fact, once we impose \utpt\ in the
form $(a-b)=3\rho\Lambda$ with $\rho^6=1$ and shift $x\to x
+{1\over 3}(2a+b)$, $P_3$ becomes $x^3 - 3 \rho^2\Lambda^2 x$
which is equal to $2\Lambda^3\rho^3{\cal T}_3({ x\over 2\rho
\Lambda})$ from section 2.3.

We now break $\CN=2$ to $\CN=1$ by turning on a superpotential
which puts the system at the point $(a,b)$ in the $\CN=2$ moduli
space.  We find it through the matrix model curve, as in \mmcg:
 \eqn\wpft{y_m^2 =(x-b) \left((x-a)^2 (x-b)  -4\eta \Lambda^3
 \right) = \left((x-a)(x-b)\right)^2 -4\eta\Lambda^3(x-b).}
{}From this we derive
 \eqn\suft{\eqalign{
 W'(x) &=(x-a)(x-b) \cr
 f(x)&=-4\eta\Lambda^3(x-b)\cr
  S &=\eta \Lambda^3. }}

For fixed superpotential, there are two values of $(a,b)$. Either
of them can be the double root of $P_3+2\eta\Lambda^2$, and the
vacuum also depends on the choice of $\eta$.  So for each tree
level superpotential, the system has four vacua with low energy
gauge group $U(1)^2$. Semiclassically, they can be interpreted as
follows. First, the unbroken $U(2)$ can be associated with either
one of the two different minima of the potential, spontaneously
breaking the global ${\bf Z}_2$ symmetry. Each of these choices
leads from the strong coupling $SU(2)$ dynamics to two vacua,
which differ by the sign of $ S =\eta \Lambda^3$.

\subsec{$U(4)$}

We now repeat the same analysis for $U(4)$.  We look for the
subspace of the $\CN=2$ theory with $N-n=4-2=2$ magnetic
monopoles. The two double roots of the $\CN=2$ curve
 \eqn\ufoc{y^2=(P_4(x)^2 + 2\Lambda^4)(P_4(x)^2 - 2 \Lambda^4)}
can be distributed between the two factors as $(s_+, s_-)=(1,1)$,
$(2,0)$ or $(0,2)$.

\bigskip
\noindent{\it Confining Branches, Monopoles Distributed As $(s_+,
s_-)= (2,0)$ or $(0,2)$}

With both double roots in the same factor of $P_4^2-4\Lambda^8$,
we get two branches, labeled by the choice of $\eta=\pm 1$:
  \eqn\ufd{\eqalign{
 &P_4(x) + 2 \eta \Lambda^4= (x^2-ax +b)^2 \cr
 &P_4(x) - 2 \eta \Lambda^4 = (x^2-ax +b)^2 -4\eta\Lambda^4.}}

The semiclassical limit of  $\Lambda \to 0$ leads to $P_4(x) \to
(x^2-ax +b)^2$, which has two double roots, so $U(4)$ is broken to
$U(2)\times U(2)$.  That this is the only classical limit for
$(s_+,s_-)=(2,0)$ or $(0,2)$ could be predicted from the reasoning
in section 2.4.

Each branch has four singular subspaces with another massless
monopole, i.e., on which \ $P_4(x) - 2 \eta \Lambda^4$ also has a
double root. This happens at the solutions of
 \eqn\moredr{(4b-a^2)^2=64\eta \Lambda^4.}
These points correspond to $(s_+,s_-)=(2,1)$ or $(1,2)$, so they
are on the branch with $n=1$ and unbroken $U(4)$.

As in \mmcg, the matrix model curve is
 \eqn\ufym{y_m^2=(x^2-ax +b)^2 -4\eta\Lambda^4}
from which we derive
 \eqn\suff{\eqalign{
 W'(x) &=x^2-ax+b \cr
 f(x)&=-4\eta\Lambda^4\cr
  S &=0.}}
We see that for each superpotential, we have two vacua (labeled by
$\eta$), one on each branch.  This can be understood
semiclassically as follows.  The low energy $U(2)\times U(2)$ is
characterized by scales which are given in the semiclassical limit
by $\Lambda_1^6=\Lambda_2^6 \approx {\Lambda^8 \over a^2-4b}$.
Gluino condensation in these groups leads to four vacua $S_1
\approx \pm \Lambda_1^3$, $S_2 \approx \pm \Lambda_1^3$. The
$\Z_2$ exchanges $S_1$ and $S_2$, and since it is an $R$-symmetry,
it acts by $S_1\leftrightarrow -S_2$.  There should therefore be
two $\Z_2$-invariant vacua, with $S_1=-S_2$ and $S=0$; these are
evidently
 the vacua that we have found on the branches with $(s_+,
s_-)=(2,0)$ or $(0,2)$.  We will clearly have to find the two
vacua with $S_1=+S_2$ and $S\not=0 $ elsewhere.

Of the four $U(2)\times U(2)$ vacua, the analysis of section 2.1
shows that two are confining and two are in a Coulomb phase. In
fact, the vacua that we have found are confining, because the
branch that we have just described can be constructed by
``multiplication by 2,'' of $P_2(x)=x^2-a x+b$ as in section 2.3,
\eqn\multw{ P_4(x) = 2\Lambda^4\rho^2 {\cal T}_2
\left({P_2(x)\over 2\rho \Lambda^2 } \right) }
with $\rho^4=1$ and $\rho^2=\eta$. This is the same as \ufd.

\bigskip
\noindent{\it Coulomb Branch, Monopoles Distributed as $(s_+,
s_-)=(1,1)$}

The other branch occurs when each of the factors $P_4(x)\pm 2
\Lambda^4$ has a single double root.  It is not too hard to show
that, modulo the freedom to add a constant to $x$, this implies
that
  \eqn\ufdt{\eqalign{
  &P_4(x) + 2 \Lambda^4= (x-a)^2\left[(x+a)^2 +{\Lambda^4 \over
  a^3}(x +2a)\right] \cr
  &P_4(x) - 2 \Lambda^4 =(x+a)^2\left[( x-a)^2 +{\Lambda^4 \over
  a^3}(x -2a)  \right]}}
for some $a$.

There now are two semiclassical limits with $\Lambda \to 0$:
\item{1.} Fixed $a$: $P_4(x) \to (x-a)^2(x+a)^2 $.
Therefore $U(4)$ is broken to $U(2)\times U(2)$.
\item{2.} $\Lambda,a\to 0$ with fixed $v=\Lambda^4/a^3 $:
$P_4(x) \to x^3(x+v)$, showing that here $U(4)$ is broken to $U(3)
\times U(1)$.

This is the first example of our new duality: the same moduli
space has two distinct semiclassical limits corresponding to
different gauge groups.  One can continuously deform a
$U(2)\times U(2)$ vacuum to a $U(3)\times U(1)$ vacuum.

Since the numbers 1 and 3 are relatively prime, these $U(3)\times
U(1)$ vacua are all in a Coulomb phase.  This whole branch is
therefore in a Coulomb phase, and the classical limit with $U(2)
\times U(2)$ will turn out to give the two $U(2) \times U(2)$
vacua that did not appear earlier. The upshot will be
 that $U(2)\times U(2)$ vacua that are
in a Coulomb phase can be smoothly transformed to $U(3)\times
U(1)$ vacua, while confining $U(2)\times U(2)$ vacua cannot make
such a transformation, since there are no confining $U(3)\times
U(1)$ vacua.

The special points on this branch where another monopole becomes
massless are at the solutions of
 \eqn\ufocc{16 a^8 = \Lambda^8.}
Explicitly, $a={\sqrt{2}\over 2}\rho \Lambda$ with $\rho^8=1$.
Note that shifting $x \to x-{\Lambda^4\over 4 a^3}$ and using the
solutions to \ufocc\ we get from \ufdt\ that $P_4(x) = 2\Lambda^4
\rho^4{\cal T}_4({ x\over 2\rho \Lambda})$. These are the same
four points that we found in \moredr.

{}From \ufdt, we find the matrix model curve, as in \mmcg:
 \eqn\ufommc{\eqalign{
 y_m^2=\ &\left[(x+a)^2 +{\Lambda^4 \over  a^3}(x +2a)\right]
 \left[( x-a)^2 +{\Lambda^4 \over  a^3}(x -2a)  \right]\cr
  =\ &\left(x^2 + x{\Lambda^4 \over a^3}-a^2 \right)^2
   - {4\Lambda^4\over a}x -{4\Lambda^8 \over a^4}.\cr}}
The superpotential and $f(x)$ are therefore
 \eqn\ufowpc{\eqalign{
 W'(x)&=x^2 + x{\Lambda^4 \over a^3}-a^2\cr
  f(x)&=-  {4\Lambda^4\over a}x -{4\Lambda^8 \over a^4}\cr
  S&={\Lambda^4\over  a}.}}

Finally, we count the number of vacua for fixed tree level
superpotential.  The most convenient way to proceed is to compute
the discriminant of the quadratic polynomial $W'=x^2+x(\Lambda^4/
a^3)-a^2$; it is \eqn\deltis{\Delta=4a^2+{\Lambda^8\over a^6}.} By
shifting $x$, we could put the superpotential in the form
\superex.  If $\Delta$ is given, then \deltis\ is equivalent to an
eighth order equation for $a$, namely
 \eqn\invo{4a^8-\Delta a^6+ \Lambda^8=0.}
So there are eight possible values of $a$ for every $\Delta$,
implying that the theory has eight vacua for each tree level
superpotential.  In the semiclassical limit of large
$\Delta/\Lambda^2$, these vacua can be understood as follows:
\item{1.} $U(3)\times U(1)$ leads to $3\times 2$ vacua.
The factor of $2$ arises from the broken global symmetry which
exchanges the two minima of the potential.  Indeed, for large
$\Delta/\Lambda^2$, \invo\ has six roots with $a\sim
(\Lambda^8/\Delta)^{1/6}$.
\item{2.} $U(2)\times U(2)$ leads to $2\times 2 -2=2$ vacua.  Here
we do not need to include the factor of $2$ which exchanges the
two minima of the potential and we subtract the two confining
vacua that we found on other branches.  To check this prediction,
we see that for large $\Delta/\Lambda^2$, \invo\ has two roots
with $a \sim(\Delta/4)^{1/2}$.

\subsec{$U(5)$}

 To find points on the Coulomb branch for $U(5)$ with three
 massless monopoles,
we want polynomials $P_5$ such that
 \eqn\ufint{ y^2 = (P_5(x) +2\Lambda^5) (P_5(x) - 2\Lambda^5) }
has $N-n=5-2=3$ double zeros.  This leads to two branches with
$(s_+, s_-)=(2,1)$ or $(1,2)$.   We will label them by $\eta=\pm
1$.  $P_5+2\eta\Lambda^5$ should have a pair of double roots, and
$P_5-2\eta\Lambda^5$ should have a single double root; we use the
freedom of shifting $x$ to place this double root at the origin.
Requiring the existence of the stated double roots leads to
 \eqn\cocp{\eqalign{
 P_5(x) + 2\eta\Lambda^5 =\ & (x^2+a x-2ac)^2(x+c)\cr
 P_5(x) - 2\eta\Lambda^5 =\ & x^2\left[x^3+(2a+c)x^2 + a(a-2c)
 x-ac(3a+4c)\right]}  }
where the parameters $a$ and $c$ are constrained to satisfy
\eqn\conss{ a^2 c^3 = \eta \Lambda^5 .}

{}From \conss\ it is clear that there are two classical limits: as
$\Lambda\to 0$, either  $a\to 0$ or $c\to 0$. In the former case,
$P_5(x)\to x^4 (x+c)$ corresponding to $U(4)\times U(1)$ while in
the latter, $P_5(x)\to x^3(x+a)^2$ giving $U(3)\times U(2)$.  This
is another example showing that the same branch can have different
classical limits.  (If we let both $a$ and $c$ to to zero, we will
get $P_5(x)=x^5$, corresponding to a classical limit with unbroken
$U(5)$.)

The points with another massless monopole correspond to
$(s_+,s_-)=(2,2)$.  They occur when $P_5(x) - 2\eta\Lambda^5$ has
two double roots.  This happens when $a^2+11 a c- c^2=0$ subject
to the constraint \conss. Solving for $c$ we get,
 \eqn\cinaa{ c = {1\over 2} (11+ 5 \epsilon \sqrt{5} ) a \qquad
 {\rm with} \qquad \epsilon^2=1 }
and
 \eqn\solcons{ a^5 = (-682+305\sqrt{5}\epsilon )\eta \Lambda^5=
 (-2+\epsilon\sqrt{5})^5 \eta \Lambda^5 .}
Equation  \solcons\ has 10 solutions for each $\eta$
 \eqn\solconss{a=\xi(-2+\epsilon\sqrt{5})\Lambda, \qquad
 \xi^5=\eta.}
However, using the freedom to shift $x$ such as to cancel the
$x^4$ term in $P_5(x)$ (bring it to the $SU(5)$ form), $x = \tilde
x-{1\over 2}a(3+\sqrt{5}\epsilon )$, \cocp\ becomes
 \eqn\unbrok{\eqalign{
 P_5(x) + 2\eta\Lambda^5 =\ & ( \tilde x+2\xi\Lambda) \left(
 \tilde x^2- \xi\Lambda \tilde x- \xi^2\Lambda^2 \right)^2\cr
 P_5(x) - 2\eta\Lambda^5 =\ &  ( \tilde x-2\xi\Lambda )\left(
 \tilde x^2 + \xi\Lambda  \tilde x - \xi^2\Lambda ^2  \right)^2
 } }
{}from which it is clear that there are only 5 different points
on each branch. Note that this is equal to $P_5(x)=
2\Lambda^5\xi^5{\cal T}_5({x\over 2\xi \Lambda })$ from section
2.3.

Why has the same $P_5$  appeared for two different values of the
pair $(a,c)$?  These values of $(a,c)$ are characterized by the
fact that $P_5(x)-2\eta\Lambda^5$ has a second double root -- but
there are two choices of which double root is the ``second'' one.

{}Using the recipe of \mmcg, we identify the matrix model curve,
 \eqn\ufimm{y_m^2=
 (x+c)\left[x^3+(2a+c)x^2+a(a-2c)x-ac(3a+4c)\right].}
{}From this we find
\eqn\soll{\eqalign{
 W'(x) =\ & x^2+(a+c)x-ac  \cr
  f(x) =\ &-4c^2 a(x+a+c)\cr
   S  = &c^2 a.}}

Finally, we count the number of vacua for fixed tree level
superpotential, by the same reasoning as in previous examples. By
computing the discriminant of $W'$, we get $\Delta = a^2+6ac+c^2$.
Using \conss\ to solve for $c$, we find that for fixed $\Delta$
there are 20 vacua, corresponding to the roots of
 \eqn\nice{ a^8(a^2-\Delta )^6 - 2a^4(17 a^2+\Delta )(1153 a^4
 + 142 a^2 \Delta + \Delta^2) \Lambda^{10} + \Lambda^{20}= 0.}
In the semiclassical limit of large $\Delta/\Lambda^2$, these
vacua can be understood as follows:
\item{1.} $U(4)\times U(1)$ leads to $4\times 2=8$ vacua,
with $a\sim \exp(\pm 2\pi i/8) (\Lambda^{10}/\Delta^3)^{1/4}$.
\item{2.} $U(3)\times U(2)$ leads to $3\times 2\times 2=12$ vacua,
which are at roughly $a\sim\pm \sqrt \Delta$.

We have not found any confining branches for $U(5)$ for the simple
reason that 5 is a prime number, so if $5=N_1+N_2$, then $N_1$
and $N_2$ are relatively prime. Hence for $U(5)$, all vacua are
 in a Coulomb phase.

\subsec{$U(6)$}

$U(6)$ is the last and richest example that we will examine in
detail. We look for points in the Coulomb branch of the ${\cal
N}=2$ theory with $N-n=6-2=4$ double zeros. These can be
distributed between the two factors of
 \eqn\uscurve{y^2= P_6(x)^2-4\Lambda^{12} = (P_6(x)-2\Lambda^6)
 (P_6(x)+2\Lambda^6)}
as $(s_+, s_-)=(3,1)$, $(1,3)$ or $(2,2)$.

\bigskip
\noindent {\it Confining $U(2)\times U(4) $ Branches, Monopoles
Distributed As $(s_+, s_-)=(1,3)$ or $(3,1)$}

We first wish to consider the branches of the moduli space of
$U(6)$ with four massless magnetic monopoles distributed as
$(3,1)$ and $(1,3)$ between the two factors. The solution of the
factorization problem \facg\ is
 \eqn\usfac{\eqalign{
 P_6(x) +2\eta \Lambda^6 & = \left[ (x-a)^2(x-b)-2\epsilon
 \Lambda^3 \right]^2 \cr
 P_6(x) -2\eta \Lambda^6 & = (x-a)^2 (x-b)\left[ (x-a)^2(x-b)
 -4\epsilon \Lambda^3 \right]
 }  }
with $\eta^2=1$ and $\epsilon^2 = \eta$.

The semiclassical limit $\Lambda\to 0$ leads to $P_6(x)\to
(x-a)^4(x-b)^2$, and therefore $U(6)$ is broken to $U(4)\times
U(2)$.  That this is the only semi-classical limit with $(s_+,s_-)
=(3,1)$ or $(1,3)$ follows from the reasoning in section 2.4.

The points with another massless monopole occur when $P_6(x)
-2\eta\Lambda^6$ has two double roots. This happens when $(a-b)^6
= (3\Lambda)^6$.  These special points on the $(s_+,s_-) =(3,1)$
branches have $(s_+,s_-)=(3,2)$; they also meet the $(s_+,s_-)
=(2,2)$ branches we will discuss below and the $U(6)$ branch with
$n=1$. The special points on the $(s_+,s_-) =(1,3)$ branches have
$(s_+,s_-) =(2,3)$ and they meet the $(s_+,s_-) =(2,2)$ branches
and the $n=1$ branch.

Using \mmcg, the matrix model curve is found from \usfac\ to be
 \eqn\mmus{y_m^2=(x-a)^2(x-b)^2 -4\epsilon \Lambda^3(x-b)  . }
{}From this we derive
 \eqn\sous{ \eqalign{ W'(x)=\ & (x-a)(x-b)\cr
                     f(x) =\ & -4\epsilon \Lambda^3 (x-b) \cr
          S =\ &\epsilon \Lambda^3  . } }

For fixed superpotential $\Delta = (a-b)^2$ there are 8 solutions.
They correspond to the four solutions of $\epsilon^4=1$ for each
choice of $(a-b) = \pm \sqrt{\Delta }$. In the semiclassical limit
these 8 vacua can be understood as follows. $U(4)\times U(2)$
gives a total of $(4\times 2)\times 2 =16 $ vacua, where $4\times
2$ reflects the strong dynamics and a factor of 2 comes from the
broken ${\bf Z}_2$ symmetry. But, following the analysis in
section 2.1, only half of these vacua (those with $b$ even) are
confining.  So we expect 8 confining vacua.

The eight vacua with $(s_+,s_-)=(3,1)$ or $(1,3)$ actually are
confining, since they can be obtained by  applying the
``multiplication by 2'' map to the polynomial
$P_3(x)=(x-a)^2(x-b)-2\rho \Lambda_0^3$ with $\rho^2=1$, which
describes the breaking $U(3)\to U(2)\times U(1)$ (and was
presented in \utd). Following the general discussion of section
2.3, the scales of the theories are related by $\Lambda_0^6 =
\sigma^2 \Lambda^6$ with $\sigma^4=1$. Using this in
$P_6(x)=2\Lambda^6\sigma^2{\cal T}_2({P_3(x)\over
2\sigma\Lambda^3})$ we recover \usfac\ with $\epsilon
=\rho\sigma$.

\bigskip
\noindent {\it Confining $U(3)\times U(3) $ Branches, Monopoles
Distributed as $(s_+, s_-)=(2,2)$}

We now consider the case of $(s_+, s_-)=(2,2)$.  In this case, the
factorization problem \facg\ has two types of solution. The first
case leads to three branches parametrized by a cube root of unity
$\rho$
 \eqn\conbranch{\eqalign{
 &P_6(x)+2\Lambda^6=(x^2+g -\rho \Lambda^2)^2 (x^2+g +
 2\rho\Lambda^2) \cr
 &P_6(x)-2\Lambda^6=(x^2 + g+\rho\Lambda^2)^2
 (x^2+g-2\rho\Lambda^2)  \cr
 &P_6(x)=(x^2 + g)[(x^2+g)^2 -3\rho^2\Lambda^4].}}
These three branches are distinct.  Since $P_6$ depends on
$\Lambda^4$ and not on $\Lambda^2$, the change $\Lambda^2 \to
-\Lambda^2$ (and therefore $\Lambda^6\to - \Lambda^6$) does not
lead to more branches. Instead, there is a global symmetry $g\to
-g$ which acts on each branch.

In the classical limit $\Lambda \to 0$ we have $P_6(x) \to
(x^2+g)^3$ and therefore $U(6)$ is broken to $U(3) \times U(3)$.
On each of these branches there are two ``$U(6)$ points'' --
points with five double roots: $g= \pm 2\rho\Lambda^2$.

We now break $\CN=2$ to $\CN=1$ by turning on a superpotential $W$
which puts the system in the ground state labeled by $g$.  It is
determined from the matrix model curve, which is found, as in
\mmcg, from \conbranch:
 \eqn\mmcurm{y_m^2= (x^2+g + 2\rho\Lambda^2) (x^2+g -
 2\rho\Lambda^2) =(x^2+g)^2 - 4\rho^2\Lambda^4.}
{}From this, we find
\eqn\uswfs{ \eqalign{ W'(x)=\ & x^2+g \cr
                       f(x)=\ & -4\rho^2 \Lambda^4 \cr
                    S  =\ & 0.  }  }
These expressions are consistent with the ${\bf Z}_2$ R-symmetry
which maps $\Phi\to -\Phi$ and $S\to -S$.

{}From the expression $W'=x^2 + g$, it is clear that $W$
determines $g$ and hence $P_N$, so for fixed $W$, there is a
single quantum vacuum on each branch. The three branches give a
total of three vacua. Semiclassically, these  can be interpreted
as coming from $U(6)\to U(3)\times U(3)$, with the two gluino
condensates in the two $SU(3)$ factors anti-aligned.  As in the
discussion after \suff\ of $U(4)\to U(2)\times U(2)$, this leads
to $S=S_1+S_2=0$, which respects the global ${\bf Z}_2$ R-symmetry
of the system.

$U(3)\times U(3)$ is expected to give a total of $3\times 3=9$
vacua, but according to the analysis in section 2.1, only 3 of
them (those with $b=0$) are confining.  We claim that the three
vacua obtained by the construction above are confining. In fact,
they can be obtained by the ``multiplication by 3'' of $P_2(x) =
x^2+g$ which is a solution to $U(2)\to U(1)\times U(1)$. One can
check that $P_6(x)=2\Lambda^6\epsilon^3 {\cal T}_3({P_2(x)\over
2\epsilon \Lambda^2})$ reproduces \conbranch\ with $\epsilon^6
=1$ and $\epsilon^2 = \rho^2$ cubic roots of unity.

The fact that these three branches are confining, with a
confinement index of 3, explains why they have no semi-classical
limit with $U(4)\times U(2)$ or $U(5)\times U(1)$.  There are
three branches for a reason explained in section 2.1; they are
distinguished by which of $W^uH$, for $u=0,1,2$, has  no area law.

\bigskip
\noindent {\it Coulomb Branch, Monopoles Distributed as $(s_+,
s_-)=(2,2)$}

The second kind of solution of the factorization problem \facg\
in this case is
 \eqn\Coubranch{\eqalign{
 &P_6(x)+2\Lambda^6= \left[x^2 + (h+g)x+{(3h+g)(9h^3+15h^2g -hg^2+g^3)
 \over 108 h^2}\right]^2\cr
 &\qquad\qquad
 \qquad\qquad \left[x^2 -{(h-g)(3h-g)^2(3h+g)\over 108
 h^2}\right]\cr
 &P_6(x)-2\Lambda^6= \left[(x+{2g\over 3})^2 + (h-g)(x+{2g\over 3})
 +{(3h-g)(9h^3-15h^2g -hg^2-g^3) \over 108 h^2}\right]^2\cr
 &\qquad\qquad \qquad\qquad \left[(x+{2g\over 3})^2
 -{(h+g)(3h+g)^2(3h-g)\over 108 h^2}\right]}}
with $g$ and $h$ satisfying the constraint
 \eqn\constraint{g^5(g^2-9h^2)^2 = 27^3h^3\Lambda^6.}
We used a parametrization which makes the global symmetry $g\to
-g$ combined with $\Lambda^6\to -\Lambda^6$ manifest.

The classical limit $\Lambda\to 0$ is obtained at:
\item{1.} $g \to 0$ with finite $h$. Here $P_6(x) \to
(x+{h\over 2})^5 (x-{h\over 2})$; i.e.\ $U(6) \to U(5) \times
U(1)$.
\item{2.} $g \to 3h$ with finite $h$.  Here $P_6(x) \to x^2
(x+2h)^4$; i.e.\ $U(6) \to U(4) \times U(2)$.
\item{3.} $g \to -3h$ with finite $h$.  Here $P_6(x) \to (x-2h)^2
x^4$; i.e.\ $U(6) \to U(4) \times U(2)$.
\item{4.} $h,g \to 0$ with $v=g^2/h$ fixed.  Here $P_6(x) \to
(x^2+{v^2 \over 108})^3$; i.e.\ $U(6)\to U(3)\times U(3)$.

This is a richer example demonstrating that the same branch can
have different classical limits.  Since 1 and 5 are relatively
prime, what we have found is clearly a Coulomb branch.

The $U(6)$ points are obtained when either $P_6(x) +2\Lambda^6$
or $P_6(x) -2\Lambda^6$ acquires another double zero.  This
happens when
 \eqn\uspts{{(h+ g)(3h+ g)^2(3h -g)\over 108 h^2}=0}
or the same equation with $g\to -g$. Taking into account
\constraint, this happens only for
 \eqn\usptss{h=g\quad {\rm with~} g {\rm ~a ~solution ~of}
 \quad (2g)^6 = 27^3\Lambda^6}
or
 \eqn\usptss{h=-g\quad {\rm with~} g {\rm ~a ~solution ~of}
 \quad (2g)^6 = -27^3\Lambda^6.}
Superficially, this leads to 12 points.  However (after an
appropriate shift of $x$) one finds that at these points $P_6(x)
\pm 2\Lambda^6$ depend only on $g^2$ and therefore there are only
6 such points. Indeed, both solutions lead to $P_6(x) =
2\Lambda^6 \rho^6 {\cal T}_6({x\over 2\rho\Lambda})$ with
$\rho^{12}=1$. At these points, this branch meets the various
confining branches that we have found.

We now break $\CN=2$ to $\CN=1$ by turning on a superpotential
$W$.  It is determined from the matrix model curve
 \eqn\mmcurveco{\eqalign{ y_m^2=\ &\left[x^2
 -{(h-g)(3h-g)^2(3h+g)\over  108 h^2}\right] \left[(x+{2g\over
 3})^2 -{(h+g)(3h+g)^2(3h-g)\over 108 h^2}\right] \cr
 =\ & \left(x^2 + {2\over 3}  g x + {g^4 - 6 g^2 h^2 - 27 h^4
 \over 108 h^2}\right)^2 - {4 g^2(g^2 - 9 h^2)\over 81h} x +{2 g^2
 (g - 3 h)^3(g + 3 h) \over 729 h^2}
  }}
and therefore
 \eqn\wcouf{\eqalign{
 W'(x) &=x^2 + {2g\over 3} x + {g^4 - 6g^2 h^2 - 27 h^4 \over
 108 h^2} =\tilde x^2 + {g^4-18g^2h^2  - 27 h^4 \over
 108 h^2} \cr
 f(x) & = - {4 g^2(g^2 - 9 h^2)\over 81h}x +{2 g^2
 (g - 3 h)^3(g + 3 h) \over 729 h^2} = - {4 g^2(g ^2- 9 h^2)
 \over 81h} \tilde x +{2 g^2 (g^4 - 81 h^4)\over 729 h^2} \cr
 S & ={g^2(g^2 - 9 h^2)\over 81h}  . }}
We used the freedom to shift $x$ by $x= \tilde x -{g\over 3}$ to
put the superpotential $W$ in a canonical form $W={\tilde
x^3\over 3} -{\Delta\over 4}\tilde x$.

We would like to parametrize the theory in terms of $\Delta$.
Including the constraint \constraint, we have two equations with
two unknowns:
 \eqn\deltas{\eqalign{
 &\Delta={27 h^4-g^4+18g^2h^2  \over 27 h^2} \cr
 &g^5(g^2-9h^2)^2 = 27^3h^3\Lambda^6.}
 }
Using these equations we can solve for $h$,
 \eqn\hsold{h= {4g^5(16g^{10}(g^2-{9\over 4}\Delta)^2 - 3({9\over
 4}\Delta) (3\Lambda)^{12} )\over 9(3\Lambda)^6 (32 g^{10}(g^2 -
 {3\over 4}\Delta)  + 3(3\Lambda)^{12})}}
and find a single equation for a single unknown $g$,
 \eqn\newcon{ g^{16}\left(g^2-{9\over 4} \Delta\right)^4  =(3
 \Lambda)^{12} \left( {9\over 8} g^{12} - {3\over 8}
 \left({9\over 4}\Delta\right)^2 g^8
 +{1 \over 4}\left({9\over 4} \Delta\right)^3g^6 +{27
 \over 256}(3\Lambda)^{12}\right).}
In deriving this equation we assumed that $g \not=0$.

Since the final constraint \newcon\ is of degree 24, for fixed
superpotential (fixed $\Delta$) there are 24 vacua. These can be
interpreted as follows:
\item{1.}  $U(1)\times U(5)$ vacua. Their number is
$1 \times 5 \times 2 =10$ (the factor of 2 is associated with the
global $\Z_2$ symmetry of the theory).
\item{2.} $U(2)\times U(4)$ vacua.  Their number is
$2\times 4 \times 2 -8=8$  (we subtracted the 8 confining vacua).
\item{3.} $U(3)\times U(3)$ vacua.
Their number is $3\times 3-3 =6$ (we subtracted the 3 confining
vacua, and we did not multiply by 2 because for $U(3)\times U(3)$
the ${\bf Z}_2$ symmetry is not broken).

It is interesting to contrast this situation with the classical
theory ($\Lambda\to 0$).  Here we find 3 solutions: $g=0,\pm
{3\over 2}\sqrt{\Delta}$.  The corresponding values of $h$ depend
on the details of the limit.  The two solutions with $g=\pm
{3\over 2}\sqrt{\Delta}$ correspond to $U(2)\times U(4)$ vacua
(compare with the analysis of the semiclassical limits above).
Each of them splits to 4 vacua in the quantum theory.  The
solution $g=0$ corresponds to $U(1)\times U(5)$ and to $U(3)\times
U(3) $.  This point splits in the quantum theory to 16 vacua
(compare with the analysis of the semiclassical limits above).

\subsec{Examples with $s_-=0$}

This completes what we will say by way of systematic analysis for
relatively small $N$. In this subsection, we study another kind of
example in which the factorization problem is easily solved
explicitly.  This is the case that all the double roots are in one
factor of the $\CN=2$ curve -- for example, $s_-=0$ and therefore
$s_+=N-n$. There is an analogous discussion for $s_+=0$. The
condition $0\le s_+\le N/2$ shows that this is possible only when
$n\le N \le 2n$. The factorization problem is solved by
 \eqn\facs{\eqalign{
 &P_N(x) + 2\Lambda^N=H_{s_+}^2(x) R_{N-2s_+}(x) \cr
 &P_N(x) - 2 \Lambda^N=H_{s_+}^2(x) R_{N-2s_+}(x) -4\Lambda^N }}
with arbitrary coefficients in $H_{s_+}^2(x) $ and
$R_{N-2s_+}(x)$.

{}From \facs, we easily find the semiclassical limit
 \eqn\semisl{P_N(x) \to H_{s_+}^2(x) R_{N-2s_+}(x).}
We see that $U(N)$ is broken to $U(2)^{s_+}\times U(1)^{N-2s_+}$.

{}From \facs\ the matrix model curve is
 \eqn\mmctz{y_m^2=R_{N-2s_+}(x)\left[H_{s_+}^2(x) R_{N-2s_+}(x)
 -4\Lambda^N \right]=\left[H_{s_+}(x) R_{N-2s_+}(x)\right]^2
 -4\Lambda^N R_{N-2s_+}(x).}

If $n+1\le N$, the degree of $R_{N-2s_+}(x)$ is $N-2s_+=2n-N\le
n-1$, and we immediately find the superpotential and $f(x)$:
 \eqn\supftz{\eqalign{
 W'(x)=\ &H_{s_+}(x) R_{N-2s_+}(x)\cr
 f(x)=\ &-4\Lambda^N R_{N-2s_+}(x)}}
If $n+2\le N$, the degree of $f(x)$ is smaller than $n-1$ and we
conclude that $S=0$.

We have seen an example of this when we studied $n=2$ with $N=4$
above with $(s_+,s_-)=(2,0)$.  As a more interesting example,
consider $N=5$ with $(s_+,s_-)=(2,0)$ and therefore $n=3$.
Following the general solution \facs, we find
 \eqn\facsf{\eqalign{
 &P_5(x) + 2\Lambda^5=(x^2+ax+b)^2(x+c) \cr
 &P_5(x) - 2 \Lambda^N=(x^2+ax+b)^2(x+c) -4\Lambda^5. }}
In the semiclassical limit, $P_5(x) \to (x^2+ax+b)^2(x+c) $, and
therefore $U(5)$ is broken to $U(2)\times U(2)\times U(1)$.

{}From  \mmctz,\supftz, we deduce
 \eqn\mmctzf{\eqalign{
 y_m^2=\ &\left[(x^2+ax+b)(x+c)\right]^2 -4\Lambda^5 (x+c)\cr
   W'(x)=\ & (x^2+ax+b)(x+c) \cr
 f(x)=\ &-4\Lambda^5  (x+c)\cr
 S=\ & 0.}}
The theory has a quartic superpotential, and therefore there is no
global symmetry.  Yet we still find $S=0$.  This is a special case
of a more general phenomenon we will see in the next section.

\subsec{Interpolation and duality}

As discussed in section 2.4, a given branch, Coulomb or confining, can have
different classical limits. We have given explicit examples of
the interpolation between classical limits $U(N_1)\times U(N_2)$ and
$U(\tilde N_1)\times U(\tilde N_2)$.  Since the $N_i$ are certain periods of
the one-form $T$, such an interpolation must involve a rearrangement of the compact
one-cycles on the Riemann surface.
With some numerical work, we have determined how this happens in all the examples
of section 3.  Here we will simply illustrate the result qualitatively, without
attempting any proofs.   In the process, we will
see that the two descriptions are related to each other through an
electric-magnetic duality.

For a general breaking pattern $U(N)\to \prod_{i=1}^nU(N_i)$, there are $2n-1$ compact
cycles, as sketched (for $n=3$) in figures 1 and 2.  There are $n$ cycles $A_1,\dots,A_n$
that in a weak coupling limit surround a pair of roots, and $n-1$ cycles $C_i$ that
connect neighboring pairs.  The intersection numbers are $A_i\cap A_j=C_i\cap C_j=0$,
 $A_i\cap C_j=\delta_{i,j+1}-\delta_{ij}$.
In particular, the sum $v=A_1+A_2+\dots +A_n$
is a null vector.  In fact, it is clear from figure 1 that this cycle is homologous to
a cycle at infinity that winds once around the whole $x$-plane; this cycle will
not be affected by the motion of the roots. So the null vector $v$ will be  invariant
under the monodromy.   If the action of the monodromies on $A_i$ and $C_j$,
$i,j<n$, is known, then the behavior of $A_n$ is determined using the fact that $v$ is
invariant.

To illustrate, we will examine examples with $n=2$. In this case, there are just three
cycles, $A_1$, $A_2$, and $C=C_1$.  If we know what happens to $A_1$ and $C$, the
behavior of $A_2$ is known.  Given the intersection relations of $A_1$ and $C$,
the monodromies must act on them by an $SL(2,{\bf Z})$ transformations plus possible
addition of a multiple of the null vector $v$.  (The null vector is
uniquely determined by the fact that the $\tilde N_i$ are  non-negative.)

Now we will sketch the  mechanism of
interpolation we found to operate in all examples.
These branches all have complex dimension two, but one complex parameter
is associated with
the freedom of shifting $x$ in $P_N(x)$. This parameter can be removed,
leaving only a space with complex dimension one.
We want to discuss how the zeroes of $P_N(x)^2-4\Lambda^{2N}$ move and the cycles evolve
as one interpolates from a classical limit with unbroken $U(N_1)\times U(N_2)$ to
a classical limit with $U(\tilde N_1)\times U(\tilde N_2)$.

We start in the semiclassical limit with the group $U(N_1)\times U(N_2)$. As
discussed in section 2, the polynomial $P_N(x)^2-4\Lambda^{2N}$ is given in
this limit in terms of the Chebyshev polynomials of the two factors $U(N_1)$
and $U(N_2)$. Each factor leads to $N_i-1$ double roots and two single roots.
This is depicted in figure 4a for $N_1=N_2=2$.

\bigskip
\centerline{\epsfxsize=1.05\hsize\epsfbox{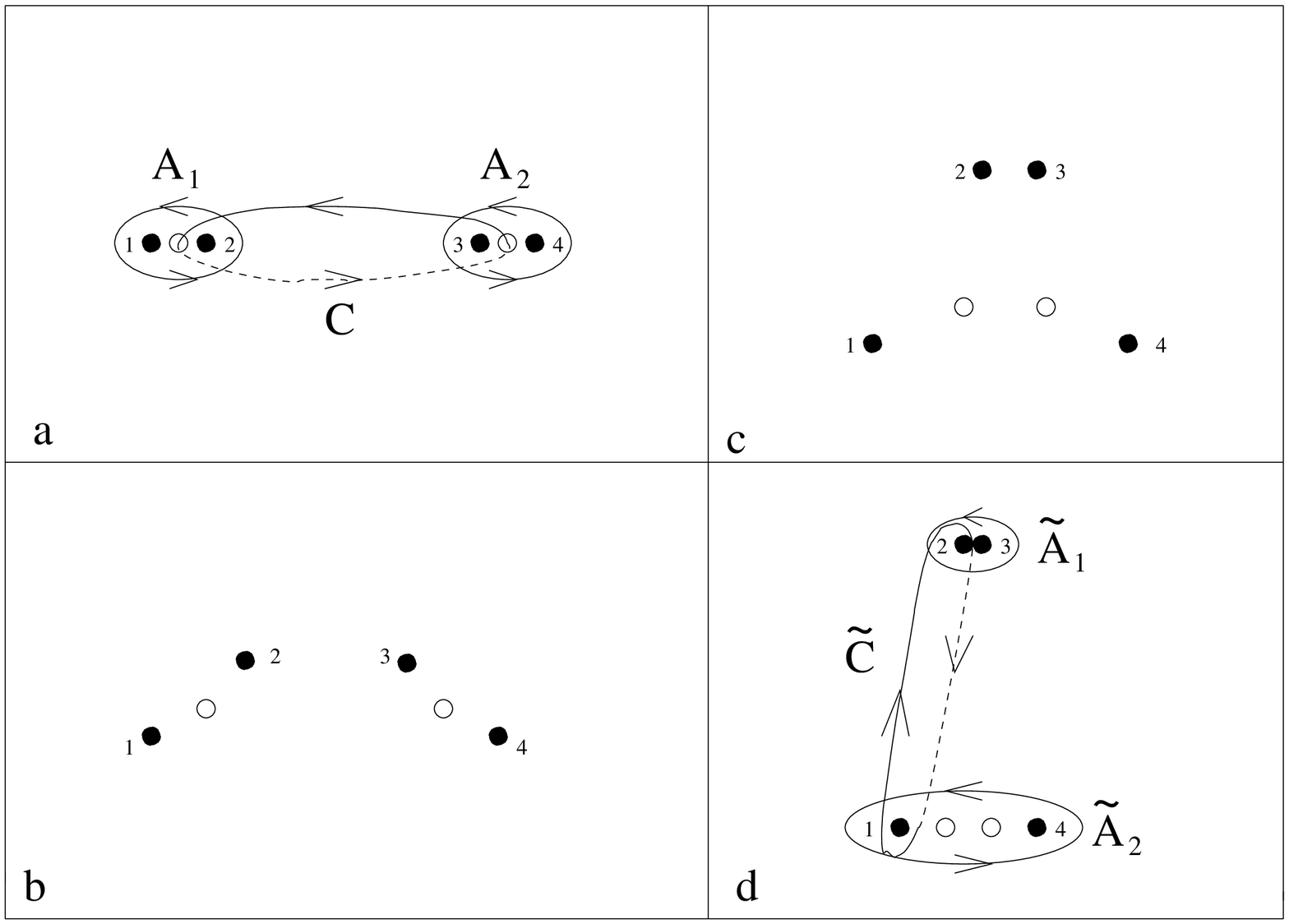}}
\noindent{\ninepoint\sl \baselineskip=8pt {\bf Figure 4}:{\sl
$\;$ Interpolation between classical limits in the Coulomb branch of
$U(4)$. The polynomial $P^2_4(x)-4\Lambda^8$  has
 four single roots (labeled 1,2,3 and 4) and two double roots; single
and double roots are depicted by black and white circles
respectively. (a) $U(2)\times U(2)$ semiclassical limit. Natural
basis of compact cycles $A_1$, $A_2$ and $C$. (b) Semiclassical
regime of $U(2)\times U(2)$ and transition to
strong coupling of $U(1)\times U(3)$.
(c) Strong coupling of $U(2)\times U(2)$ and transition to
semiclassical regime of
$U(1)\times U(3)$. (d) $U(1)\times U(3)$ semiclassical limit. Natural
basis of compact cyles  $\tilde A_1$, $\tilde A_2$ and $\tilde C$}}
\bigskip

As we vary the superpotential to go to a strongly coupled region,
 the two groups mix (figures 4b and 4c)
and the zeroes then separate
into two new groups (figure 4d).  One group has two single roots and
$\tilde N_1$ double roots, and the other has two single roots and $\tilde N_2$ double
roots.

It is easy to see the way the cycles are transformed.  Following the steps from
figure 4a to figure 4d, $A_1$ evolves into $-\tilde C$ and $C$ evolves into $\tilde A_1$.
We can describe this by the monodromy matrix
\eqn\transcy{ \pmatrix{\tilde A_1 \cr  \tilde C }=
\pmatrix{ 0 & 1 \cr -1 & 0 } \pmatrix{A_1 \cr C }. }
In this example, with this choice of cycles, the monodromy is in $SL(2,{\bf Z})$, with no
addition of a null vector.  In fact,
the transformation matrix in \transcy\
is the $S$ generator of $SL(2,{\bf Z})$, revealing the electric-magnetic nature
of the duality.  Since
  $ A_1+ A_2 = \tilde A_1 + \tilde A_2$, it follows from the above that
\eqn\rcy{ \tilde A_2 = A_1 + A_2 - C.}

Since $N_i=\oint_{A_i}T$ and $c=\oint_CT$, we have
\eqn\hcy{\eqalign{ \tilde N_1 & = c \cr
                   \tilde N_2 & = N_1+N_2-c \cr
                    \tilde c  & = -N_1.\cr}}
For example, if we start with the Coulomb vacuum $N_1=N_2=2$, $c=1$,
then we get $\tilde N_1=1$, $\tilde N_2=3$, $\tilde c=-2$. This is the
case depicted in figure 4. For producing such an interpolation, we used
$P_4(x)$ in \ufdt.
It is convenient to introduce a dimensionless parameter $
{\Lambda\over a}$. The figures, from $4a$ to $4d$, show the roots of
$P_4(x)^2-4\Lambda^8$
as the parameter changes from $0$ to $\infty$. Other examples from section 3
can be described similarly.

Let us now interpret this in the $\N =1$ theory, where superpotentials of the
form \tuggu\ describe the low energy dynamics.
To construct the $\N=1$ effective superpotential for the gluino bilinears $S_i$,
we need the periods of the differential $R(x)$ as well as those of $T$.
We have as in \CIV\ and eqn. \noggo\
\eqn\nsno{\eqalign{S_i&=\oint_{A_i}R,\cr
                   {1\over 2\pi i}{\partial{\cal F}\over \partial S_i}&=
                               \oint_{B_i}R.}}
To evaluate these expressions, we need the transformation laws of the noncompact
cycles $B_i$.  These are uniquely determined modulo possible addition of a null
vector by the intersections with the compact cycles.  Since the monodromy must
also preserve $B_1\cap B_2=0$, it is fixed modulo addition of a common null vector to
$B_1$ and $B_2$, and this is irrelevant in the sense that it is equivalent to
a $2\pi$ shift in the theta angle.  So we get
\eqn\hno{\eqalign{\tilde B_1 & = A_1+B_2\cr
                  \tilde B_2 & = B_2.}}
Consequently, the transformation laws are
\eqn\tanpp{\eqalign{ \tilde S_1 = &\
-{1\over 2\pi i}\left( {\del {\cal F}\over \del S_1} -
{\del {\cal F}\over \del S_2} \right) \cr
\tilde S_2 = &\ S_1+S_2 + {1\over 2\pi i}\left( {\del {\cal F}\over \del S_1} -
{\del {\cal F}\over \del S_2} \right)\cr
{1\over 2\pi i}{\partial\tilde {\cal F}\over \partial \tilde S_1}&= S_1+{1\over 2\pi i}
{\partial {\cal F}\over \partial S_2}\cr
{1\over 2\pi i}{\partial\tilde {\cal F}\over \partial \tilde S_2}&={1\over 2\pi i}
{\partial {\cal F}\over \partial S_2}\cr}
}

\newsec{Equations Characterizing Components Of Moduli Space}

With the exception of section 3.6, we have discussed so far
examples in which the rank $n$ of the low energy gauge group is
small and the number $w=N-n$ of condensed monopoles is
correspondingly large.  Here we will concentrate on examples with
relatively small $w$, and correspondingly a large rank of the low
energy gauge group $U(1)^n$.

If $U(N)$ is broken classically to $\prod_iU(N_i)$, then
$w=\sum_i(N_i-1)$.  This implies that if $w<N/2$, a typical range
in the following discussion, then some of the $N_i$ are 1.  This
means that all vacua are Coulomb vacua and confinement is not
relevant in distinguishing different branches.  Nevertheless,
there are different branches of the moduli space of vacua, for a
reason explained in section 2.4.  The different branches
correspond to decompositions $w=s_++s_-$, where $s_+$ and $s_-$
are, respectively, the number of double roots of the two
polynomials $P_N(x)+2\Lambda^{2N}$ and $P_N(x)-2\Lambda^{2N}$.

Not having confinement as a useful order parameter and with only a
very limited role for discrete symmetries, there appear to be no
conventional order parameters that explain the existence of $w+1$
branches labeled by $s_+$.  Instead, as sketched in the
introduction, we will seek here to characterize each branch by
describing holomorphic functions of chiral operators that have a
vanishing expectation value on a given branch.  The functions with
this property will depend on $s_+$ and $s_-$.

Certainly, there is no loss of essential generality in focusing on
the case $s_+\geq s_-$.  If $s_+=s_-$ or $s_+=s_-+1$, then the
analysis below does not reveal any unusual equations obeyed on a
given branch.   For each $w$, precisely one of these possibilities
is realized, and  we know of no unusual equations obeyed on this
branch. We will find such equations in all the other cases
\eqn\inpo{s_+\geq s_-+2,} and these relations will depend on
$s_-$.

To make explicit the double roots, we write
\eqn\dblroots{\eqalign{P_N+2\Lambda^{2N} & = H_{s_+}^2
R_{N-2s_+}\cr
                       P_N-2\Lambda^{2N} & = \tilde H_{s_-}^2
\tilde R_{N-2s_-}.\cr}} The curve of the ${\cal N}=2$ theory is
\eqn\ntwocu{y^2=(P_N+2\Lambda^{2N})(P_N-2\Lambda^{2N}) =(\tilde
H_{s_-}H_{s_+})^2 R_{N-2s_+}\tilde R_{N-2s_-}.} However, in the
matrix model we remove the quadratic factor $(\tilde
H_{s_-}H_{s_+})^2$, setting $y_m=y/\tilde H_{s_-}H_{s_+}$, so
\eqn\mblroots{y_m^2=R_{N-2s_+}\tilde R_{N-2s_-}.} We write this
\eqn\nblroots{y_m^2={R_{N-2s_+}(H_{s_+}^2R_{N-2s_+}-4\Lambda^{2N})\over
\tilde H_{s_-}^2}=\left({H_{s_+}R_{N-2s_+}\over \tilde
H_{s_-}}\right)^2\left({1-{4\Lambda^{2N}\over
H_{s_+}^2R_{N-2s_+}}}\right).} As shown in \CDSW, the order
parameters $t_r=-(1/32\pi^2)\Tr\, \Phi^r W_\alpha^2$ are
\eqn\ordpar{t_r=\oint_\infty {dx\,x^r y_m}=\oint_\infty {dx\,x^r
}{H_{s_+}R_{N-2s_+}\over \tilde
H_{s_-}}\sqrt{1-{4\Lambda^{2N}\over H_{s_+}^2R_{N-2s_+}}}.} Here
the contour is over a large circle near infinity in the complex
$x$-plane (and we recall that a factor of $1/2\pi i$ is included
in the definition of the symbol $\oint$). A standard argument
shows that the factor $\sqrt{1-{4\Lambda^{2N}/
H_{s_+}^2R_{N-2s_+}}}$ can be replaced by 1 if \eqn\boroots{r\leq
s_++s_--2=w-2} as then the terms obtained by expanding the square
root vanish too rapidly at infinity to contribute.  The number of
values of $r$ for which this inequality is obeyed is $s_++s_--1$;
given our assumption \inpo, this is at least $2s_-+1$.

For $r$ such that \boroots\ is obeyed, we have simply
\eqn\nordpar{t_r =\oint_\infty {dx\,x^r }\,{H_{s_+}R_{N-2s_+}\over
\tilde H_{s_-}}.} The rational function $H_{s_+}R_{N-2s_+}/\tilde
H_{s_-}$ has poles only at the $s_-$ zeroes of $\tilde H_{s_-}$,
so it has the form \eqn\toto{{H_{s_+}R_{N-2s_+}\over \tilde
H_{s_-}}=\sum_{i=1}^{s_-}{b_i\over x-a_i}+{\rm polynomial}} for
some complex numbers $a_i$, $b_j$. Now we evaluate \nordpar\ to
give \eqn\boto{t_r=\sum_{i=1}^{s_-} b_ia_i^r,} valid for $r\leq
w-2$.

This equation expresses the $s_++s_--1$ quantities $t_r$, $0\leq
r\leq s_++s_--2$, in terms of the $2s_-$ quantities $a_i$ and
$b_j$. So the $t_r$ cannot be independent; they will obey at least
$s_+-s_--2$ independent algebraic equations. Before working out
these equations in general, let us consider the first few cases.
If $s_-=0$, we have simply $t_r=0$, $0\leq r \leq s_++s_--2$.  If
$s_-=1$, we have $t_r=ba^r$ for some $a$ and $b$, whence
$t_it_j=t_kt_l$ whenever $i+j=k+l$ (and $i,j,k,l\leq w-2$).  The
first such relation (and the only example if $s_-=1$ and
$s_+=s_-+2=3$) is the equation $t_0t_2-t_1^2=0$. We can write this
equation as the vanishing of the determinant of the $2\times 2$
matrix \eqn\invo{M_2=\left(\matrix{ t_0 & t_1 \cr t_1 & t_2
}\right).} One way to prove that $\det \,M_2=0$ for $s_-=1$ is to
write \eqn\hinvo{M_2=\left(\matrix{1 & 1 \cr a_1 & a_2 \cr}\right)
\left(\matrix{ b_1 & 0 \cr 0 & b_2\cr}\right)\left(\matrix{1 & a_1
\cr 1 & a_2 \cr}\right),} where we have written a formula for
$M_2$ that is valid if $s_-=2$.  But if we set $s_-=1$, we should
take $b_2=a_2=0$ (and then $a_1=a, b_1=b$) and clearly $\det
\,M_2=0$ since the second factor in $M_2$ has vanishing
determinant.

In general, for any $n$, we set $M_n$ to be the matrix whose $i,j$
matrix element is $t_{i+j-2}$.  For example, the next case is
\eqn\pinvo{M_3=\left(\matrix{t_0 & t_1 & t_2 \cr t_1 & t_2 &
t_3\cr
 t_2 & t_3 & t_4 \cr}\right).}
If $s_-=n$, $M_n$ can be written as a product of three matrices,
namely \eqn\prodma{M_n=A_nB_nA_n^T} with
 \eqn\rodma{A_n=\left(\matrix{ 1 & 1 & \dots & 1\cr a_1 & a_2 &
 \dots & a_n\cr
   a_1^2 & a_2^2 & \dots & a_n^2\cr
   \vdots & \vdots &  & \cr
    a_1^{n-1}&a_2^{n-1} &\dots &a_{n}^{n-1}\cr}\right),}
and $B_n$ a diagonal matrix with eigenvalues $b_1,b_2,\dots
,b_n$. (Thus, the $i,j$ matrix element of $A_n$ is $a_j^{i-1}$.)
$A_n$ is a Vandermonde matrix, whose determinant is
\eqn\juno{\pm\prod_{i<j}(a_i-a_j).} The determinant of $M_n$ is
hence \eqn\muno{\det M_n=\prod_{i<j}(a_i-a_j)^2\prod_kb_k} and in
particular is generically nonzero for $s_-=n$.

$\det M_n$ depends on the $t_r$ with $r\leq 2n$.  Since our
derivation has assumed that $r\leq w-2$, the maximum value of $n$
that we should consider is \eqn\buno{n=\left[\half w - 1\right],}
where for any real $x$, $[x]$ denotes  the greatest integer less
than or equal to  $x$. In what follows we set $n$ to this value.
If $s_-<n$, the formula \prodma\ for $M_n$ remains valid, but we
should set all but $s_-$ of the $b_i$ (and $a_i$) to zero.  Hence
the matrix $B_n$, and consequently also $M_n$, has rank $s_-$.  It
follows that the rank $s_-+1$ minors of $M_n$ (the determinants of
$(s_-+1)\times (s_-+1)$ matrices obtained by omitting some rows
and columns from $M_n$) vanish. These are our relations.

Each minor is a homogeneous polynomial in the $t_i$ of degree
$s_-+1$.  For example, for $s_-=0$, the rank $s_-+1$ minors are
simply the $t_i$ themselves, and for $s_-=1$, they are the
quadratic functions $t_it_j-t_kt_l$ (for  $i+j=k+l$) whose
vanishing we found  earlier for $s_-=1$.

Since $M_n$ is an $(n+1)\times (n+1)$ matrix, it has minors of all
ranks up to $n+1$ (the rank $n+1$ minor being simply $\det M_n$).
The relations we have found are hence non-trivial whenever $n\geq
s_-$, or equivalently whenver $s_+\geq s_-+2$. For $s_+=s_-+2$, we
have $n=s_-$, and the unique relation of this type is $\det
M_{s_-}=0$.

 For any given
$s_-$, the rank $s_-$ minors do not generically vanish, since one
of them (the determinant of the upper left $s_-\times s_-$
submatrix of $M_n$) is $\det M_{s_--1}$, which is evaluated by
setting $n=s_-$ in \muno.  So a branch with given $s_-$, and
$s_+\geq s_-+2$, is characterized by saying that the rank $s_-$
minors of $M_n$ are generically nonvanishing, while the rank
$s_-+1$ minors vanish identically. We have accomplished our goal
of finding holomorphic functions of chiral order parameters  whose
vanishing on some branches and not others distinguishes the
different branches, at least modulo the exchange
$s_+\leftrightarrow s_-$.

\centerline{\bf Acknowledgements}

It is a pleasure to thank M.~Douglas for helpful
discussions.
This work was supported in part by DOE grant
\#DE-FG02-90ER40542
and NSF grant \#NSF-PHY-0070928 to IAS.

\appendix{A}{Generalization to Superpotentials of Arbitrary Degree.}

In section 2.2 we restricted the degree of the superpotential to
be less than $N+1$. The main reason was that only the first $N$
coordinates $u_i's$ are independent. In this section we will
generalize the discussion by allowing a superpotential of any
degree. The idea is to regard all $u_i$'s as independent
coordinates on a larger space subject to constraints.
Classically, a simple way of obtaining the constraints is by
looking at the generating function of $u_i's$,
\eqn\reon{ \Tr {1\over x-\Phi_{cl} } = \sum_{l=0}^\infty {\Tr\,
\Phi_{cl}^l\over x^{l+1}} =
 {N\over x} + \sum_{l=1}^\infty  {l u_l \over x^{l+1}} . }
Notice that this object is also written in terms of $P_N(x) =
\det (x-\Phi_{cl})$ which is a function only of $u_i's$ for $i=1,
\ldots ,N$ as follows,
\eqn\relt{  \Tr {1\over x-\Phi } = {P'_N(x)\over P_N(x)} =
{d\over dx}\log P(x). }
Integrating \relt\ and \reon\ with respect to $x$, we get,
\eqn\coge{ P(x) = x^N \exp \left( - \sum_{i=1}^{\infty} {u_i\over
x^i}\right) . }
where the integration constant was fixed by matching the $x^N$
terms.

The constraints we are after come from imposing \coge\ i.e, the
Laurent series has to terminate at order $x^0$. Since the
coefficient of $x^{-l}$ only depends on $u_i$ with $i\leq l$ and
is linear in $u_l$, imposing that to vanish gives a system of
equations for $u_l$ with $l\geq N+1$ in triangular form that can
be solved in terms of the first $N$ coordinates.

In writing the effective superpotential \const\ one has to use
$\langle \Tr\,\Phi^l \rangle$ instead of $\Tr\,\Phi^l_{cl}$. In
the first section we did not make any distinction since  $\langle
\Tr\,\Phi^l \rangle =\Tr\,\Phi^l_{cl}$ for $l\leq 2N-1$. For
$l\geq 2N$ the relation receives instanton corrections. This
implies that a modification to the constraints is needed. We now
turn to this issue.

Consider the quantum mechanical analog of \reon,
\eqn\reonq{\left\langle  \Tr {1\over x-\Phi } \right\rangle =
 {N\over x} + \sum_{l=1}^\infty  {l U_l\over x^{l+1}}  }
where we have denoted ${1\over l}\langle \Tr\, \Phi^l\rangle$ by
$U_l$.

The full quantum mechanical analog of \relt\ is,
\eqn\retq{\left\langle  \Tr {1\over x-\Phi } \right\rangle =
{P'_N(x)\over \sqrt{P_N^2(x)-4\Lambda^{2N}}} =
 {d\over dx}\log \left( P_N(x)+\sqrt{P_N^2(x)-4\Lambda^{2N}} \right) }

Integrating with respect to $x$ both \reonq\ and \retq\ we get,
\eqn\kjh{ {1\over 2}\left( P_N(x)+\sqrt{P_N^2(x)-4\Lambda^{2N}}
\right)  =
 x^N \exp \left( -\sum_{i=1}^{\infty} {U_i\over x^i}\right) }
where the constant of integration was determined to be $\half$ by
taking $\Lambda$ to zero.

Finally we can find the analog of \coge\ by solving for $P_N(x)$
in \kjh,
\eqn\kol{ P_N(x) =  x^N \exp \left( -\sum_{i=1}^{\infty}
{U_i\over x^i}\right) + {\Lambda^{2N}\over x^N}  \exp \left(
\sum_{i=1}^{\infty} {U_i\over x^i}\right). }
Now we are ready to write the generalization of \const\ for a
superpotential of any degree.
\eqn\supergen{ \eqalign{ W_{eff} = & \sum_{r=0}^{k}g_r U_{r+1} +
\oint V_{k-N}(x) \left(   x^N \exp \left( -\sum_{i=1}^{\infty}
{U_i\over x^i}\right) + {\Lambda^{2N}\over x^N}  \exp \left(
\sum_{i=1}^{\infty} {U_i\over x^i}\right)  \right) dx \cr & +
\sum_{i=1}^{N-n} \left( L_i(\oint {P_N(x)\over x-p_i}
 dx -2\epsilon_i \Lambda^{N})
 +  B_i\oint {P_N(x)\over (x-p_i)^2}dx \right) } }
where $V_{k-N}(x)$ is a polynomial of degree $k-N$ whose
coefficients are to be thought of as Lagrange multipliers imposing
constraints determining $U_l$ for $l=N+1,\ldots ,k+1$ in terms of
$U_l$ with $l=1,\ldots , N$ and $\Lambda$.

The next step is to follow the computations in section 2. The
derivative of $W_{eff}$ with respect to $U_{r+1}$ is
\eqn\finn{\eqalign{ g_r + & \oint {V_{k-N}(x)\over x^{r+1}} \left(
 - x^N \exp \left( -\sum_{i=1}^{\infty} {U_i\over x^i}\right) +
{\Lambda^{2N}\over x^N}  \exp \left( \sum_{i=1}^{\infty} {U_i\over
x^i}\right)  \right) dx \cr + & \oint {Q_{N-n-1}(x)\over
H_{N-n}(x)} {\del P_N(x) \over \del U_{r+1}} dx  = 0 \ .} }
Using \pnideni\ and the fact that in the relevant range of $r$ we
have $U_r=u_r$,
$${\del P_N(x)\over \del U_{r+1}} =  \cases{ -{P_N(x)\over
x^{r+1}} & $r+1\leq N$ \cr
 0 & $r+1 > N$ } . $$
However, inside the integral we can write ${\del P_N(x)\over \del
u_{r+1}} =  -{P_N(x)\over x^{r+1}}$ for any $r$ since for $r+1>N$
the integral vanishes.

In order to simplify \finn\ note that after imposing the
$V_{k-N}(x)$ constraints,
$$ x^N \exp \left( -\sum_{i=1}^{\infty} {U_i\over x^i}\right) =
 {1\over 2}\left( P_N(x) + \sqrt{P_N^2(x)-4\Lambda^{2N}} \right) + {\cal O}(x^{k+1}) .$$
It is easy to see that the $ {\cal O}(x^{k+1})$ terms do not
contribute to the integrals. Therefore, we can write the integral
in \finn\ that contains $V_{k-N}(x)$ as,
$$\eqalign{ & \oint  {V_{k-N}(x)\over x^{r+1}} \left( -{1\over
2}\left( P_N(x)+\sqrt{P^2_N(x)-4\Lambda^{2N}}\right) +
 {2\Lambda^{2N} \over P_N(x)+\sqrt{P^2_N(x)-4\Lambda^{2N}} } \right)dx = \cr
 - & \oint  {V_{k-N}(x)\over x^{r+1}}\sqrt{P^2_N(x)-4\Lambda^{2N}} dx .}
$$

Using these results we can multiply \finn\ by $z^r$ and sum over
$r$ from zero to $k$ to get,
$$ W'(z) = \oint  {V_{k-N}(x)\over
(x-z)}\sqrt{P^2_N(x)-4\Lambda^{2N}} dx + \oint {Q_{N-n-1}(x)\over
H_{N-n}(x)} {P_N(x) \over (x-z)} dx .$$

Finally, using \fact\ we can replace
$\sqrt{P^2_N(x)-4\Lambda^{2N}}$ by $\sqrt{F_{2n}(x)} H_{N-n}(x)$
in the first integral and $P_N(x)$ by $\sqrt{F_{2n}(x)}H_{N-n}(x)
+ {\cal O}(x^{-N})$ in the second. As in section 2.2, the ${\cal
O}(x^{-N})$ does not contribute to the integral. Note that had we
not included instanton corrections in the constraints, we would
have obtained $P_N(x)$ instead of $\sqrt{P^2_N(x)-4\Lambda^{2N}}$
in the first integral and the ${\cal O}(x^{-N})$ could not have
been dropped for $k\geq 2N$.

The final result is thus
\eqn\curm{  W'(z) = \oint \sqrt{F_{2n}(x)} \left( V_{k-N}(x)
H_{N-n}(x) + Q_{N-n-1}(x) \right) {dx\over (x-z)}. }
This agrees with the matrix model equation of motion \eqn\urm{
W'(z)=\oint{y_m(x)dx\over x-z}} if the matrix model curve is
$$ y^2_m(x) = F_{2n}(x){\tilde Q}_{k-n}^2(x) . $$ where ${\tilde
Q}_{k-n}(x) =V_{k-N}(x) H_{N-n}(x) + Q_{N-n-1}(x)$. Moreover,
\curm\ implies that $y^2_m$ is known up to a polynomial
$f_{k-1}(x)$ of degree $k-1$, i.e.
$$y^2_m  = F_{2n}(x){\tilde Q}_{k-n}^2(x) = W'^2_k(x) +
f_{k-1}(x),  $$
providing the generalization of \simp\ and the result \appfin\
discussed in section 2.2.

Let us consider some special cases:

\item{1.} No massless monopoles $n=N$: Then ${\tilde Q}_{k-N}(x)
=V_{k-N}(x)$.

\item{2.} Degree of superpotential equal to $N+1$, i.e.\ $k=N$:
${\tilde Q}_{k-n}(x) = V_{0} H_{N-n}(x) +  Q_{N-n-1}(x)$. In
particular, for $n=N$, i.e.\ $U(N)$ completely broken to $U(1)^N$,
${\tilde Q}_{0}$ is a constant and $y^2_m(x) = F_{2N}(x) =
P^2_N(x) -4\Lambda^{2N}$.

\appendix{B}{Proof of the Generalized Konishi
Anomaly from Strong Coupling Analysis.}

In this appendix, using the results from section 2.2 we show that
the generalized Konishi anomaly equation
\eqn\appkoni{ \left\langle \Tr {W'(\Phi)\over z-\Phi}
\right\rangle  = 2 R(z) \left\langle \Tr {1\over z-\Phi}
\right\rangle }
follows from the effective superpotential \const.  For simplicity
we will assume that the degree of $W$ is less than $2N+1$ such
that $\langle \Tr W'(\Phi) \rangle = \Tr W'(\Phi_{cl})  $.

Instead of viewing $u_r$ as the coordinates on the ${\cal N}=2$
Coulomb branch, we can use $\Phi_{cl}$ and mod out by $U(N)$. This
is valid except at points where some of the eigenvalues of
$\Phi_{cl}$ coincide. Then, instead of varying $W_{eff}$ with
respect to $u_r$ as in \Ucons, we vary with respect to $\phi_I$,
the eigenvalues of $\Phi_{cl}$ (recall that
$P_N(x)=\prod_I(x-\phi_I)$), to get
 \eqn\phiclvar{W'(\phi_I)=\sum_{i=1}^{N-n} L_i \oint {P_N(x) \over
 (x-\phi_I)(x-p_i)} dx}
 (We have used the result $B_i=0$ from section 2.2.)
{}From this we derive that with $z$ outside the contour of
integration
 \eqn\phiclvarz{\eqalign{
 \Tr {W'(\Phi_{cl}) \over
 z-\Phi_{cl}}=&\sum_{I=1}^N\sum_{i=1}^{N-n} L_i \oint {P_N(x) \over
 (z-\phi_I) (x-\phi_I)(x-p_i)} dx\cr
 =&\sum_{I=1}^N\sum_{i=1}^{N-n} L_i
 \oint {P_N(x) \over (z-x)(x-p_i) } \left({1 \over x-\phi_I }
 -{1 \over z-\phi_I } \right) dx\cr
=&\sum_{i=1}^{N-n} L_i
 \oint {P_N(x) \over (z-x)(x-p_i) } \left(\Tr {1 \over x-\Phi_{cl} }
 -\Tr {1 \over z-\Phi_{cl} } \right) dx\cr
 }}
Using \biva\ $P_N(p_i)\Tr {1 \over p_i-\Phi_{cl}} =0$, and the
fact that $P_N(\phi_I)=0$ the first term does not contribute. The
second term can be simplified using \Qdefi
 \eqn\phiclco{
 \Tr {W'(\Phi_{cl}) \over z-\Phi_{cl}}= \! - \Tr {1
 \over z-\Phi_{cl} }\sum_{i=1}^{N-n}\! L_i \oint\! {P_N(x) \over (z-x)
 (x-p_i) } dx=\! - \Tr {1
 \over z-\Phi_{cl} }\oint\! {P_N(x) Q_{k-n}(x) \over H_{N-n}(x) (z-x) }dx
 }
where we used manipulations as in \grs. It is important to keep
in mind that we can not replace $P_N(x)$ by
$\sqrt{F_{2n}(x)}H_{N-n}(x)$ as we did in \rec. The reason in that
the terms of order ${\cal O}(x^{-N})$ can not be dropped in the
integral. To see this, note that $|z| > |x|$ for any $x$ inside
the contour of integration. Therefore, $1/(z-x)$ can be expanded
as $(1/z)\sum_{l=0}^{\infty} (x/z)^l$.

A convenient way to deal with this is to write,
$$ \oint_{z\;out} {P_N(x) Q_{k-n}(x) \over H_{N-n}(x) (z-x) }dx =
\oint_{z\; in} {P_N(x) Q_{k-n}(x) \over H_{N-n}(x) (z-x) }dx
-\oint_{C_z} {P_N(x) Q_{k-n}(x) \over H_{N-n}(x) (z-x) }dx$$
where $C_z$ is a small contour around $z$. ``$out$'' and ``$in$''
refer to the point $x=z$ being outside or inside the contour of
integration. The first integral on the rhs gives $W'(z)$ as \grs\
indicates. The second one can be evaluated at the pole.

Using \fact\ to write,
$$ H_{N-n}(z) = { \sqrt{P^2_N(z)-4\Lambda^{2N}}\over
\sqrt{F_{2n}(z)}}. $$
and $y_m(z)$ from \ymmq, we get \phiclco\ to be,
 \eqn\hgs{  \Tr {W'(\Phi_{cl}) \over z-\Phi_{cl}}=\Tr {1
 \over z-\Phi_{cl} }\left( W'(z)  - y_m(z) {P_N(z)\over
 \sqrt{P^2_N(z)-4\Lambda^{2N}}} \right)}
The left hand side becomes a polynomial in $z$ when combined with
the term proportional to $W'(z)$ on the right hand side. The
right hand side can also be simplified by using that $\Tr {1
\over z-\Phi_{cl} }  ={P_N'(z) \over P_N(z)} $. \hgs\ is then,
\eqn\konn{\Tr {W'(\Phi_{cl}) -W'(z)  \over z-\Phi_{cl}}= y_m(z)
{P_N'(z)\over \sqrt{P^2_N(z)-4\Lambda^{2N}}}}
In the right hand side we use the relation in the chiral ring of
${\cal N}=2$
$$ {P_N'(z)\over \sqrt{P^2_N(z)-4\Lambda^{2N}}}  = \left\langle
\Tr {1\over z-\Phi} \right\rangle. $$
For the left hand side we use the fact that
 $$ \Tr {W'(\Phi_{cl}) -W'(z)  \over z-\Phi_{cl}} $$
contains only $\Tr\Phi_{cl}^l$ with $l\leq k$ which for $k<2N$ is
equal to $\langle \Tr \Phi^l \rangle$,
$$ \Tr {W'(\Phi_{cl}) -W'(z)  \over z-\Phi_{cl}} =
 \left\langle \Tr
 {W'(\Phi)-W'(z)\over z-\Phi} \right\rangle .$$
Finally, we can write \konn\ as follows,
\eqn\finn{ \left\langle \Tr {W'(\Phi)\over z-\Phi} \right\rangle
= \left\langle \Tr {1\over z-\Phi} \right\rangle (W'(z) - y_m(z))
}
By definition, the resolvent of the matrix model $R(z)$ satisfies
$2R(z) = W'(z) - y_m(z)$. Therefore \finn\ becomes the
generalized Konishi anomaly equation,
$$  \left\langle \Tr {W'(\Phi)\over z-\Phi} \right\rangle  = 2
R(z) \left\langle \Tr {1\over z-\Phi} \right\rangle$$

\appendix{C}{Chebyshev polynomials}

In this appendix we list the first six Chebyshev polynomials of
first and second kind. These are used in the examples discussed
in section 3.

The definition given in section 2.3 of Chebyshev polynomials of
the first and second kind of degree $t$ and $t-1$ respectively is
the following. Set $x=\cos\theta$ and let
\eqn\chb{ {\cal T}_t(x) =\cos (t\theta )  \qquad  {\cal
U}_{t-1}(x) ={1\over t} {d{\cal T}_t\over dx}(x) = {\sin
(t\theta)\over \sin \theta} . }
{}From this it is simple to compute the first of then,

 $$\vbox{\settabs
 \+   $6 \qquad $ &   $32 x^6 -48x^4 +18 x^2 -1 \qquad$& $64 x^6 -
 80 x^4 +  24 x^2 -1 $\cr
 \+    $t$  & $ {\cal T}_t(x)  $ &   ${\cal U}_t(x) $\cr
  \+   $1$  &   $x$           &    $ 2x$  \cr
  \+   $2 $ &   $2x^2-1 $    &    $4 x^2-1$  \cr
  \+   $3$  &   $4 x^3-3 x $   &  $ 8 x^3-4 x $  \cr
  \+   $4 $ &   $8 x^4-8 x^2+1 $ & $ 16 x^4-12 x^2-1$  \cr
  \+   $5 $ &   $16 x^5 - 20 x^3 -5 x $  & $ 32 x^5 -32 x^3 + 6x
  $\cr
  \+   $6 $ &   $32 x^6 -48x^4 +18 x^2 -1 $& $64 x^6 - 80 x^4 + 24
  x^2 -1 $\cr} $$

\appendix{D}{A Magnetic Index}

Here we want to explain an interesting result that was obtained
in an unsuccessful attempt to find a new order parameter for
these models.  The attempt was unsuccessful because the object $
\nu$ that we defined turned out to be computable in terms of
objects that are already known.

We will examine closely the physical spectrum of particles in a
theory in which classically $U(N)$ is broken to a low energy
subgroup such as  $U(N_1)\times U(N_2)$. (Eventually we will
generalize to an arbitrary breaking pattern
$U(N)\to\prod_iU(N_i)$.)  Allowing for the strong quantum
dynamics, the true low energy gauge group is $U(1)\times U(1)$.
One $U(1)$ is the center of $U(N)$ and decouples from the
theory.  We will disregard it and focus on the $U(1)$ that is not
decoupled, whose generator we call $Q$.

Physical particles in the theory are in general electrically and
magnetically charged with respect to this $U(1)$.  The product of
the smallest nonzero electric charge $q$ (of an unconfined
magnetically neutral particle) times the smallest physical
magnetic charge $m$ is, according to Dirac, an integer $\nu$. (We
measure $m$ as a multiple of the Dirac quantum $2\pi \hbar c$.) We
will call this integer the magnetic index, and we claim that
always \eqn\togo{\nu t=N,} where $t$ is the confinement index
defined in section 2.1.

First let us verify this in examples.  We consider first the case
of $U(4)$ broken to $U(3)\times U(1)$.  We embed $U(3)$ as the
upper left $3\times 3$ block in $U(4)$.  For the monopole, we
consider an 't Hooft-Polyakov monopole embedded in $U(4)$ in such
a way that (in a unitary gauge) the magnetic charge is
\eqn\uttu{\left(\matrix{0 &   &   & \cr
                         & 0 &   & \cr
                         &   &  1& \cr
                         &    &  & -1 \cr}\right),}
where in other words the ``1'' acts in $U(3)$ and the ``$-1$'' in
$U(1)$.  To find the smallest electric charge that is not
confined, we start with the massive $W$-bosons, which are a
$U(3)$ triplet embedded as follows in $U(4)$:
\eqn\bbutu{\left(\matrix{0 & & & W^1\cr & 0 & & W^2\cr & & 0 &
W^3\cr & & & 0 \cr}\right).} The components $W^1$, $W^2$, and
$W^3$ have, respectively, electric charges $1,1,$ and $2$, with
respect to the electric charge generator given in \uttu.\foot{In
evaluating the electric charge of an $SU(3)$ singlet, we can
simply use the charge matrix that appears in \uttu;  subtracting
an $SU(3)$ generator to get a multiple of $Q$ would not change
the result. } The $SU(3)$ singlet combination
$\epsilon_{ijk}W^iW^jW^k$ is unconfined.  Its electric charge is
$q=1+1+2=4$.  Since the minimum magnetic charge $m$ in this
theory is 1 (in Dirac units), the product $\nu=qm$ equals 4.  The
breaking of $U(4)$ to $U(3)\times U(1)$ gives a Coulomb branch,
so $t=1$, and $\nu t = 4$, as claimed in \togo.

Now we consider the breaking of $U(4)$ to $U(2)\times U(2)$.
 We embed the two factors of $U(2)$ as the upper
left and lower right blocks in $U(4)$, and the monopole charge as
\eqn\uttu{\left(\matrix{0 &   &   & \cr
                         & 1 &   & \cr
                         &   &  -1& \cr
                         &    &  & 0 \cr}\right).}
We focus on the upper right $2\times 2$ block of $W$ bosons,
transforming as $({\bf 2},{\bf \overline 2})$ of $U(2)\times
U(2)$: \eqn\nuttu{\left(\matrix{W^{11} & W^{12}\cr W^{21} &
W^{22}\cr}\right).} The components $(W^{11},W^{21},W^{12},W^{22})$
have (with respect to the charge generator in \uttu) respectively
electric charges $(1,2,0,1)$, and the basic $SU(2)\times SU(2)$
singlet
$\epsilon_{ij}\epsilon_{i'j'}W^{ii'}W^{jj'}=W^{11}W^{22}-W^{12}W^{21}$
hence has charge $q=2$.  In the confining branch, with $b=0$ in
the notation of section 2.1, the confining index is $t=2$, the
minimum magnetic charge is again $m=1$, and $\nu=qm=2$. So again
$\nu t=4$, as claimed.

What, however, happens in the branch with classical breaking to
$U(2)\times U(2)$ and with $b=1$?  In this case, there is no
confinement, and $t=1$.  What about $\nu$?  Shifting the value of
$b$ changes us from confinement to oblique confinement in one of
the $SU(2)$'s, but this does not affect which combinations of $W$
bosons are confined.  So the minimum electric charge, in the same
units, is still $q=2$.  However, in section 2.1, we described a
process of magnetic screening that occurs in this model when an
external Wilson loop in the fundamental representation of $SU(4)$
is considered; it is screened by the nucleation of an 't
Hooft-Polyakov monopole with unit charge, and this is the
mechanism that leads to $t=1$. The fact that the 't
Hooft-Polyakov monopole screens an external Wilson line means
that, in the absence of the Wilson line, it would have infinite
energy and would be confined (we explain the mechanism in more
detail below).  So the minimum unconfined magnetic charge when
$b=1$ actually has $m=2$.  Since $q$ is still $2$, this gives
$\nu=4$, which, with $t=1$, is again consistent with $\nu t=4$.

At least in hindsight, it should not come as too much of a
surprise that we had to consider  here the confinement of certain
magnetic monopoles. We have taken account of electric confinement
to evaluate $q$, so we should expect to consider magnetic
confinement, that is confinement of some magnetic monopoles, to
evaluate $m$.

More directly, the confinement for $b=1$ of the minimum charge
monopole arises as follows.  The monopole has a collective
coordinate whose quantization leads to the Julia-Zee dyon.  There
is only a single collective coordinate, and quantizing it gives
the monopole  equal color charges in the two
$SU(2)$'s.\foot{Global $SU(2)$ charge is not defined in the field
of a magnetic monopole \vari. What we here call the color
electric charge (for a given $SU(2)$) is the quantity that is
well-defined.  It originates from the projection of the charge
generator in $\uttu$ into the $SU(2)$ of interest; this charge is
an integer for $SU(2)$ generators, and would be a half-integer
for external quarks.} In $SU(2)$ gauge theory, ordinary
confinement means that a monopole is unconfined if its color
electric charge is an integer multiple of the $W$ boson color
electric charge, and oblique confinement means that the color
electric charge of an unconfined monopole is a half-integral
multiple of the $W$ boson color electric charge. Setting $b=1$
means that one $SU(2)$ has ordinary confinement and the other has
oblique confinement. Given this, the states obtained by
quantizing the collective coordinate are all confined, as we have
claimed.

\bigskip\noindent{\it Proof Of The Relation}

Now that we have given an idea of what the relation $\nu t=N$
means semiclassically, we proceed to a mathematical proof of this
relation.  In the process, we will also define the index $\nu$
for arbitrary rank of the low energy group; above, we assumed
rank 1.

We start with the ${\cal N}=2$ Coulomb branch.  The low energy
gauge group (omitting the center of $U(N)$) is $U(1)^{N-1}$.  The
massive gauge bosons and monopoles have charges that fill out a
rank $2(N-1)$ lattice that we will call $L$.  On $L$ there is an
integer-valued antisymmetric inner product that we will call
$\omega$.  For example, if $N-1=1$, there is a single $U(1)$ and a
vector in $L$ is specified by giving electric and magnetic charges
$(e,g)$. Given two vectors $l=(e,g)$ and $l'=(e',g')$, the inner
product is then $\omega(l,l')=eg'-e'g$. In general, for any
number of $U(1)$'s, $\omega$ is defined by a similar formula,
summing over all of the $U(1)$'s. The Pfaffian of $\omega$ is in
absolute value \eqn\ugu{|{\rm Pfaff}(\omega)|=N.} This relation
reflects the fact that if we incorporate electric charges in the
fundamental representation of $SU(N)$, then $\omega$ would become
unimodular.

Now, after perturbing the theory by a superpotential (or in any
other way, for that matter), some charges may condense.  The
condensed charges generate a sublattice $M$ of $L$.  The
condensed objects are always ``mutually local,'' in the sense that
for $m,m'\in M$, \eqn\guxo{\omega(m,m')=0.} A particle is
confined if its charge vector $l$ is such that $\omega(l,m)\not=
0$ for some $m\in M$.  So the charge vectors of unconfined
charges lie in $M^\perp$, the sublattice of $L$ consisting of $l$
such that $\omega(l,m)=0$ for all $m\in M$. Physically, the
charge of an unconfined particle can only be measured modulo $M$,
since particles with quantum numbers in $M$ have condensed.  So
the measureable charges in the low energy theory take values in
$L'=M^\perp/M$.  The antisymmetric product $\omega$ induces an
analogous antisymmetric product $\omega'$ for the physical,
unconfined charges in the low energy theory.  The general
definition of $\nu$ is \eqn\edo{\nu=|{\rm Pfaff}(\omega')|.}
$\nu$ determines how many extra types of charge could be added to
the low energy theory (keeping fixed its gauge group) without any
inconsistency.

$L'$ is the lattice of charges that are not confined, and hence
the quotient $L''=L/L'$ might be regarded as the lattice of
charges that are confined.  There is a well-defined inner product
$\omega'':M\times L''\to {\bf Z}$; for any $m\in M$, $l''\in
L''$, we simply lift $l''$ to an element  $l\in L$ in an
arbitrary fashion, and define $\omega''(m,l'')=\omega(m,l)$.  The
choice of $l$ is well-defined modulo an element of $M^\perp $;
adding such an element to $l$ does not change $\omega(m,l)$.  We
can extend $\omega''$ to an antisymmetric form on $M\oplus L''$ by
defining $\omega''(l'',m)=-\omega(m,l'')$.  The confinement index
$t$ is $t={\rm Pfaff}(\omega'')$.  In fact, the confinement index
measures how many types of charge are conceivable (they have
integer inner products with vectors in $M$) but are absent in the
theory and can be represented instead by external Wilson loops.

Now it is a fact of linear algebra that in this situation
\eqn\yogo{{\rm Pfaff}(\omega)= {\rm Pfaff}(\omega')\cdot {\rm
Pfaff}(\omega'').}  Putting all of this together, we get the
promised relation $N=\nu t$.

If we are willing to make some unnatural choices, we can define
the lattices and explain the key fact \yogo\ in a way that might
be easier to follow. There is not any natural way to embed $L'$
as a sublattice of $L$, because vectors in $L'$ can only be
interpreted as elements of $L$ modulo $M$. However, this
indeterminacy is unimportant for us; we simply embed $L'$ as a
sublattice $L_1\subset L$ by picking a basis of $L'$ and lifting
each basis vector to a vector in $L$, making a choice that (for
each basis vector) is unique up to adding an element of $M$. If
we do this, then $\omega'$ can be identified with $\omega_1$, the
restriction of $\omega$ to $L_1$.

Having identified $L_1$ as a sublattice of $L$, we can without
further arbitrary choices  define another sublattice
$L_2=L_1^\perp$; that is, $L_2$ consists of elements $l_2\in L$
such that $\omega(l_1,l_2)=0$ for all $l_1\in L_1$.  With respect
to the decomposition $L=L_1\oplus L_2$, $\omega$ is
block-diagonal, \eqn\unto{\omega=\left(\matrix{\omega_1 & 0 \cr 0
& \omega_2\cr}\right).} From this, \yogo\ is clear.

 \listrefs
\end